\documentstyle[psfig]{report} 
 
\def\dd{\mbox{d}}
\def\ddd{\mbox{\sm d}}
\def\O{\Omega}  
\def\o{\omega} 
\def\bra{\langle}  
\def\ket{\rangle}  
\def\a{\alpha}  
\def\b{\beta} 
\def\d{\delta}          
\def\D{\Delta}           
\def\LL{\triangle} 
\def\g{\gamma}           
\def\G{\Gamma}  
\def\e{\epsilon} 
\def\ve{\varepsilon}  
\def\et{\eta}  
\def\f{\phi}
\def\F{\Phi}
\def\vf{\varphi}  
\def\k{\kappa} 
\def\l{\lambda}         
\def\L{\Lambda}  
\def\m{\mu} 
\def\n{\nu}  
\def\s{\sigma}  

\def\o{\omega} 
\def\p{\pi}
\def\r{\rho} 
\def\t{\tau}
\def\th{\theta} 
  
\def\ra{\rightarrow}  
 
\def\pa{\partial}  
  
\def\Pl{\ell_{\sm{Pl}}} 
\newcommand{\ti}[1]{\tilde{#1}} 
 
\newcommand{\sm}[1]{\mbox{\scriptsize #1}}  
\newcommand{\tn}[1]{\mbox{\tiny #1}} 
\renewcommand{\@}[1]{\sqrt{#1}} 
\newcommand{\Tr}{{\mbox{Tr}}\,} 
\def\be{\begin{eqnarray}} 
\renewcommand{\le}[1]{\label{#1}\end{eqnarray}}  
\def\eea{\end{eqnarray}} 
\def\ee{\end{eqnarray}}
\newcommand{\eq}[1]{(\ref{#1})}  
\def\nn{\nonumber\\}

\def\ffract#1#2{\raise .35 em\hbox{$\scriptstyle#1$}\kern-.25em/ 
\kern-.2em\lower .22 em \hbox{$\scriptstyle#2$}} 
\def\dts{\d(\ti\s-\ti\s')}  
\def\ts{(\ti\s-\ti\s')} 
 
\def\fts{f(\ti\s-\ti\s')} 
\def\tx{(\ti x-\ti x')}

\def\GN{G_{\mbox{\tn N}}}

\def\Ric{{\mbox{Ric}}}

\def\na{\nabla}

\def\half{{1 \over 2}}
\def\in{{\sm{in}}}
\def\out{{\sm{out}}}
\def\sin{{\tn{in}}}
\def\sout{{\tn{out}}}

\def\>{\rangle} 
\def\<{\langle} 
\def\nonu{\nonumber \\{}}
\def\ca{{\cal A}}
\def\gzero{g_{(0)}}

\def\gi{g_{(0)}^{-1}}
\newcommand{\tnnn}[1]{\mbox{\tiny #1}} 
\def\bea{\begin{eqnarray}}

\begin{document}
\rm\large
\pagestyle{empty}
\pagenumbering{roman}

\begin{center}
{\huge
\setlength\baselineskip{30pt}
{\bf Quantum Gravity \\
and the\\
Holographic Principle}
\par}
\vspace{2.5\baselineskip}
{\Large
Quantumgravitatie en het holografische beginsel
\par}
\vspace{3\baselineskip}
(met een samenvatting in het Nederlands)

\vspace{7\baselineskip}

{\large PROEFSCHRIFT}
\\[2\baselineskip]

{\small
ter verkrijging van de graad van doctor aan de 
Universiteit Utrecht, op gezag van de Rector 
Magnificus, Prof. dr. W.H. Gispen, ingevolge 
het besluit van het College voor Promoties 
in het openbaar te verdedigen op maandag 
18 juni 2001 des namiddags te 16.15 uur}

\vspace{4\baselineskip}

door
\\[2.5\baselineskip]
{\Large Sebastian de Haro Oll\'{e}}
\\[2\baselineskip]
geboren op 23 november 1973 te Barcelona

\end{center}

\clearpage
\newpage
\large
{\setlength\tabcolsep{15pt}
\noindent
\begin{tabular}{@{}ll}
Promotor:
 &  Prof.\,Dr.~G.\,'t Hooft\\
\\ 
 & Spinoza Instituut en\\
 & Instituut voor Theoretische Fysica\\
 & Universiteit Utrecht\\
\end{tabular}

\vfill
\noindent
\newpage
$$ $$
$$ $$
$$ $$
$$ $$
$$ $$
$$ $$
$$ $$
$$ $$
$$ $$
$$ $$
$$ $$
$$ $$
$$ $$
$$ $$
$$ $$
$$ $$
$$ $$
$$ $$
$$ $$
$$ $$
$$ $$
$$ $$
$$ $$
\rightline{\it Politics is for the moment. An equation is for eternity}
\rightline{Albert Einstein}
$$ $$
$$ $$
\rightline{\it Voetbal is simpel. Het is echter moeilijk om simpel te 
voetballen}
\rightline{Johan Cruijff}

\newpage
\large
\pagestyle{plain}
\tableofcontents
\newpage
\newpage
$$ $$

\newpage
\pagenumbering{arabic}

\chapter{Introduction}\label{Intro}

Ever since in 1974 Hawking discovered that black holes emit radiation 
\cite{Hawking}, there has been great controversy about the fact that black 
holes can evaporate, and about their fate after have they have done so. 
Indeed, as is well-known, pairs of particles and anti-particles can form in 
vacuum. These particles however tend to recombine and, under usual 
circumstances, they will annihilate each other after a very short time. 
However, when these pairs form in the vicinity of a black hole, there is a 
small chance for the particle to have just enough energy to escape to 
infinity, whereas its partner with negative energy is doomed to fall into the 
black hole. Obviously this is a small effect, as the probability for a 
Hawking particle to have enough energy to escape to infinity, where we can 
measure it, is extremely small. Yet the mere idea that such a process is 
possible is a great challenge for theoretical physics, for it raises the 
question what would happen if we were able to isolate a black hole (in our 
minds) so that nothing falls in but it only can emit particles and hence 
evaporate. In fact, small black holes will evaporate very fast, as the 
temperature of Hawking radiation is inversely proportional to the mass:
\be
T={\hbar c^3\over8\p k\GN M}=6\cdot 10^{-8} (M_\odot/M)\,\mbox{K},
\ee
where $M_\odot$ is the solar mass. For a black hole as heavy as the sun this 
is a very tiny effect, but small black holes, like the ones that were formed 
in the early universe, may have a mass small enough to emit strong radiation.

The controversy we alluded to above is not so much concerned with the fact 
that black holes emit radiation, but rather with the nature of the radiation: 
it is purely thermal. Its spectrum is that of black-body radiation, which 
means that it contains little information about the initial state of the 
black hole. Take for example a page of this thesis and burn it (this is just 
a thought experiment). After the paper is completely burned, all the precious 
information that was in it is lost. A close look at the few ashes that are 
left behind or an analysis of the radiation that is emitted will not help us 
puzzling out what was written on the paper. We can recover a great deal of 
the information about it --- its chemical composition, etc.---, but not the 
detailed information about how molecules were precisely arranged on the 
surface of the sheet.

With black holes the situation is very similar. The Hawking radiation that is 
emitted is coarse-grained, it does not contain precise information about for 
example how the black hole exactly formed and all its past history. This is 
why information is lost in the process of evaporation.

Yet for the burned page we know this is not completely true: if we were able 
to keep track of each single molecule after the page is burned, applying the 
laws of physics (and chemistry) we would be able to give their precise 
configuration when the thesis was still intact, and so we would succeed at 
recovering the lost information. This is nothing else than the statement that 
thermodynamics can be derived from microscopic physics by a coarse-graining 
procedure. In quantum mechanical terms, if the final state is pure, the 
initial state must be pure as well unless there is a violation of quantum 
mechanics. Now if Hawking's argument is correct, black holes violate quantum 
mechanics, as their final state is mixed and not pure.

One can hardly overestimate the importance of Hawking's paradox for our 
understanding of nature. If true, it points to a fundamental discrepancy 
between general relativity and quantum mechanics, and so it is extremely 
important to find out whether there is a mistake in the formulation of 
Hawking's argument, or whether we have to change the fundamental laws of 
physics by allowing quantum mechanics to be even less ``classical". Indeed, 
conventional quantum mechanics already leads to conceptual difficulties, but 
a theory where transitions between pure and mixed states are allowed would be 
even less transparent and, what is worse, it would be very unlikely to 
respect basic principles of physics like energy conservation.

A few years before Hawking radiation was discovered, Jacob Bekenstein 
developed the laws of black hole thermodynamics, based on the analogy between 
black hole mechanics and thermodynamics \cite{Bekenstein}. He argued that, up 
to a constant, the entropy of a black hole must be proportional to its area:
\be
S_{\tn{BH}}={kc^3A\over4G\hbar},
\le{BHentropy}
and the precise proportionality factor of $1/4$ was only determined when 
Hawking radiation was discovered. Based on the analogy with statistical 
mechanics, this suggests that the area of the black hole is a measure for the 
number of microscopical states that give rise to the same macroscopic black 
hole of mass $M$, charge $Q$ and angular momentum $J$.

In string theory, several microscopic countings have been made that confirm 
the area-law \eq{BHentropy} with the right proportionality coefficient. 
Though performed for so-called extremal and near-extremal black holes, which 
are presumably not of much astrophysical relevance, these countings give, for 
the first time, a microscopical explanation of the black hole entropy 
formula.

Motivated by the above relation between entropy and area, in 1993 't Hooft 
conjectured that at Planckian energies our world is not three-, but 
two-dimensional \cite{ghologr}. The argument, in simplified form, was as 
follows. Consider a closed region of space-time of volume $V\sim R^3$ and 
energy $E$ and ask how many physical states there are in this region. For the 
states to be physical and thus measurable for an outside observer, we must 
require that the radial size of the region we consider is larger than the 
size of its Schwarzschild radius. Otherwise the surface would lie within its 
own horizon and would be hidden to the observer outside. Since the 
Schwarzschild radius is given by the energy inside, we get the bound
\be
2E<R
\le{Schwrad}
i.e. the Schwarzschild radius should always be smaller than the actual 
radius, and so the energy density inside the volume is not allowed to be too 
large.

Given ordinary quantum field theory, the most probable state would be a gas 
at some temperature $T$. Its energy would be given by Boltzmann's law,
\be
E\sim VT^4.
\le{energy}
In what follows we suppress all multiplicative constants of order 1. The 
total entropy is
\be
S\sim VT^3,
\le{VT3}
and so combining \eq{Schwrad} with \eq{energy} one gets a bound on the 
temperature. This gives, for the entropy,
\be
S< V^{1\over2} \sim A^{3\over4},
\le{tHbound}
which for large area does not exceed the entropy of a black hole of the same 
size. Thus, black holes have the largest entropy ordinary matter can possibly 
have. In fact, they have a larger entropy than what is suggested by the 
stronger bound \eq{tHbound}. This is not surprising, as any form of matter 
will form a black hole if we increase its energy density more and more. 

What is surprising is that the limit on the entropy is set by the area, 
\eq{BHentropy}, and not by the volume. 't Hooft's explanation was that most 
of the states of field theory are not physical, for their energy is so large 
that they are confined inside their own Schwarzschild radius. So, the 
expectation is that gravitational physics reduces the number of physical 
degrees of freedom: states with energy corresponding to a Schwarzschild size 
larger than the size of the physical system are not physical and so should be 
disregarded, hence the number of states grows exponentially with the area 
instead of the volume. It was then conjectured that quantum gravity should be 
described by a topological field theory, in the sense that all its degrees of 
freedom live on the boundary. This is called the {\it holographic 
hypothesis}.

There have been various generalisations of the holographic principle which we 
will not go in detail into, as in this thesis we will only consider the, from 
a geometrical point of view, most simple cases. In general, one has to define 
the boundary of a certain region, and its inside and outside. This can be 
done by looking at the propagation of light rays from a certain region 
\cite{Bousso}.

Much progress in the understanding of the holographic principle came from 
very different considerations when in 1997 Maldacena conjectured the 
so-called AdS/CFT correspondence \cite{Malda}. The AdS/CFT correspondence 
goes back to the long-ago conjectured relationship between gauge theories and 
strings \cite{glargeN}. It relates string and gravity theories in a certain 
back-ground (so-called ``anti-de Sitter", AdS for short) to certain field 
theories which do not contain gravity (CFT stands for ``conformal field 
theory"). AdS space is a space with a timelike boundary, and in this sense it 
can be compared with a ``box" (one can think of it as a cylinder of circular 
base). The field theory is defined at the boundary of the space, which 
corresponds to the wall of the cylinder. Thus, the field theory lives in a 
space of one dimension less. The AdS/CFT correspondence thus gives a simple 
realisation of the holographic principle: the gravitational degrees of 
freedom in the bulk can be arranged in such a way that they describe a 
non-gravitational theory living on the boundary of the space.

The holographic principle is not only a statement about the number of 
microstates of the theory. It also implicitly assumes that these degrees of 
freedom reorganise on the boundary in a somehow physically meaningful way. 
This implies that the boundary theory should at least respect causality. The 
AdS/CFT correspondence is a nice arena to perform tests of causality, and in 
fact some non-trivial tests have been performed with black holes and 
collisions between massless particles. Although some bizarre behaviour has 
been found \cite{PST,SuTo,LoTh} from the boundary point of view, so far no 
contradictions have been perceived with the causality principles of quantum 
field theory. Perhaps even more surprising than the fact that the theory 
lives on the boundary is the fact that the AdS/CFT correspondence relates 
bulk gravity to one of the field theories that were already known.

In this thesis we are mainly concerned with two different approaches to 
holography. The first one is an analysis of the eikonal regime of quantum 
gravity, where the theory reduces to a topological field theory. This is the 
regime where particles interact at high energies but with small momentum 
transfer. We also consider quantum gravity away from the extreme eikonal 
limit and find indications that the theory remains topological. The second 
approach we pursue is the AdS/CFT correspondence, where one can ask very 
precise questions about the way the geometry of the bulk and the matter 
fields are encoded in the boundary theory. We also study warped 
compactifications, where our $d$-dimensional world is regarded as a slice of 
a $d+1$-dimensional space-time, and analyse in detail the question as to 
where the $d$-dimensional observer can find the information about the extra 
dimension. Much of what we do does not assume string theory directly, 
although most of our results can be embedded in string theory, and in fact we 
think string theory is probably the best way to understand and think about 
our results. In particular, the discussion of the AdS/CFT correspondence does 
assume string theory. Even though in this thesis we investigate two 
apparently very different approaches, our aim is in fact to apply them to 
situations where both can be used. In this way we are naturally led to 
considering Planckian scattering in AdS. This will be studied in chapter 
\ref{GSM}, where we make a few preliminary remarks about the relation between 
both.

The thesis is organised as follows. The first chapter is introductory: we 
first explain the sorts of problems related to black holes and Hawking 
radiation which motivate this work. We explain why the assumption of the 
holographic principle can be a way to solve them. Then we review the features 
of quantum gravity in the eikonal regime, string theory and the AdS/CFT 
correspondence. Particular emphasis is laid on how holography arises in the 
context of quantum gravity and of the AdS/CFT correspondence. In chapter 
\ref{HEscattering} we study in detail high-energy scattering between massless 
particles: classical and quantum mechanical features of gravitational 
scattering, and how to go beyond the eikonal approximation. In chapter 
\ref{GSM} we generalise some of these results to spaces with a cosmological 
constant (positive or negative) and find the corresponding dual theories. A 
particularly interesting case is that of a positive cosmological constant. We 
believe our results are relevant to the discussions in \cite{Bousso,HKS,FKMP} 
on the possibility of describing holographic duals of de Sitter space. In 
chapter \ref{reconstruction} we study the reconstruction of space-time and of 
space-time fields from the CFT. We do this perturbatively in the distance to 
the boundary. We develop a systematic method to regularise and renormalise 
the bulk action, and interpret our results from the CFT point of view. In 
chapter \ref{warped} we reinterpret the counter-terms of the gravitational 
action as generating the dynamics from the point of view of an observer 
living on a brane of codimension 1. We analyse the cases of asymptotically 
AdS, dS and flat space-time.

\section{Holography in Quantum Gravity}\label{QG}

The most clear and astonishing example of a holographic map between a 
gravitational and a non-gravitational theory is perhaps the AdS/CFT 
correspondence. It remains very mysterious, however, how holography may work 
if the bulk space-time is not AdS but asymptotically flat. In particular, a 
satisfactory description of the four-dimensional Schwarzschild black hole is 
still lacking.

As a matter of fact there exists a holographic description, if not of an 
evaporating Schwarzschild black hole, of a Rindler space-based model that is 
to mimic the most important features of the near-horizon region of the 
four-dimensional black hole. This is the S-matrix description discussed by 't 
Hooft \cite{g9607}, which we are going to examine in detail in this thesis. 
However, even if this is truly a holographic model, the quantum mechanical 
properties of the model are not well understood beyond the eikonal 
approximation, and no entropy formula has been derived. Nevertheless, it is 
quite remarkable that this model does exhibit explicitly how the information 
that falls into the black hole is stored into  the outgoing radiation without 
violating any no-quantum-copying-machine  principle. In particular, one can 
compute an approximated S-matrix. It furthermore has a striking similarity 
with string theories  and non-commutative geometry. It also gives interesting 
insights in the non-perturbative regime of quantum gravity in the eikonal 
approximation. For these reasons, we think that the model is worth studying, 
the more because it is applicable in the context of AdS where we also have a 
dual CFT description. It would be extremely interesting if one could 
``compare" both holographic duals, and we will make a few preliminary remarks 
in that direction. It is clear that a cross-fertilisation between the 
S-matrix model, where the issue of unitarity is exhibited explicitly, and the 
CFT description, for which there exists an extraordinarily precise 
dictionary, is most desirable (for a discussion of the issue, see, e.g., 
\cite{LoTh,SussS-m,PolS-m}).

The next sections are an introduction to some aspects of the eikonal regime 
of quantum gravity, first in the specific context of point-like particles on 
a fixed background, and later in a more general set-up. We review in 
particular how holography arises in the context of quantum gravity. There are 
many other relevant papers on the subject (see, e.g., \cite{KO1,KO2}), but 
for the purpose of this thesis we restrict ourselves to the ones that will be 
used in later sections.

\subsection{Quantum Gravity in the Eikonal Regime}\label{eikonal}

The main ingredient of the S-matrix Ansatz is the gravitational interactions 
between in-going particles and out-coming radiation on a black hole horizon. 
These interactions are not taken care of in the derivation of Hawking 
radiation, and because of the extreme high frequencies of the in-falling 
modes these interactions cannot be neglected.

If quantum gravity would be non-predictable in the way originally discussed 
by Hawking, we would have to enlarge the uncertainty in quantum mechanics to 
allow for an uncertainty in the state of the wave-function: on top of the 
statistical description of observables postulated by quantum mechanics, there 
would be an uncertainty in the quantum state \cite{Hawking2}. However, there 
are strong reasons to believe that gravity can be reconciled with quantum 
mechanics without giving up unitarity. String theory, and in particular the 
AdS/CFT correspondence, supports such a view. Nevertheless it is important 
for the understanding of quantum gravity to be able to point to a loophole in 
the original argument. Although the contents of this section have already 
been discussed at length in \cite{g9607} and other publications by 't Hooft, 
we will review the S-matrix Ansatz once more because it is the starting point 
for other considerations in the next chapters. 

The basic idea is to take into account the fact that in-going and out-coming 
particles interact gravitationally at the horizon. If the black hole was 
formed by some in-falling matter configuration, there will be traces of its 
initial state on the geometry near the horizon, and so, when Hawking 
radiation is emitted, it will be scattered off that non-trivial surrounding 
geometry. 

Consider a Schwarzschild black hole in a typical state, say a superposition 
of in-going and out-going particles. States for the Schwarzschild (Rindler) 
observer are related to the Kruskal (Minkowski) vacuum by the well-known 
Bogolyubov transformation,
\be
a_{\ti k,\o}&=&{1\over\@{1-e^{-2\pi\o}}}\left[b_{\ti k,\o} 
+e^{-\pi\o}b^\dagger_{-\ti k,-\o}\right].
\le{bogolyubov}
The operator $a$ annihilates a particle of energy $\o$ and momentum $\ti k$ 
in Rindler space, whereas $b$ is directly related to the annihilation 
operator in Minkowski space. One can easily check that this mixing between 
creation and annihilation operators gives rise to the following relation 
between states:
\be 
|0\ket_{\sm{M}}=\prod_{\ti k,\o}\@{1-e^{-2\pi\o}}\sum_{n=0}^\infty 
e^{-n\pi\o}|n,n\ket_{\ti k,\o}.
\le{vacuum} 
Consider now the process of ``purifying" such a state by removing first one 
particle and subsequently all the others, until we are left with the vacuum. 
For the Kruskal observer, more and more particles are being added to his 
state, with such tremendous energies that they will interact gravitationally, 
eventually forming small and even big black holes. It is clear that in such a 
situation the Bologyubov transformation \eq{bogolyubov} will not be correct, 
as gravitational interactions were neglected in its derivation. So, for the 
Kruskal observer, the vacuum of the Rindler observer is not at all a vacuum 
state nor a thermal bath of particles, but it will rather be a highly 
complicated, gravitationally interacting state. If we had a way of adding or 
removing particles from our state, keeping track of correlations with other 
particles, we could then reach any state in Fock space if only we had one 
reference state.

To realise this in practise goes beyond present knowledge, but we can give an 
approximated picture. In the next section we will consider an arbitrary state 
of out-going particles and add one in-going particle to see how the state 
changes. Repeating this procedure many times, we can compute the S-matrix of 
the whole process, up to an unknown phase which is the transition element 
between those reference in- and out-states. Notice that in this context it is 
not possible to compute this phase because in the eikonal approximation which 
we will be considering there is no black hole formation. To describe the 
creation of small black holes one has to consider the full transfer of 
momentum. We will not discuss black hole creation, but we will discuss how to 
go beyond the eikonal approximation. Black hole formation is a very important 
issue which has been considered in a simplified 2+1-dimensional set-up in 
\cite{hans-juergen}. Important related discussions in the context of the 
AdS/CFT correspondence and string theory can be found in \cite{BBKVR,Kirill}.

The natural objects to have falling into a black hole are massless objects, 
since any massive object that is falling into the black hole will be boosted 
to the speed of light with a tremendous energy. Therefore, we concentrate on 
massless point particles. The momenta of in-falling particles grow 
exponentially with Schwarzschild time, whereas momenta of out-coming 
particles decrease exponentially. A time lapse $\d t=4M\g$ in Rindler 
co-ordinates corresponds to a Lorentz-boost in Kruskal co-ordinates, 
\be 
u&\rightarrow& e^\g\, u\nn 
v&\rightarrow&e^{-\g}\,v\nn 
p_u&\rightarrow&e^{-\g}\,p_u\nn 
p_v&\rightarrow&e^\g\, p_v, 
\le{Lorentz}
in co-ordinates where the future horizon is at $v=0$, and the past horizon at 
$u=0$. So the momentum of in-falling particles grows exponentially as they 
approach the horizon. 
 
We anticipate that the gravitational effect of such a massless particle on 
the trajectories of the out-going Hawking particles takes the form of a 
shift, 
\be 
u\rightarrow u+p_v^\in(\th',\phi')\,f(\th,\phi,\th',\phi'),
\le{00}
so also the horizon shifts and out-coming particles come out at time 
$u=p_v^\in f$. This means that the size of the black hole has become larger. 
Notice that, according to \eq{Lorentz}, this shift grows larger and larger as 
Schwarzschild time goes by, and so at some point it will not be negligible. 
The point of view we advocate in this thesis is that this is a relevant 
effect that should be taken into account in the unitarity argument. Indeed, 
as explained in \cite{g9607}, there seems to be a hidden assumption in the 
derivation of the Hawking spectrum. This derivation performs a co-ordinate 
transformation from Minkowski to Rindler co-ordinates in the asymptotic 
region, where the energy of particles is rather low, and so this 
transformation seems a good approximation, at least as long as one computes 
macroscopic properties like the intensity of the emitted flux. However, when 
it comes to microscopic correlations between the radiation and the in-going 
particles, this approximation fails because it does not take into account the 
fact that particles collided at very high energy near the horizon and so they 
remain correlated afterwards.

In this thesis we will concentrate on the effect of the shift \eq{00}. Since 
the shift only takes into account the in-going momentum and not other 
possible charges of the particle, this is only an approximation to the real 
problem. However, notice that in a world with no other charges momentum would 
be enough to recover the information about the particle that was sent in. In 
realistic models this is also a good approximation because at those energies 
gravity is the dominant interaction. Electromagnetic interactions are 
subdominant and they can be easily incorporated in this model, but other 
charges are more difficult to account for. For a discussion of this issue we 
refer to \cite{g9607,JaKaOr}. 

Several objections have been raised against the existence of an S-matrix with 
such properties. The strongest one seems to be the no-quantum-copying-machine 
principle \cite{SuThUg}, which can be formulated as follows. Imagine sending 
some pure state into a black hole, and assume there is some linear operator 
$X$ copying this information on an outgoing state. Since the Hilbert space 
decomposes into an in- and an out-component, ${\cal H}={\cal 
H}_{\sm{in}}\otimes{\cal H}_{\sm{out}}$, the operator $X$ acts as
\be
X\left(|\psi\ket_{\sm{in}}\otimes|\phi\ket_{\sm{out}} \right) = 
|\psi\ket_{\sm{in}}
\otimes|\psi\ket_{\sm{out}}.
\ee
However, by letting $X$ act on a superposition $|\psi\ket_{\sm{in}}
=|\a\ket_{\sm{in}}+|\b\ket_{\sm{in}}$ one easily sees that an operator 
defined as above would not be linear and so would violate one of the basic 
principles of quantum mechanics, the linear evolution of states. Therefore, 
according to this argument there is no such a thing as a quantum copying 
machine.

This argument assumes that Hilbert space can be separated into an in- and an 
out-component, but that turns out not to be true in the S-matrix Ansatz. 
Actually both Hilbert spaces are complementary just like position and 
momentum space are in quantum mechanics. So we are forced to describe physics 
in either one Hilbert space or in the other, but not in both at the same 
time. The operator that will do the job of ``copying" the information of the 
in-going waves to out-coming radiation will be 
$X=e^{ip_{\tn{in}}fp_{\tn{out}}}$, where $p$ is the momentum of the in- or 
out-going waves, and $f$ is a function of the impact parameter. This operator 
certainly acts linearly on wave-functions, and it is actually directly 
related to the S-matrix. The fact that the in- and out-Hilbert spaces are 
complementary means that we
cannot do measurements on outgoing waves without influencing the outcomes of 
measurements done on in-going waves: if we choose to measure certain 
observables outside the black hole, this will imply an uncertainty for the 
outcomes of measurements inside the black hole.

It is easy to see how the shift \eq{00} comes about. An in-falling massless 
particle with momentum $p_\in$ is described by the following shock-wave 
metric \cite{AiSe}
\be
\dd s^2=2\dd u(\dd v-p_\in\d(u)\,f\,\dd u)+\dd x^2+\dd y^2.
\le{shock0}
The geodesics of massless test particles in this metric are easy to compute 
and give
\be
v(u)&=&v_0 +p_\in\,\th(u)\left(f+u\,{\pa x^i\over\pa u}\,\pa_if\right)\nn
x^i(u)&=&x^i_0-{1\over2}\,p_\in\,\pa_if\,u\,\th(u)
\le{geodesics}
where $u$ parametrises the null geodesic. The function $f$ is given by
\be
f=-4\GN\,\log(x^2+y^2).
\le{shift}
The impact parameter is the transverse distance between both particles, 
$b=\@{x^2+y^2}$ (in co-ordinates where one of the particles is at the 
origin). Therefore, for large transverse separations as compared to the 
Planck length, the derivatives of $f$, $\partial_if\sim{1\over b}$, can be 
neglected. More precisely, we have the following small dimensionless 
parameter: $\ve=\GN p_\in/b$. The approximation where this parameter is taken 
to be small is called the eikonal approximation. In that approximation, we 
see from the above formulae that $v$ is modified purely by a shift as the 
test particle crosses the world-line of the in-going particle, whereas the 
transverse co-ordinates remain unchanged. So, after the collision the 
particle continues along the same straight line, and the only effect of the 
collision is a time delay. This in turn means that the momentum transfer 
during the collision is negligible.

Next we briefly summarise the considerations leading to the black hole 
S-matrix \cite{g9607}. Take some reference state $|p_\in\ket$ of particles 
falling into a black hole, distributed over the horizon as $p_\in=p_\in(\O)$. 
Then assume that we have an element of the $S$-matrix that describes the 
formation and evaporation of the black hole,
\be
{\cal N}=\bra{\mbox{in}}_0|{\mbox{out}}_0\ket =\bra 
p_{\in,0}(\O)|p_{\out,0}(\O)\ket.
\ee
If we perturb the in-going state by adding some momentum $\d p_\in$, 
$p_\in\rightarrow p_\in+\d p_\in$, out-going particles will be shifted 
according to \eq{geodesics}:
\be
\d v=f\d p_\in.
\le{shift20}
So the out-state is modified by:
\be
|p_\out'\ket=e^{i\d v\,\^p_\sout}|p_{\out,0}\ket,
\ee
the caret meaning that we are generating a shift. So we get a new $S$-matrix 
element
\be
\bra p_\out'|p_\in\ket={\cal N}\, e^{-i\d p_\sin\,p_\sout f}.
\ee
This way we can reach any state $|p_\out\ket$ from a known state 
$|{\mbox{in}}_0\ket$ by the successive addition of infinitesimal amounts of 
momentum, and so we get
\be
\bra p_\out|p_\in\ket={\cal N}'e^{-ip_\sin p_\sout f}
\le{amplitude}
where we filled in the expression for the shift. The magnitude of ${\cal N}'$ 
is fixed by unitarity, but its phase is arbitrary and may depend on the 
details of the formation of the black hole. We refer to \cite{g9607} for 
further details.

When computing the scattering amplitude from \eq{amplitude}, one finds 
\cite{g87} the Veneziano amplitude for scattering between strings, with an 
imaginary string constant related to Newton's constant.

A Fourier transform of the above gives
\be
\bra p_\out(\O)|p_\in(\O)\ket&=&\int{\cal D}u_\in{\cal D}u_\out\,\exp 
\left[{\over}i\int\dd^2\O\,(\pa u_\in\pa u_\out+\right.\nn
&&+\left.p_\in u_\in-p_\out u_\out +u_\in u_\out){\over}\right]
\le{S-m}
which resembles very much the path integral over the world-sheet action of a 
string. Notice that there is a mass term that breaks conformal invariance. 
This term, however, is absent if instead of a black hole we consider a 
Minkowski background\footnote{When talking about a {\it background} in the 
context of shock-wave solutions, we mean a shock-wave on some background 
space-time.}.

The fields $u_\in$ and $u_\out$ are introduced as the Fourier transforms of 
$p_\in$ and $p_\out$, and so at the quantum level we have
\be
[u_\in(\O),p_\in(\O')]&=&i\d(\O-\O')\nn
{}[u_\in(\O),p_\in(\O')]&=&i\d(\O-\O')\nn
{}[u_\in(\O),u_\out(\O')]&=&if(\O-\O').
\le{commutators}
We see that gravity drastically changes the structure of space-time as seen 
by massless particles. Co-ordinates between particles become mutually 
non-commuting operators.

This has far-reaching consequences for the interpretation of Minkowski space 
as the near-horizon region of Kruskal space. The positions of particles that 
fall into a black hole are correlated with the positions of the emitted 
particles, and so Hilbert space does not reduce to a direct product of in and 
out Hilbert spaces. In other words, modifying the state of in-falling 
particles does modify the state of the Hawking radiation that is sent out. 
This obviously reduces the dimensionality of Hilbert space drastically, 
although we haste to add that every state $u$ still depends on a continuous 
variable, the angular variable $\O$, and so a transverse cutoff is still 
needed in this crude approximation.

It should now be clear why it is claimed that high-energy scattering presents 
holographic features. The theory that one gets is the sigma model \eq{S-m}, 
whose fields are defined on a two-dimensional surface, the two-sphere for the 
case of a Kruskal back-ground. This can be best understood in the context of 
the results of \cite{VV}, which we will review in the next section.

\subsection{Quantum Gravity as a Topological Field Theory}\label{TFT}

In the previous section we saw that collisions of massless particles at high 
energies exhibit great similarity with strings, the reason being the extended 
nature of the gravitational shock-wave. One can wonder whether this is a 
specific feature of the shock-wave solution, or a general property of gravity 
at high energies.

In references \cite{KO1,KO2,VV} it was shown that most of the features of the 
S-matrix model can be understood as specific properties of the eikonal limit 
of quantum gravity. Indeed, in this regime quantum gravity can be shown to 
have zero bulk degrees of freedom, all the degrees of freedom living purely 
on the boundary. So in that regime quantum gravity reduces to a topological 
field theory. The boundary here is the usual asymptotic null boundary of 
Minkowski space if we are talking about asymptotically flat spaces, but in 
chapter \ref{GSM} we will see that it can also be the boundary of dS and AdS 
space. This result is at first extremely puzzling, as in general one would 
expect gravity to have a nonzero number of degrees of freedom in the bulk. 

The derivation by Verlinde and Verlinde also sheds light on the validity 
regime of the S-matrix ansatz. As we will see, the eikonal regime is a 
perturbative regime as far as transverse processes are concerned, but is 
non-perturbative in the longitudinal length scale. We will review this 
argument in some detail here, as it will be the starting point of our 
generalisation in chapter \ref{GSM}.

The basic argument involves dimensional analysis of the different length 
scales in the problem. This is a usual argument used in field theory to 
derive the perturbation expansion. We can set all the dependence on 
dimensionful quantities into the metric by a rescaling of co-ordinates. 
Imagine that the typical length scale of the problem is given by some 
quantity $\ell$, then the metric scales like
\be
G_{\m\n}=\ell^2\^G_{\m\n},
\ee
where $\^G_{\m\n}$ is dimensionless. In four dimensions, the Einstein-Hilbert 
action scales like
\be
S_{\tn{EH}}[\ell^2\^G]=\ell^2S_{\tn{EH}}[\^G].
\ee
Now although in the path integral one integrates over all metrics, one 
expects that the dominant contribution will be given by those configurations 
whose size is that of the physical system, and so it seems reasonable to 
expect that $\^G_{\m\n}$ is typically of order 1. With this assumption, the 
coupling constant multiplying the action is
\be
g={\Pl\over\ell}.
\ee

This argument is commonly used to argue that when energies are of the 
Planck size, the theory is strongly coupled and so one needs the full quantum 
gravity theory to make sensible predictions.

Consider, however, a process where particles collide with Planckian energies 
but almost head-on. In such a collision, the longitudinal variables 
$x^\a=(t,x)$ fluctuate rapidly, whereas fluctuations in the transverse plane 
$y^i=(y,z)$ are much slower. In such a situation we have not one but rather 
two relevant length scales, namely, the longitudinal and the transverse 
scales.
Therefore we can form two dimensionless ratios:
\be
g_\parallel&=&{\Pl\over\ell_\parallel}\sim1\nn
g_\perp&=&{\Pl\over\ell_\perp}\ll 1.
\le{couplings}
From now on, the first few Greek characters $\a,\b,\dots$ refer to the 
longitudinal space, and middle Latin letters $i,j,\dots$ refer to the 
transverse plane. 

Taking $\ell_\parallel$ to be of order $\Pl$, we are left with one 
dimensionless coupling:
\be
\k={\Pl\over\ell_\perp}\ll1.
\ee
Performing the rescaling in the action explicitly, in four dimensions the 
Einstein-Hilbert action splits into three terms:
\be
S[G]_{\tn{EH}}={1\over8\p\GN}\int\dd^4x\,\@{-G}\,R[G]&=&{1\over\k^2}\,
S_0[\hat G] +{1\over\k}\,S_1[\^G] +S_2[\^G].
\ee

Thus, part of the action is strongly coupled, whereas $S_0$ is weakly 
coupled. The important conclusion is that for the weakly coupled piece we can 
use the saddle-point approximation. As far as this part of the action is 
concerned, the leading contribution is given by the classical configurations. 
Therefore, in the limit of low-momentum transfer, high-energy amplitudes can 
be computed using semi-classical techniques.

Considering perturbations around a classical background, 
\be
g_{\m\n}=g_{\m\n}^{\sm{cl}}(x) +\k\,h_{\m\n},
\ee
the authors of \cite{VV} found that the action reduces to
\be
S_{\tn{EH}}=\int\@{-g_{\sm{cl}}}\,[h^i_iK^{\a\b}h_{\a\b} 
+{1\over4}\e^{ik}\e^{jl}\na_\a h_{ij}\na^\a h_{kl} 
-{1\over2}(R_i+\e^{\a\b}\pa_\a h_{i\b})^2] +\mbox{tot. der.}
\le{EH}
where $R_i=\e^{\a\b}\pa_\a\pa_iX^a\pa_\b X_a$ and 
$K^{\a\b}=\na^\a\na^\b-g_{\sm{cl}}^{\a\b}\na^2$, and the fields $X^a$ are to 
be defined below.

The field equations for the metric $g_{\m\n}^{\sm{cl}}$ are determined by the 
term of the action linear in the perturbation, $h_{\m\n}$. The lowest order 
term $S_0$ vanishes identically for solutions of the equations of motion. 
Verlinde and Verlinde found the following solutions to the equations of 
motion:
\be
g_{\a\b}^{\sm{cl}}&=&\et_{ab}\,\pa_\a X^a\pa_\b X^b\nn
g_{ij}^{\sm{cl}}&=&g_{ij}(y)\nn
g_{i\a}^{\sm{cl}}&=&0,
\le{solutionssaddle}
and one also has $R_i=0$. Notice that the action \eq{EH} contains no 
$y$-derivatives, and so it is like a dimensionally reduced 2-dimensional 
action. The $X$-fields then represent $y$-dependent displacements of the 
longitudinal plane into itself.

It is now convenient to define a vector field $V_i^\a$ with the following 
properties
\be
\pa_iX^a&=&V_i^\a\pa_\a X^a,\nn
\pa_ig_{\a\b}&=&\na_\a V_{i\b}+\na_\b V_{i\a}.
\le{vectorfield}
This vector field describes the flow of the $X^a$-fields in the 
$y$-direction. The action then reduces to
\be
S_{\tn{EH}}=\int\@{g_\parallel 
g_\perp}\left(R[g_\perp]-\e_{\a\g}\e_{\b\d}\na^\a V^\b_i\na^\g V^{i\d} 
-\half(\e^{\a\b}\pa_\a V_{i\b})^2\right).
\le{EHred}
This theory is topological: the first two terms can be written as a total 
derivative, and the last term is set to zero by the constraint $R_i=0$.

Therefore, the action \eq{EHred} reduces to the following boundary term:
\be
S_{\tn{EH}}=S_{\pa M}[\bar X]=\int_{\pa M}\dd x^\a\int\@{g_\perp}\, 
\e_{ab}\left(R[g_\perp]\bar X^a\pa_\a\bar X^b +\pa_i\bar X^a\pa_\a\pa^i\bar 
X^b\right)
\le{EHred2}
where $\bar X$ are the boundary values of $X$. The boundary here corresponds 
to the four asymptotic null regions of the 2-dimensional Minkowski plane.

For full details, we refer to \cite{VV}. After including point particles, it 
turns out that the $X^a$'s couple to the longitudinal momenta of the 
particles. The S-matrix computed from \eq{EHred} with point particles gives 
exctly the amplitude computed by 't Hooft. In fact, quantisation of the model 
gives rise to the following commutator:
\be
[X^a(y),X^b(y')]=i\e^{ab}f(y,y')
\le{comm-1}
where $f$ is the Green's function.
\be
(\LL_h-\half R[h])\,f(y,y')=\d^{(2)}(y-y')
\ee
This is obviously 't Hooft's result \eq{commutators}.

The important conclusion of \cite{VV} is that, in the eikonal regime, quantum 
gravity is a topological field theory: its degrees of freedom live on the 
boundary, and its only physical perturbations are the global variations of 
the fields $X^a$. When coupled to point particles, the saddle-point of these 
variations correspond to shock waves. Indeed, after inserting the solutions 
\eq{solutionssaddle}, the full four-dimensional metric is :
\be
\dd s^2&=&\et_{ab}\,\pa_\a X^a\pa_\b X^b\,\dd x^\a\dd x^\b +g_{ij}(y)\,\dd 
y^i\dd y^j,
\le{VVshock}
with
\be
X^-=x^-+p^-\th(x^+)f(y),\nn
X^+=x^+-p^+\th(x^-)f(y),
\ee
and this is obviously a generalisation of the Aichelburg-Sexl metric 
\eq{shock0} for the case of two shock-waves\footnote{In four dimensions, 
there are no exact two-particle solutions known. Equation \eq{VVshock} is 
only valid at the linearised level.}.

In chapter \ref{GSM} we will perform a systematic study of the eikonal 
regime, valid for spaces with a cosmological constant and of any dimension.

\section{String Theory}\label{StringTheory}

Although still unsolved, Hawking's information paradox has proven to be a 
very useful scenario to obtain new insights that can help us construct a 
consistent theory of quantum gravity. Discussions about black holes have led 
to the discovery of several guiding principles that should be present in 
quantum gravity. Holography, complementarity, some sort of extendedness 
beyond the point particle approximation, and non-commutativity, seem to be 
some of the features that quantum gravity should meet. All of these are 
present in the eikonal regime of quantum gravity which we studied in the 
previous sections. In general, however, quantum gravity as a theory of point 
particles is quite intractable and one may need to make some additional 
assumption like the assumption that particles have a string-like extension. 
This leads us to string theory.

There are several reasons to think that making such an assumption is a good 
idea. Suffice it to say that string theory seems to have built in some of the 
above principles, in particular the principle of holography, as we will 
discuss in the next section.

The action of a point particle is simply given by the invariant length of its 
world line:
\be
S=-m\int_\g\dd s\@{-G_{\m\n}(z)\,\dot z^\m\dot z^\n}
\ee
where $\g$ is the world line of the particle and $z^\m$ its trajectory along 
this world line. It is, however, more convenient to have a quadratic action. 
This can be done by introducing an auxiliary field:
\be
S=\half\int\dd s\,({1\over e}\,G_{\m\n}\dot z^\m\dot z^\n -e\,m^2).
\ee
For a quantum mechanical particle, one integrates over all possible 
trajectories and also over the auxiliary field:
\be
\int{\cal D}z{\cal D}e\,e^{iS[z,e]}.
\ee
The saddle point approximation to the path integral selects the classical 
trajectory with minimal length.

For strings the situation is analogous to the point particle case. The path 
integral now contains the exponentiated area of the string,
\be
\int{\cal D}X{\cal D}h\,e^{iS[X,h]}
\ee
where $X(\t,\s)$ denotes the embedding of the string into target space and 
$h_{ij}$ is an auxiliary field representing the metric on the string. The 
action is given by
\be
S&=&-{T\over2}\int\dd^2\s\, [\@{h}h^{ij}G_{\m\n}(X)\pa_iX^\m\pa_jX^\n 
+\e^{ij}B_{\m\n}(X)\pa_iX^\m\pa_jX^\n]\nn
&&+{1\over4\pi}\int\dd^2\s\@{h}\,\f(X)\,R[h]
\le{MaSt}
where we are allowing for additional background fields apart from the metric: 
the dilaton $\f(X)$ and an antisymmetric tensor field $B_{\m\n}(X)$.

It is well known that at low energies string theory reproduces gravity. The 
vanishing of the $\b$-functions of the sigma-model \eq{MaSt} imposes, at 
first order in $\a'$ \cite{CFMP}:
\be
R_{\m\n} -{1\over4}H_\m^{\,\,\,\a\b}H_{\n\a\b} +\na_\m\na_\n\f&=&0\nn
\na^\a H_{\a\m\n} -2\na^\a\f\,H_{\a\m\n}&=&0\nn
4(\na\f)^2 -4\Box\f -R+{1\over12}\,H^2&=&0,
\le{betaf}
where $H_{\m\n\a}$ is the field-strentgh constructed from $B_{\m\n}$. The 
expansion parameter $\a'$ is proportional to the string length and is 
inversely proportional to the string tension $T$. The $\b$-function equations 
at lowest order determine the space-time dimension, $D=26$ for the bosonic 
string, and $D=10$ for the superstring.

These equations can be integrated to the following effective action:
\be
S=-{1\over\k^2}\int\dd^Dx\,\@{G}\,e^{-2\f}[R+4(\na\f)^2 -{1\over12}\,H^2].
\le{effact}
By a field redefinition of the metric one can bring the action to the 
Einstein frame. It is clear that higher order terms in the expansion of the 
$\b$-functions \eq{betaf} will show up as $\a'$-corrections in the effective 
action \eq{effact}. These are typically of order $R^2$ and higher, and they 
predict specific corrections to Einstein's theory.

The action \eq{effact} does not contain all of the massless supergravity 
fields. Let us for example concentrate on type IIB string theory. In this 
case there are additional terms one can add to the effective action. One of 
these is a self-dual 5-form $F_5$, which then gives rise to an extremal 
3-brane solution of the following form: 
\be
\dd s^2&=&f^{-1/2}(-\dd t^2+\dd\vec{x}^2_3) +f^{1/2}(\dd r^2+r^2\dd\O_5^2)\nn
f&=&h+{R^4\over r^4}
\le{extremal}
and $h=1$. This solution has a constant dilaton, a covariantly constant 
5-form flux along the $S^5$ and $H=0$. The strentgh of the flux and the value 
of the dilaton are absorbed in the definition of $R$. However, 3-branes can 
also be viewed from a different point of view: they are the hyperplanes on a 
10-dimensional flat space on which open (and closed) strings can end and they 
are called Dirichlet branes. From this point of view, one can effectively 
describe the physics by the effective action on the D3-brane by describing 
its collective modes, which are the excitations of the open strings. The 
effective action in the case of $N$ D-branes placed on top of each other is 
the Dirac-Born-Infeld action, which generalises the world-volume action of a 
single D-brane and accounts for the strings being stretched between the 
branes.

One can also consider the above solution for $h=0$. The space-time is then 
AdS$_5\times S^5$. The D3-brane and the AdS metrics agree at $r/R\ll1$, which 
is precisely the near-horizon limit considered in the AdS/CFT correspondence. 
In other words, near the horizon of the D3-brane the space looks locally like 
AdS$_5\times S^5$, just as the near-horizon geometry of the Schwarzschild 
black hole is Rindler space times a two-sphere of constant radius. 

AdS$_5\times S^5$ is an exact solution of string theory, but the above 
extremal D3-brane metric is not. $\a'$-corrections to the effective above 
action \eq{effact} become important as the energy increases. Here we again 
concentrate on the case of type IIB, which is where these corrections are 
best known. Keeping only the 5-form in the RR sector, the action at next 
order in $\a'$ is given by \cite{GrZa,FPSS,GrWi}:
\be
S=\int\dd^{10}x\@{g}\,[e^{-2\f}(R +4(\pa\f)^2 +\g W) -{1\over2\cdot 
5!}\,F_5^2],
\ee
where $\g$ is a number of order $\a'^3$. The self-duality of $F_5$ ensures 
that there are no higher order corrections in $F$. $W$ is a sum of certain 
contractions of four Weyl tensors, $W\sim C^4$. Terms of order $R^2$ and 
$R^3$ are removed by a field redefinition. The Einstein frame is reached by a 
redefinition $g\rightarrow e^{\f/2}g$, and the action becomes:
\be
S=\int\dd^{10}x\@{g}\,[R-\half(\pa\f)^2 -{1\over2\cdot5!}\,F_5^2 +\g 
e^{-{3\over2}\,\f}W].
\ee
Branes play an essential role in the arguments leading to the AdS/CFT 
correspondence. Since the D3-brane is not an exact solution of the 
$\b$-function equations, it would be very interesting to analyse 
$\a'$-corrections to the metric \eq{extremal}. The analysis of these 
corrections will be presented elsewhere \cite{AKS}.

\section{The AdS/CFT Correspondence}\label{AdS/CFT}

The AdS/CFT correspondence is the most concrete example of a holographic 
duality. It states that {\it string theory in an AdS space-time is equivalent 
to a certain conformal field theory formulated on the boundary of AdS}. So, 
for example, if the bulk is AdS$_5$, the dual CFT on the boundary is ${\cal 
N}=4$ SYM on the boundary of AdS$_5$ which can be thought of as a cylinder. 
Of course, as it stands the formulation of this duality is still too vague. 
Later on we will give more details about the correspondence.

One of the surprising things about the Maldacena or AdS/CFT conjecture is 
that string theory contains gravity, whereas the field theory does not. This 
suggests that gravitational theories have redundant degrees of freedom 
\cite{Gdeterm} or, at least, they can be reorganised in a more economic way. 
This gives rise to a theory that is non-gravitational and, furthermore, is 
defined on a manifold with one space dimension less. In other words, gravity 
does not contain as many degrees of freedom as one would naively think.

The relationship between gauge theories and theories containing gravity like 
string theory is long standing \cite{glargeN}. However, a precise connection 
exists only since the discovery of the AdS/CFT correspondence 
\cite{Malda,Wit,Gubs}. There is a large literature on checks of the 
correspondence between supergravity in AdS and the large $N$ limit of 
conformal field theories. In this thesis we will concentrate on rather 
generic but precise questions concerning the holographic map between both 
theories. Indeed, it is important to have a precise understanding of how 
quantities in the bulk and on the boundary are mapped into each other in 
order to understand how holography works.

As said, the focus will be on generic questions concerning the duality. 
Mostly we will not specify the details of the CFT that we are studying but 
assume that it exists and require minimal knowledge about it, like which 
sources are turned on. Then we try to reconstruct the bulk theory as far as 
we can with this information, until new information from the CFT is required. 
That we are interested in generic properties of the holographic map is due to 
the fact that we would like to understand holography in general, i.e. also 
for other backgrounds than AdS. Hopefully this will give more insight in why 
the duality works. In the case of the AdS/CFT correspondence, the duality 
between open and closed strings lies at the heart of the holographic relation 
\cite{KhVe}.

Among the many phrases that can be found in the holographic dictionary, a 
very important notion is that of the UV/IR connection \cite{SuWi,PePo,BKLT}, 
i.e. the duality between high and low energies on both sides of the duality. 
More precisely, the renormalisation group scale in the gauge theory is 
interpreted as the compactification radius of the gravity theory. Radial 
evolution is then related to the renormalisation group equations 
\cite{JdBVV,Erik,VVRGflow}. 

Another, related aspect one would like to understand precisely is the 
geometry. How is the information about the geometry of the bulk precisely 
encoded in the boundary theory? More precisely, we can ask: given a certain 
boundary theory, how does one reconstruct the classical bulk space-time and 
the fields on this space-time? This an other questions will be addressed in 
chapter \ref{reconstruction}. Some of those results will be extended in 
chapter \ref{warped} to asymptotically flat and asymptotically de Sitter 
space-times: the information about the bulk geometry is encoded in certain 
specific ``holographic" stress tensors. 

For a review of the arguments motivating this duality, see \cite{AdSreview}. 
One important issue that one has to address with any duality is its limits of 
validity. Indeed, string theory in AdS is usually too complicated to be dealt 
with in detail. One of the tractable limits is the supergravity limit where 
$\Pl<l_s\ll R$, which implies $1\ll g_sN<N$. The condition $R\gg\Pl$ is 
needed in order for higher curvature corrections to be small. $\Pl<l_s$ is 
equivalent to $g_s<1$ which is needed in order to avoid string loop 
corrections in the string coupling $e^\phi$, which are not well defined in 
supergravity which is a non-renormalisable theory. On the SYM side this 
corresponds to the large $N$, strong 't Hooft coupling $\l=g_{\sm{YM}}^2N$ 
limit of the theory. There are of course other intersting limits that one can 
look at but we will not consider those here.

Let us now discuss how to make the AdS/CFT correspondence more precise. In 
particular there is an important issue about boundary conditions at infinity 
that needs to be considered \cite{Gubs,Wit}. AdS has a timelike boundary at 
infinity. Therefore, fields on this space can propagate to the boundary and 
so one has to supplement them with certain boundary conditions. There is a 
precise 1-1 correspondence between the boundary values of fields on AdS and 
operators on the CFT. We collectively denote bulk fields by $\Phi$, and their 
boundary values by $\phi_{(0)}$. The string partition function is then a 
functional of the boundary values of the fields:
\be
Z_{\sm{string}}[\f_{(0)}]=\int_{\f_{(0)}}{\cal D}\F\, \exp(-S[\F]).
\le{Zstring}
According to the proposal in \cite{Wit}, this should be equal to the 
generating functional of correlation functions in the CFT,
\be
Z_{\sm{string}}[\f_{(0)}]=Z_{\sm{CFT}}[\f_{(0)}]=\bra \exp[\int_{\pa 
X}\dd^dx\@{g}\,\f_{(0)}(x)O(x)]\ket_{\sm{CFT}}
\le{correspondence}
where $O(x)$ is a specific composite operator in the CFT and $\pa M$ is the 
boundary of the manifold $M$. Thus, the boundary values of the fields act as 
sources for computing correlation functions of operators in the CFT.

The partition function \eq{Zstring} is an intractable object to deal with, so 
one has to consider some limit like for example the supergravity limit. In 
this limit, one of the fields that will be integrated over in \eq{Zstring} is 
the metric. Thus, we are strictly speaking not considering AdS space, but any 
Einstein manifold with fixed metric at infinity. Therefore, the metric in the 
bulk is allowed to fluctuate as long as it preserves the boundary conditions.

Obviously, at low energies we are interested in the supergravity limit of 
\eq{Zstring} where the dominant contribution to the path integral is given by 
the saddle-point approximation. The partition function then reduces to:
\be
Z_{\sm{sugra}}[\f_{(0)}]=\exp(-S[\F_{\sm{cl}}(\f_{(0)})]),
\le{correspondence2}
where $\F_{\sm{cl}}$ are now fields that satisfy the low-energy equations of 
motion with fixed boundary values  $\F(r,x)|_{r=0}=\f_{(0)}(x)$.

In general, massive scalar fields that solve the equations of motion behave 
differently from $\F(r,x)\rightarrow\f_{(0)}(x)$ as they approach the 
boundary. They can either decay more rapidly or develop singularities. A more 
detailed analysis gives:
\be
\F(r,x)=r^{d-\D}\f_{(0)}(x)+\cdots,
\ee
where $\D$ satisfies
\be
\D(\D-d)=m^2
\le{mass}
and $m$ is the mass of the scalar field. So $\D$ has two possible values, 
$\D=d/2\pm\@{d^2/4+m^2}$ which satisfy $\D_++\D_-=d$. This means that the 
expansion in general has the following asymptotic form:
\be
\F(r,x)=r^{d-\D}(\f_{(0)}+{\cal{O}}(r^2)) +r^\D(\vf(x)+{\cal{O}}(r^2)).
\ee
where $\f_{(0)}$ and $\vf$ are two independent modes. The unitarity bound on 
the mass implies $\D>(d-2)/2$. The existence of two independent solutions to 
the equations of motion reflects the fact that usually one needs to impose 
two boundary conditions on the fields: initial conditions for the positions 
and the momenta\footnote{However, quantisation in AdS is subtle due to the 
fact that there is no complete Cauchy surface \cite{AvIsSt,Gibbons}.}. In 
AdS, usually one of these two modes will vanish if we also impose some 
regularity condition in the centre of AdS or some global condition like the 
vanishing of the Weyl tensor.

The correspondence between the gravity and the CFT computations has been 
tested for 2-, 3- and 4-point functions of several operators 
\cite{FMMR,DHoFrSk,AdSreview}.

Klebanov and Witten have argued \cite{KleWit} that for $-d^2/4<m^2<-d^2/4+1$ 
the existence of two independent modes for fields in this mass range implies 
the existence of two conformal field theories dual to the same bulk metric. 
These are called the $\D_+$ and the $\D_-$-theory. In the $\D_+$-theory, the 
lowest-order mode\footnote{Lowest order in $r$. This mode is the first mode 
to appear in a perturbative expansion in terms of $r$.} $\f_{(0)}$ has the 
usual interpretation as an external source that couples to an operator $O(x)$ 
of conformal dimension $\D_+$, whereas $\vf(x)$ (which appears at order 
$\D_-$) is related to the expectation value of $O(x)$. In the $\D_-$-theory, 
on the other hand, $\f_{(0)}$ is interpreted as an expectation value whereas 
$\vf$ is the source. Both theories are related by a Legendre transformation. 
The case $\D_+=\D_-$ is special and corresponds to the tachyon of minimal 
mass.

As it stands, the correspondence \eq{correspondence2} is meaningless as both 
sides suffer from divergences. These, however, can be regularised and 
renormalised by adding appropriate counter-terms \cite{HS,BK,KLS,KSS1}. It 
has been shown \cite{SuWi,HS} that the IR divergences on the gravitational 
side correspond to UV divergences on the gauge theory side. In chapter 
\ref{reconstruction} we will develop a systematic method to regularise and 
renormalise the on-shell supergravity action. Although from the gravity point 
of view the divergences are purely classical and related to the infinite 
volume of AdS, it is essential to remove them in order for gravity solutions 
to have a sensible interpretation in terms of mass, entropy, etc. 
\cite{BrownYork,BK,KLS}.

In chapter \ref{reconstruction} we will study these issues in detail for 
scalar fields, for the metric and for the coupled gravity-matter system.

\chapter{Holography in High-Energy Scattering}\label{HEscattering}

In this chapter we consider collisions between massless particles at very 
high energies. We do this perturbatively in the eikonal approximation, where 
collisions are almost head-on and the impact parameter is large. In this 
regime, gravity reduces to a topological field theory with zero bulk degrees 
of freedom. We discuss how to go beyond the extreme eikonal regime as well as 
first and second quantisation of gravitationally interacting particles.

The contents of this chapter are based mainly on \cite{SdHCQG} and 
\cite{SdHJHEP}. The chapter is organised as follows. The first section 
reviews  massless particle solutions of Einstein's equations for various 
back-ground geometries. In section \ref{ClscPlen} we discuss classical 
scattering between these particles at very high energies, and in section 
\ref{The S-matrix} we give a covariant generalisation (in transverse space) 
of 't Hooft's S-matrix, discussed in section \ref{eikonal} of the 
introduction. A first step towards the restoration of covariance in the 
longitudinal plane is taken in section \ref{beyondeikonal} where we compute 
the transfer of momentum during collitions at high energies. In the next 
section, section \ref{Quantisation}, we discuss the quantum theory and find a 
closed algebra between momenta and a gravitational correction to Heisenberg's 
uncertainty which is nothing but an expression of this momentum transfer. In 
section \ref{2+1} a precise link is proven between transfer of momentum and 
covariance. In section \ref{2ndquant} we discuss second quantisation of 
gravitationally interacting particles and find that they satisfy an exchange 
algebra that is very much reminiscent of the Moyal product. We close the 
chapter with a discussion and some conclusions in section \ref{discconc}.

\section{Pointlike massless particles in Einstein's 
theory}\label{Pointlikepart}

When energies are so high that gravity becomes the dominant force and 
particles start interacting gravitationally, one needs to take into account 
the back-reaction of particles on the back-ground geometry. That is, one 
cannot trust the free Einstein equations, but one has to couple them to the 
matter fields of the particles. Our main focus will be massless particles, as 
these are the relevant excitations when we discuss scattering in the 
neighbourhood of a black hole.

Massless particles are included in Einstein's theory as follows. The 
gravitational action is given by
\be
S=S_{\mbox{\tn{EH}}} +S_{\mbox{\sm{matter}}},
\ee
where 
\be
S_{\mbox{\tn{EH}}}&=&{1\over16\p\GN}\int_X\dd^dx\,\@{-G} \left(R+2\L\right)
\eea
and $S_{\mbox{\sm{matter}}}$ is the matter action belonging to the massless 
particle. In spaces with a boundary, as is the case when the cosmological 
constant is negative, the action \eq{1} may be supplemented with additional 
boundary terms to ensure a well-defined variational problem. This point will 
be discussed in detail in later chapters.

As is well-known, the matter action for a massless particle includes an 
auxiliary field $e$:
\be
S_{\sm{matter}}=-\half\int\dd s\,e(s)\,G_{\m\n}\,\dot{z}^\m\dot{z}^\n.
\eea
Making use of the gauge invariance of the action, the equations of motion of 
the matter fields $z^\m$ and of the auxiliary field give the usual geodesic 
equation together with the constraint that the particle is massless. 
Einstein's equations then take the following form:
\be
R_{\m\n}-\half\,G_{\m\n}\,R-\L\,G_{\m\n}=-8\p\GN\,T_{\m\n},
\le{Einstein0}
where the stress-energy tensor is given by:
\be
T^{\m\n}(x)=-{p\over\@{-G(x)}}\int_\g\dd s\, \d^{(d)}(x-z(s))\, 
\dot{z}^\m\dot{z}^\n,
\le{10}
and $\g$ is the world-line of the particle, parametrised by $s$, $z^\m(s)$ 
the trajectory of the particle along the world-line, $p$ the momentum of the 
particle along the light-cone, and $d$ the space-time dimension. To find 
solutions describing massless particles, one solves \eq{Einstein0} coupled to 
the geodesic equation and the constraint.

A useful technique to obtain solutions describing massless particles from 
existent vacuum solutions is Penrose's cut-and-paste method \cite{Penrose}. 
As explained in the introduction, massless particles can be seen as 
space-time defects of dimension 1. Penrose's method provides a space-time 
with a delta-function singularity with support on the null line along the 
particle's trajectory, given two flat pieces of Minkowski space. As shown by 
Dray and 't Hooft \cite{gnp85}, this kind of gravitational solution 
generalises to a much larger class of asymptotically flat space-times. It can 
be further generalised to spaces with either positive or negative 
cosmological constant.

Let us start with the following rather general class of metrics:
\be
\dd s^2=G_{\m\n}\,\dd x^\m\dd x^\n=2A(u,v)\,\dd u\dd v+g(u,v)\,h_{ij}(x^k)\dd 
x^i\dd x^j,
\le{8.0}
in co-ordinates $x^\m=(u,v,x^i)$. Let us assume that this metric is a vacuum 
solution of Einstein's equations. In this space-time, the stress tensor takes 
the following form:
\be
T_{vv}=4pA^2\d(v)\d^{(d-2)}(x)
\ee
for a particle travelling along the null line $v=0$, $x^i=0$. The 
cut-and-paste method suggests the following ansatz for the metric:
\be
\dd s^2=2A(u,v)\,\dd v(\dd u -f(x^k)\d(v)\dd v) +g(u,v)\,h_{ij}(x^k)\dd 
x^i\dd x^j,
\le{ansatz}
and indeed by direct computation (see Appendix \ref{appA3}) one finds that 
Einstein's equations reduce to the following equations at $v=0$:
\be
\LL_h f-{1\over A}\,\pa_u\pa_vg\,f&=&\frac{gA}{\@{-G}}\, 
\d^{(d-2)}(x),\label{9}\\
\pa_uA=\pa_ug&=&0\label{9a}.
\ee
The first equation is a junction condition for gluing together both parts of 
the metric along the null line $v=0$. The second equation is the requirement 
that the metric has a Killing vector along the null trajectory of the 
particle. So, two regions of space-time can be glued together only along a 
direction with a Killing vector.

The metric \eq{ansatz} is singular at $v=0$. However, this singularity can be 
removed by a discontinuous co-ordinate transformation
\be
\ti u=u-f(x^k)\,\th(v).
\ee
In these co-ordinates, the metric is finite but not continuous. Geodesics are 
continuous but not differentiable. With a further (continuous but not 
differentiable) co-ordinate transformation one can make the metric 
continuous.

One first remark is that the $d$-dimensional Einstein equations reduce to the 
equation of motion of a massive scalar, coupled to a source, on the 
space-time defect. In this case the space-time defect is a null surface of 
dimension 1. This result underlies the S-matrix Ansatz. In later chapters we 
will see that when the defect is timelike and of codimension 1, the induced 
equations are Einstein's equations coupled to certain stress-tensors. It 
would be interesting to perform a similar analysis for other types of 
defects, like e.g. null defects of codimension 1.

The solutions to \eq{9} and \eq{9a} are easy to find if the background is 
Minkowski space. In four dimensions we find the logarithmic solution given in 
the introduction, equation \eq{shift}. In other dimensions the solution 
generically goes like $f\sim{1\over|x|^{d-4}}$. For more general backgrounds, 
like the Schwazschild black hole, some solutions are given in \cite{gnp85}.

It is not difficult to extend the above analysis to space-times including a 
cosmological constant \cite{HI}. Let us just give the solution for pure AdS 
space with a massless particle travelling from the boundary to the bulk. We 
write pure AdS$_d$ in the following co-ordinates (see Appendix \ref{appA1} 
for the transformation to Poincare co-ordinates):
\be
\dd s^2={4\over(1-y^2/\ell^2)^2}\,\et_{\m\n}\dd y^\m\dd y^\n,
\le{pureAdS}
where $\ell$ is the AdS radius and $y^2=\et_{\m\n}y^\m y^\n$. The stress 
tensor of a massless particle can straightforwardly be computed and gives 
$T_{uu}=-p\,\d(u)\d(\r)$, where $\r$ is the radial co-ordinate 
$\r=\sum_{i=1}^{d-2}y_i^2$.

This metric is not of the class considered above. However, it gives the 
following solution of Einstein's equations with a massless particle:
\be
\dd s^2={4\over(1-y^2/\ell^2)^2}\,\left(\et_{\m\n}\dd y^\m\dd y^\n 
+8\p\GN\,p_u \d(u)(1-\r^2/\ell^2)f(\r)\dd u^2\right)
\le{HI0}
provided
\be
\LL_h f -4\,{d-2\over\ell^2}\,f=\d(\r).
\le{shiftads0}
$\LL_h$ is the Laplacian on the transverse hyperbolic space,
\be
\dd s^2={\dd\r^2+\r^2\dd\O^2_{d-3}\over(1-\r^2/\ell^2)^2}.
\ee 
The solutions to \eq{shiftads0} are given in chapter \ref{GSM} and they of 
course reduce to the Minkowski solutions $f\sim{1\over|x|^{d-4}}$ in the 
limit when the AdS radius goes to infinity, $\ell\rightarrow\infty$.

This metric can also be obtained with Penrose's method because the conformal 
factor has no dependence on the longitudinal co-ordinates at the locus of the 
shock-wave. In fact, the condition \eq{shiftads0} is very similar to \eq{9a} 
and it is very likely that one can easily generalise the construction of Dray 
and 't Hooft to space-times with a cosmological constant that have the 
Horowitz-Itzhaki shock-wave as a special case. For this case one needs to 
introduce a dependence on the transverse length $\r$ in the Ansatz \eq{8.0}. 
Note that the effect on outgoing massless particles takes the form of a shift 
also in this case (see Appendix \ref{appA}). This can be shown either by 
direct computation or by using the fact that massless geodesics are invariant 
under Weyl rescalings of the metric. The latter fact relates the trajectories 
in AdS to trajectories in flat space.

It is interesting to note that shock-wave solutions are exact solutions of 
string theory. Indeed, in \cite{AmKl2} it has been shown that shock-wave 
backgrounds are solutions to all orders in the sigma-model perturbation 
theory. In \cite{HI}, it was shown that also the AdS shock-wave does not 
receive any $\a'$-corrections from a geometrical argument used in 
\cite{Kallosh,Horowitz}. The argument uses the fact that all scalar 
combinations that can be formed from the contribution to the Riemann tensor 
due to the shock-wave vanish. Thus, corrections to the supergravity action 
can only come from the AdS part of the metric, but these are known to be 
equally zero. Thus, shock-waves are among the few known examples of exact 
backgrounds of string theory. Another interesting fact is that the amplitude 
computed by 't Hooft agrees, at large distances, with the amplitude of a free 
string in the shock-wave background generated by another string. The latter 
also agrees with the (infinite genus) amplitude of two interacting strings in 
a flat background. So, the shock wave can be regarded as a non-perturbative 
effect coming from the resummation of flat-metric string contributions 
\cite{ACV,AmKl1}. At small distances, however, the string amplitudes do not 
exhibit the singular behaviour of the point particle case. Let us however 
point of that to our knowledge no amplitude valid beyond the eikonal regime 
has been computed so far for the point particle case, and so there is not 
much one can conclude from the discrepancy.

\section{Classical scattering at Planckian energies}\label{ClscPlen}

Next we compute the effect of shock-waves on the trajectories of test 
particles. This is a straightforward computation if one is careful 
\cite{SdHJHEP}, although there are mathematical subtleties on has to take 
into account \cite{steinbauer,KuSt}. We illustrate this for the case of a 
Minkowski background, but the computation generalises straightforwardly to 
other spaces. Take the metric
\be
\dd s^2=2\dd v\left(\dd u-{\sf f}_v(\ti x)\,\d (v)\dd 
v\right)+\dd x^2+\dd y^2,
\le{4ab}
where the shift function is ${\sf f}_v(\ti x)\equiv-\frac{1}{T}\int\dd^2\ti 
x'P_v(\ti x')\,f(\ti x-\ti x')$. This is a straightforward generalisation for 
the case that the total momentum is not concentrated at one point, but is a 
distribution over the shock-wave. This allows to describe an arbitrary amount 
of left-movers (see Figure \ref{fig1}) all sitting on a plane of constant $v$ 
with total momentum distribution $P_v$. The in-going momentum distribution 
$P_v(\ti x)$ is typically equal to
\be
P_v(\ti x)=\sum_{i=1}^Np_v^i\d (\ti x-\ti x^i),
\le{4b}
if there are $N$ particles with transverse positions $x^i$ on the plane of 
the shock-wave. The right-moving particles have initial momentum $p_u^0$. All 
particles satisfy the mass-shell condition $p_\m^2=0$.

The first geodesic equation in the metric \eq{4ab} gives
\be
\ddot{v}=0,
\le{5b}
where the dot denotes the derivative with respect to the affine parameter 
$\l$ 
along the geodesic. This equation allows us to use $v$ as a time co-ordinate. 
The other equations are solved as follows:
\be
u(v)&=&u(0)-\frac{1}{2T}\,\mbox{sgn}\,(v)\int\dd^2\ti x'\,P_v(\ti 
x')\left(f(\ti x_0-\ti 
x') 
+v\frac{\pa x^i}{\pa v}(0)\,\pa_if(\ti x_0-\ti x')\right)\nn
x^i(v)&=&x^i(0)+p^i_0v+\frac{1}{2T}\,v\,\mbox{sgn}\,(v)\,\int\dd^2\ti 
x'\,P_v(\ti x')\,\pa_if(\ti x_0-\ti x'),
\le{7}
where $\ti x_0\equiv \ti x(0)$. As a
boundary condition, we have chosen that the initial momentum in the 
$u$-direction is zero,
and in the transverse $i$-direction\footnote{The latter will be set to zero 
in the following.} it is $p^i$.
\begin{center}
        \begin{figure}[h] \hspace{4cm}
        \psfig{figure=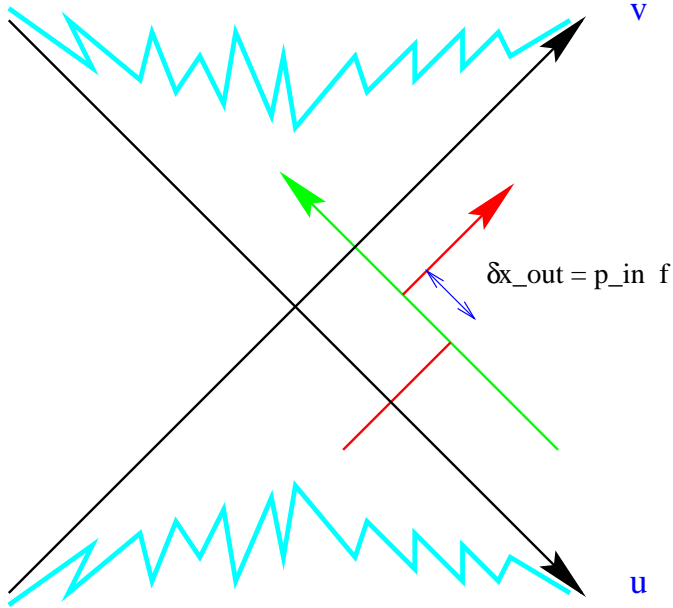}
        \caption{Effect of a shock-wave in the lightcone 
directions}\label{fig1}
        \end{figure} 
\end{center}

If we now concentrate on the $y-v$ plane, differentiating
\eq{7} yields
\be
\frac{\pa y}{\pa v}=\frac{1}{2T}\,\mbox{sgn}\,(v)\, \int\dd^2\ti x'P_v(\ti 
x')\,\pa_yf(\ti 
x_{\sm{0}}-\ti x').
\le{18}
This agrees with a standard computation by Dray and 't Hooft \cite{gnp85} 
where massless geodesics are obtained from massive ones by boosting a black 
hole to the speed of light while sending its mass to zero.

As mentioned in the introduction, we will be working in the first few orders 
in the eikonal approximation. We introduce the expansion parameter $\ve\equiv 
G\,p_{\sm{in}}b$, where $p_{\sm{in}}$ is the in-going momentum and 
$b$ the impact parameter, given by the transverse separation between the 
colliding particles. $\ve$ can be taken to be small in the eikonal regime, 
and it will control our perturbative expansion. Notice that, since $f$ is 
logarithmic in the transverse distance, $\pa_if\sim\frac{1}{b}$.

The first of \eq{7} gives us the shift \eq{1} in the longitudinal co-ordinate 
$u$ as a consequence of the in-going particle plus a correction that is 
${\cal O}(\ve^2)$ and can be neglected as long as the in-going transverse 
momentum is small, $p_\perp\ll p_\parallel$. The second of \eq{7} can be 
represented by a kink in the trajectory of the out-coming particle, see 
Figure \ref{fig2}. This is a higher-order effect.

As mentioned, these are also the trajectories in AdS with a shock-wave, 
\eq{HI0}, with ${\sf f}$ replaced by the corresponding shift \eq{A1}.
\begin{center}
        \begin{figure}[h] \hspace{4cm}
        \psfig{figure=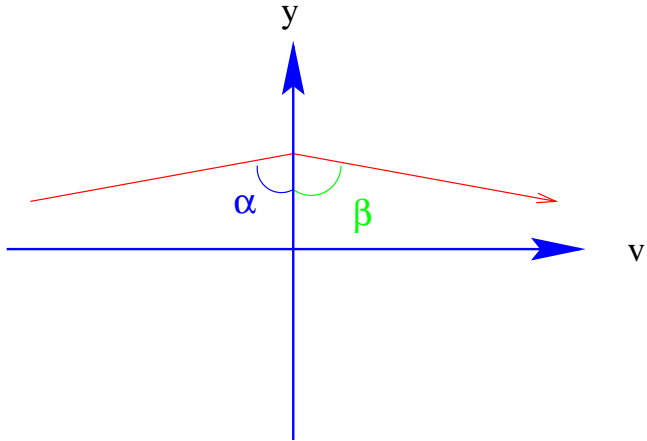}
        \caption{Effect of the shock-wave in the transverse direction}
	  \label{fig2}
        \end{figure} 
\end{center}

So far we have discussed how the trajectories of out-coming particles are 
modified by the shock-waves of in-going particles. Next we will consider the 
momentum transfer involved.

In Figure \ref{fig2} the trajectories in the $y-v$ plane are shown. These 
follow from \eq{7}. We learn from the figure that
\be
\tan\g =\frac{p_y}{p_u},
\le{-1}
where the angle $\g$ is defined by $\g =\pi-\a-\b$, and $\a$ and $\b$ are 
defined 
as in the
figure. $p_y$ and $p_u$ are the momentum 
of the out-coming particle in the $y$ and $v$-directions, respectively, {\it 
after} it passes the shock wave. These quantities are different from the 
momenta before the interaction, which we denote by $p_\m^0$. We do not 
explicitly write the superscripts in or out, as it should be clear from the 
context whether the momentum refers to the in-going or out-coming 
particle\footnote{In the remainder of this section we assume there is only 
one particle coming in.}.

We now take the initial transverse momentum to be zero, $p_y^0=0$. This means 
that $\a=\pi/2$ and hence, from \eq{18},
\be
\cot\a+\cot\b=\tan\g =\frac{1}{T}\,p_v\,\pa_yf(y_0).
\le{-3}
One can easily check that the exchange of momentum in the $v$-direction, to 
first order in $\ve$, is equal to zero and hence $p_u\simeq p_u^0$. 
Therefore, we have
\be
p_y=\frac{1}{T}\,p_up_v\pa_yf.
\le{-4}
Since $p_y\sim\frac{\pa y}{\pa v}$, this can also be directly deduced from 
\eq{18}.

If the initial transverse momentum is nonzero, differentiating \eq{7} once 
yields, at $v>0$,
\be
\frac{\pa u}{\pa v}=-\frac{1}{2T}\,\frac{\pa x^i}{\pa v}\,p_v\pa_if(\ti x_0),
\le{-5}
so for the out-coming particle we have
\be
p_v^\out=-\frac{1}{T}\,p^{0,\out}_ip^\in_v\,\pa^if(\ti x_0).
\le{-6}
The same relation is obtained from the mass-shell condition $p_\m p^\m=0$. 

From \eq{-4} and \eq{-6} we find that, roughly speaking, $\d p_\perp\sim 
p_\parallel\,\ve$ and $\d p_\parallel\sim p_\perp\,\ve$, and so if 
$p_\perp\ll p_\parallel$, the transfer of momentum in the transverse plane is 
much larger than in the longitudinal plane.

Furthermore, as the transfer of momentum in both the longitudinal and the 
transverse plane are ${\cal O}(\ve)$, they are negligible for large 
transverse separations (compared to the Planck length). That is the regime 
where the eikonal approximation is valid.

\section{The S-matrix}\label{The S-matrix}

The classical trajectories found in section \ref{ClscPlen} are enough to 
obtain the scattering amplitude of two particles in the eikonal 
approximation. In this approximation, the net effect of the presence of a 
shock-wave on another particle is a shift of the corresponding wave-function. 
Naively one would think that since the whole effect is only a shift, it can 
be gauged away with a suitable choice of co-ordinates. However, as argued in 
section \ref{TFT}, although locally on both sides of the shock-wave there is 
no effect, there is an important global effect which is the shift. This shift 
cannot be removed by a co-ordinate transformation, despite the suggestive 
form of the metric \eq{VVshock}. This can be more easily understood in 
analogy with the electromagnetic case \cite{g9607,JaKaOr}. When a charged 
particle is boosted towards the speed of light, the electromagnetic field 
$A_\m$ of the particle is pure gauge outside the light-cone of the particle, 
i.e. $A_\m=\half\pa_\m\L$ and so has no net physical effect there. However, 
the gauge field is discontinuous along the world line of the particle, and so 
the transformations needed to gauge it away are different on the future and 
past light-cones, $A_\m^\pm=\pm\half\,\pa_\m\L$ with $\L=Q/2\p\log |x|$, $Q$ 
being the charge of the particle. Therefore, the total effect is physical.

As explained in the introduction, the scattering amplitude computed from the 
shift \eq{shift20} is the Veneziano amplitude. The effective action that one 
finds after a Fourier transformation of the amplitude is
\be
S=\int\dd^2\ti x\left(-T\pa_iu\pa^iv +P_uu-P_vv\right).
\le{Smatrix}
This is nothing but a rewriting of \eq{S-m} for a flat background. As 
remarked in the introduction, this is the action of a non-linear sigma model 
with a coupling to an external source $P_\m$. The coupling constant is 
$T={1\over8\p\GN}$. The equations of motion following from this action 
directly lead to the geodesic equation in the eikonal approximation:
\be
\pa_i^2u(\ti x)&=&{1\over T}\,P_v(\ti x)\nn
\pa_i^2v(\ti x)&=&-{1\over T}\,P_u(\ti x),
\le{02}
which are solved by
\be
u(\ti x)&=&u_0-{1\over T}\int\dd^2\ti x'P_v(\ti x')\,f\tx\nn
v(\ti x)&=&v_0+{1\over T}\int\dd^2\ti x'P_u(\ti x')\,f\tx.
\le{03}

We first write these equations according to a 2+2-splitting of space-time. 
This is easy to do in the longitudinal plane. We find:
\be
X^a(\s)&=&x^a -{1\over T}\,\e^{ab}\int\dd^2\s'P_b(\ti\s)\,\fts.
\le{X}
The quantisation of this model has been discussed in \eq{commutators}. We 
get:
\be
[X^a(\s),X^b(\s')]=-{1\over T}\,\e^{ab}\fts.
\le{comm0}

Notice that the minus sign difference in \eq{03} is crucial to obtain the 
epsilon tensor. Indeed, had we guessed a relation of the type $\pa_i^2x^a\sim 
p^a$, then the right-hand side of \eq{comm0} would not have been 
antisymmetric and the model would have been inconsistent at the quantum 
level\footnote{Of course, the sign can be reabsorbed in the definition of 
momentum, but this leads to non-standard commutation relations and is 
therefore not very useful for the discussion of covariant generalisations.}. 
Indeed, due to the complete symmetry between $u$ and $v$, one's naive guess 
would have been a geodesic equation where both terms in \eq{02} have the same 
sign. However, the minus sign is directly linked to causality: one of the 
particles is in-going, whereas the other is out-going. We will comment some 
more on this in the conclusion.

The presence of an epsilon-tensor is also proven in \cite{VV} from the 
manipulations of the Einstein-Hilbert action coupled to massless particles in 
the eikonal limit, as reviewed in the introduction. 

Recall that the equation of motion \eq{02} is only valid in Minkowski space. 
Indeed, when the manifold is curved the shift function $f$ gets a mass term 
as shown in \eq{9}, and this is the equation one has to take as a starting 
point for more general backgrounds. It can be expressed in terms of the 
induced metric $h_{ij}$ if one considers that the second term on the 
left-hand side of \eq{9} is a relic of the two-dimensional Ricci-tensor. From 
the computation outlined in Appendix \ref{appA3}, we find that the Ricci 
tensor of the vacuum metric \eq{8.0} equals
\be
R_{ij}[G]=R_{ij}[h]- \frac{\pa_u\pa_vg}{A}\,h_{ij},
\le{51}
where $R_{ij}[G]$ is the transverse part of Ricci tensor obtained from the 
full four-dimensional Riemann tensor, see \eq{B3}, and $R_{ij}[h]$ is the 
Ricci tensor corresponding to the two-dimensional metric $h_{ij}$. Since 
$R_{ij}[G]$ satisfies the vacuum Einstein equations, the constraint reduces 
to
\be
R[h]=\frac{2}{A}\,\pa_u\pa_vg
\le{53}
for a two-dimensional metric $h$. This obviously gives the metric on the 
sphere if $g=r^2$ and $A=1$. We can write equation \eq{9} as
\be
\left(\LL_h-\frac{1}{2}\,R[h]\right)f&=&\frac{1}{\@{h}}\,\d ^{(2)}(\ti x-\ti 
x_{\sm{0}}).
\le{green}
It is now obvious how to include this extra term in \eq{02}:
\be
(\LL_h-\frac{1}{2}\,R[h])\,X^a= \frac{1}{2T}\,\e^{ab}\,P_b.
\le{eom1}
This equation is reminiscent of the focusing theorem. It is solved exactly as 
before,
\be
X^a(\s)=x^a+ {1\over2T}\int\dd^2\ti\s\,\@{h}\,\e^{ab}\,P_b(\s')\fts,
\le{eom2}
where $f$ is now the solution of the generalised Green equation \eq{green}.

It is now straightforward to find a ``covariant" generalisation of the action 
\eq{Smatrix} in the eikonal limit:
\be
S=-{T\over2}\int\dd^2\s\,\@{h}\,[h^{ij} \pa_iX^a\pa_jX_a +\half\,R[h]X^aX_a 
+{1\over T}\,\e^{ab}X_aP_b].
\le{covaction}
We put the word ``covariant" between quotation marks because the fields $X^a$ 
are still two-dimensional as we are still in the eikonal limit. Thus, 
covariance here is only with respect to the transverse co-ordinates.

An alternative way to derive this equation is by performing a Fourier 
transformation of the amplitude \eq{amplitude} with a generalised shift that 
satisfies \eq{green}. In the case that the metric $h$ is the metric on the 
unit sphere, we get the amplitude \eq{amplitude} computed for the 
Schwarzschild background.

Let us consider the symmetries of \eq{covaction} for a moment. First of all 
there is the Lorentz symmetry which we just referred to. It is interesting to 
note that this symmetry is induced by time translations in Rindler time, as 
shown in \eq{Lorentz}. Thus, one can say that time translations in the bulk 
induce Lorentz boosts on the boundary. This is very reminiscent of the 
relation between radial translations in the bulk and conformal 
transformations on the boundary for the case of AdS, although the groups are 
obviously different. It will be interesting to investigate the symmetries of 
the boundary action in the case of AdS.

The original 't Hooft action \eq{Smatrix} was invariant under Weyl rescalings 
of the boundary metric. In the general case we find that the term in 
\eq{covaction} proportional to the curvature explicitly breaks this symmetry 
and we are left with a global symmetry only.

\section{The eikonal limit and beyond}\label{beyondeikonal}

We have mentioned that shock-wave solutions are exact solutions of Einstein's 
equations, even if one includes any higher-curvature corrections, like for 
example the ones that appear in string theory. These come from the conformal 
invariance of the sigma-model. Furthermore one can compute the exact effect 
of the shock-wave on outgoing particles and the transfer of momentum. 
Therefore one can ask the question: what happens when one increases the 
energy up to the Planck scale and perhaps beyond? In other words, how does 
one go beyond the eikonal approximation? This is the question we are going to 
analyse in detail in this section.

't Hooft has suggested \cite{g94,g9607} that a covariant generalisation of 
the equations of motion \eq{eom1} should automatically account for the 
transfer of momentum\footnote{By covariant we really mean covariant with 
respect to 4-dimensional diffeomorphisms.}. However, as we will explain 
later, it is extremely difficult to find a consistent generalisation of this 
formula in four dimensions.

Instead, we will choose another approach here. In \eq{-4} and \eq{-6} we 
found the exact momentum transfer. These formulae indeed hold without any 
approximations. So we will write these formulae in a manifestly covariant 
form, and will then discuss quantisation. The covariant expression will 
automatically account for the transfer of momentum. In this section we will 
study this formula in detail, and in later sections we show that it is 
consistent with quantisation. Let us first give the expression:
\be
P^\m_\out(\ti\s)=(g^{\m\n}+A^{\m\n})P_\n^{0,\out}(\ti\s),
\le{eomP}
where
\be
A^{\m\n}(\ti\s)=-\frac{1}{T}\,\e^{\m\n\l\r}
\e^{ij}\pa_iX_\l(\ti\s)\int\dd^2\ti\s'\,P_{\r,\in}(\ti\s')\,\pa_jf(\ti\s-\ti\
s').
\le{AAmunu}
For the in-operators, one interchanges the labels in-out in the above 
expression. The quantities $P^0_\out$ and $P_\out$ are the momenta of the 
out-coming particle before and after the interaction, respectively.

\begin{center}
        \begin{figure}[h] \hspace{4cm}
        \psfig{figure=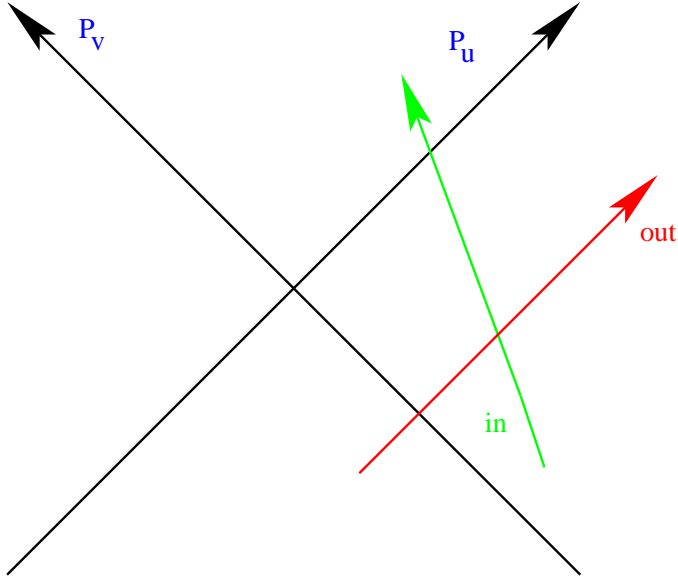}
        \caption{Collision at non-zero angle}
	  \label{fig3}
        \end{figure}
\end{center}

Let us now check that this covariant expression reproduces the momentum 
transfer computed before. To that end, we first have to set up some 
notation. We use the notation of reference \cite{VV}, explained in section 
\ref{TFT} of the introduction. Four-dimensional fields $X^\m$ split into a 
longitudinal and a transverse component, $X^\m=(X^a,Y^m)$. The internal 
co-ordinates $x^\a$ and $y^i\equiv\s^i$ are (in the usual gauge) the zero 
modes of $X^a$ and $Y^m$, respectively, roughly: $X^a=x^\a+\cdots$ and 
$Y^m=\s^i+\cdots$. In view of this, and since we will be making a distinction 
between $(X,Y)$ and $(x,y)$, the indices $a$ and $\a$ can be identified, and 
also $m$ and $i$ (but notice that $X^\a\not=x^\a, Y^i\not=\s^i$). We analyse 
the case when the background is Minkowski. Both for the out- and the 
in-particles we have $P_\parallel=P_\a=(P_u,P_v)$, $P_\perp=P_i$. We will 
consider the change of momentum for the out-coming particles produced by the 
in-going particles, but  the expressions for the in-going particles are 
trivially obtained by exchanging the labels ``in" and ``out". The kinematics 
is illustrated in Figure \ref{fig3}. 

There is still a point in using this $(2{+}2)$-splitting of space-time even 
if the transverse momentum is not zero, because the longitudinal and 
transverse momenta behave differently in the first few orders in the eikonal 
approximation. We get:
\be
P_i^\out(\ti\s)&=&P_i^{0,\out}(\ti\s) +\frac{1}{T}\, 
P^{0,\out}_v(\ti\s)\int\dd^2\ti\s'
\,P^\in_u(\ti\s')\,\pa_i\fts\nn
&-&\frac{1}{T}\, P^{0,\out}_u(\ti\s)\int\dd^2\ti\s'\,
P^\in_v(\ti\s')\,\pa_i\fts +{\cal O}(\ve^2)\nn
&=&P_i^{0,\out}(\ti\s)+\frac{1}{T}\,\e^{ab}P^{0,\out}_a\int\dd^2\ti\s'\,P^\in
_b\,\pa_i\fts.
\le{99}
Notice that if the operator on the left-hand side of \eq{99} carries an 
out-label, then the operator on the right-hand side of \eq{100} which is 
evaluated at $\ti\s$ corresponds to the same out-particle, whereas the 
operators which are integrated over give the contributions from the 
in-particles. The same is true if one reverses the labels.

Note that even if the initial transverse momentum is zero, like in head-on 
collisions, it will be non-vanishing after the interaction. The two particles 
will spin around each other for a short time. This agrees with equations 
\eq{-4} and \eq{-6}, which were obtained from kinematical  considerations. If 
the momentum of the out-going particle satisfies $p_\parallel^\out\gg 
p_\perp^\out$, using the equation for the shift
\be
X^a(\ti\s)=x^a+\frac{1}{T}\int\dd^2\s'\,\e^{ab}P_b(\s')\,f(\s-\s')
\le{100b}
we find from \eq{99}
\be
P_i(\ti\s)=P_i^0(\ti\s)+P_a^0\pa_iX^a,
\le{100}
to first order in $\ve$. This expression was found in \cite{g9607} from the 
consideration that the transverse momentum is not an independent variable, 
together with the requirement that the transverse momentum generates 
transverse translations. Here we see that it straightforwardly follows from 
the transfer of momentum during the collision.

The transverse momentum \eq{99} can also be written as
\be
P_i(\ti\s)&=&P_i^0(\ti\s)+\frac{\e}{T}\,P^0_a(\ti\s)\int\dd^2\ti\s'\,P_0^a(\t
i\s')\,\pa_if(\ti\s-\ti\s'),
\le{100c0}
where $\e=1$ if $P_i$ is an operator corresponding to the in-going particles 
and 
$\e=-1$ for the out-operators. This is the usual sign convention, where all 
in-going momenta are defined to be positive, and out-coming momenta to be 
negative \cite{g9607}. Indeed, if initially the in-going particles only have 
momentum $P_v$, and the out-coming ones only momentum $P_u$, \eq{100c0}
gives
\be
P_i^{\sm{in}}(\ti\s)&=&P_i^{0,\in}(\ti\s)+\frac{1}{T}\,P_{v,0}^\in(\ti\s)\int
\dd^2\ti\s'\,P^{\sm{out}}_u(\ti\s')\,\pa_i\fts\nn
P_i^{\sm{out}}(\ti\s)&=& 
P_i^{0,\out}(\ti\s)-\frac{1}{T}\,P_{u,0}^\out(\ti\s)\int\dd^2\ti\s'\,
P_v^{\sm{in}}(\ti\s') \,\pa_i\fts.
\le{100c}

The next nontrivial check concerns the longitudinal momentum transfer. Using 
\eq{eomP}, we find
\be
P_u(\ti\s)^\out&=&P_u^{0,\out}(\ti\s) 
-\frac{1}{T}\,P^i_0\int\dd^2\ti\s'\,P_u\,\pa_if 
+\frac{1}{T}\,P_u^0\int\dd^2\ti\s'\,P^i\,\pa_if,\nn
P_v^\out(\ti\s)&=&P_v^{0,\out}(\ti\s) 
+\frac{1}{T}\,P^i_0\int\dd^2\ti\s'\,P_v\,\pa_if 
-\frac{1}{T}\,P_v^0\int\dd^2\ti\s'\,P^i\,\pa_if.
\le{101}
In covariant $(2{+}2)$-notation,
\be
P^a_\out(\ti\s)&=&P^a_{0,\out}(\ti\s) 
+\frac{1}{T}\,\e^{ab}\,P_b^{0,\out}(\ti\s) \int\dd^2\ti\s' 
P^i_\in(\ti\s')\,\pa_if(\ti\s-\ti\s')\nn
&-&\frac{1}{T}\,\e^{ab}\,P^i_{0,\out}(\ti\s) \int\dd^2\ti\s' 
P^\in_b(\ti\s')\,\pa_if(\ti\s-\ti\s')\nn
&=&P^a_{0,\out}(\ti\s) -P^i_{0,\out}(\ti\s)\,\pa_iX^a(\ti\s)\nn
&+&{1\over T}\,\e^{ab}P^0_{b,\out}(\ti\s)\int\dd^2\ti\s' P^i_\in(\ti\s') 
\pa_if(\ti\s-\ti\s').
\le{102}
Again, it perfectly agrees with \eq{-6} in the corresponding limit, 
$P^i_\in=0$. Notice that the transfer of longitudinal momentum is zero if the 
initial transverse momentum is zero. So, although for vanishing initial 
transverse momentum there is still a transverse momentum transfer, in the 
longitudinal plane this transfer is zero to first order in $\ve$.

Notice that in the situation that is usually considered, $P^i_\in=0$, 
equations \eq{100} and \eq{102} can be rewritten as
\be
\d P_i&=&+W_i^aP_a^0\nn
\d P^a&=&-W_i^aP^{i,0}.
\ee
where we defined $W_i^a=\pa_iX^a$. Since $W_i^a$ is proportional to the 
vector field $V_i^\a$ in formula \eq{vectorfield} by $W_i^a=V_i^\a\pa_\a 
X^a$, we see that the latter is responsible for the transfer of momentum. The 
analogy with fluid dynamics suggested in \cite{VV} becomes more transparent 
from this computation: this vector field accounts for the ``flow" or 
``vorticity" of momentum during collisions at high energies.

\section{Quantisation}\label{Quantisation}

A full quantum theory for this non-linear four-dimensional model is extremely 
difficult to write down away from the eikonal limit. To quantise the theory 
we have to give a complete set of observables and the way they act on states 
in Hilbert space. Now in this model the space-time co-ordinates are not 
independent, but are related by shift equations. Upon expanding the fields 
into eigenmodes of this equation of motion with the corresponding creation 
and annihilation operators, we will find that co-ordinates do not commute. 
This result has been known for a long time (see \cite{g9607} and references 
therein). Quantisation in the eikonal limit is quite straightforward and 
gives rise to non-commuting co-ordinates \eq{comm0}. However, beyond the 
eikonal approximation we encounter non-linearities which are difficult to 
deal with. We anticipate that we will not be able to give an exhaustive list 
of commutation rules among all operators beyond the eikonal limit, for 
basically the same reason it was not found in earlier works 
\cite{g9607,SdHJHEP}. Instead, we will give a complete set of commutators 
between the momenta, $P$. These commutators of course satisfy the Jacobi 
identity. It is however not clear how to derive a well-defined commutator 
between the $X$'s. One would think it can be obtained from a generalisation 
of \eq{eom2} or \eq{comm0}, but this is not straightforward as these 
equations become highly non-linear at low impact parameter. Part of the 
problem also stems from the fact that the commutator is non-local and so does 
not transform properly under non-linear co-ordinate transformations of the 
world-sheet co-ordinates. In fact, it is not even clear whether $X^\m$ is a 
good starting point to define a quantum theory that incorporates non-linear 
effects, as it does not transform as a vector in target space. 't Hooft has 
stressed \cite{g9607} that a consistent quantisation scheme can perhaps be 
found if one introduces variables that are better behaved. In the next 
section we will see the derivative of $X$, $\pa X$, is a better physical 
observable. In this section we will concentrate on the operator $P$, which is 
also well behaved as it is a natural object of the tangent space.

As said, here we will assume the canonical commutator between position and 
momentum operators, and find commutation relations for the momenta from the 
momentum transfer equation \eq{eomP}. We will see that one does find a set of 
commutation rules that is consistent, where the momentum operator has the 
usual interpretation as the generator of translations. We will find that the 
commutators appearing in \cite{g94}, postulated from the condition that 
momentum operators generate translations, automatically follow from our 
equations of motion. We also find new commutators which close the algebra of 
momenta.

As for the commutators between the $X$'s, these should follow from the Jacobi 
identity. Indeed, in three dimensions we have solved the Jacobi identity and 
recovered the results which were found in \cite{g9805} and \cite{SdHJHEP} 
directly from the shock-wave equations of motion. However, we have not been 
able to integrate the equation in four dimensions.

We first consider the action of the operators $\^P^\m$ and $\^X^\m$ on state 
vectors $|P_0\ket$ and $|X\ket$. We obviously have
\be
\^P^\m|P_0\ket=P_0^\m|P_0\ket;\nn
\^X^\m|X\ket=X^\m|X\ket.
\le{79}
Furthermore, the operators $\^P$ and $\^X$ satisfy the usual commutation 
relation
\be
&[\^X^\m(\ti\s),\^P^\n(\ti\s')]=ig^{\m\n}\,\dts.
\le{80}
Indeed, at the level of the path integral and in the eikonal limit these 
operators were related by a Fourier transformation \cite{SdHCQG}. From now on 
we drop the carets on operator-valued quantities.

The quantum theory will however be an interacting theory, and has to take 
\eq{eomP} into account. Therefore, just as in the eikonal limit \eq{X} was 
promoted to an operator identity, leading to \eq{comm0}, our assumption will 
be that \eq{eomP} is also a relation between a free operator $P_0$ and the 
interacting field $P$. We get the following modified commutator:
\be
&[X^\m(\ti\s),P^\n(\ti\s')]= i\left(g^{\m\n}+A^{\m\n}\right)\dts.
\le{Amunu}
This simply means that, due to the back-reaction, the number of independent 
measurements one can do simultaneously is reduced according to:
\be
\D x \D p\geq\frac{\hbar}{2}+{\cal 
O}\left(\frac{\Pl^2\,p_{\sm{in}}}{b}\right).
\le{92}
The modification of the canonical commutation relation \eq{80} in the 
presence of gravitational interactions has also been predicted (although in 
different contexts) by several authors \cite{Maggiore,Kempf,max}.

The generalised commutator \eq{Amunu} has a simple interpretation if we go 
back to the underlying shock-wave picture. Before the interaction takes 
place, the different momenta are independent variables. However, {\it after} 
the interaction, they are coupled through the momentum transfer equation 
\eq{eomP}. Then the longitudinal momenta generate sideways displacements as 
well. So it is natural to identify the canonical momentum $P^\m _{\sm{can}}$ 
with the momentum before the interaction, which we denote by $P^\m_0$, and 
$P^\m$ with the momentum after the collision. The latter describes the 
momentum transfer, and can be seen to be a measure for the recoil of the 
particles. This holds both for the in-going and the out-coming particles. 
Although $P^\m$ is not a canonical operator, when writing \eq{Amunu} out in 
components we will see that it generates translations in the sense of field 
theory. The situation here is similar to cases with background 
electromagnetic fields, take for example a particle in an electromagnetic 
field. In that case, the kinetical momentum, which is the operator that (by 
Ehrenfest's theorem) satisfies the classical equation of motion, is not the 
canonical momentum operator.

Let us now take a closer look at the commutation relation \eq{Amunu},
\be
[X^\m(\ti\s),P^\n(\ti\s')]=iG^{\m\n}\,\dts,
\le{104}
where the ``generalised metric" is defined as
\be
G^{\m\n}\equiv g^{\m\n}+A^{\m\n}.
\le{105}
Writing \eq{104} out in components, we find
\be
{}[u(\ti\s),p_i(\ti\s')]&=&i\pa_iu\,\dts;\nn
{}[v(\ti\s),p_i(\ti\s')]&=&i\pa_iv\,\dts;\nn
{}[Y^m(\ti\s),p_u(\ti\s')]&=&i\pa_uY^m\,\dts;\nn
{}[Y^m(\ti\s),p_v(\ti\s')]&=&i\pa_vY^m\,\dts.
\le{106}
In the 2+2 splitting, this can be reexpressed as
\be
{}[X^a(\ti\s),p_i(\ti\s')]&=&i\pa_iX^a(\ti\s)\,\dts\nn
{}[Y^m(\ti\s),p_\a(\s')]&=&i\pa_\a Y^m(\ti\s)\,\dts.
\le{107}
In the gauge where longitudinal indices $\a$ are along $X^a$, and transverse 
indices $i$ are along $Y^m$, defining $\d X^a=X^a-x^a$ and $\d Y^m=Y^m-\s^m$, 
we have:
\be
\pa_\a\d Y_i+\pa_i\d X_\a=0.
\ee

Note that the operators $P_\m$ are not usual translation operators. They 
rather generate translations of the fields $X^\m$ along the internal 
directions.

In quantum mechanics, co-ordinates are independent of each other, and so the 
right-hand side of \eq{104} reduces to the canonical commutator 
$ig^{\m\n}\,\dts$. But in our case we have a two-dimensional field theory 
where the longitudinal and the transverse co-ordinates become mutually 
dependent fields. This renders \eq{104} non-vanishing even if the indices 
$\m$ and $\n$ are different (notice that, for $\m\not=\n$, \eq{104} is 
nonzero if one of the indices is transverse, say $i$, and the other one is a 
longitudinal index $\a$). So $p$ generates translations just as in field 
theory, as one directly sees from \eq{106}.

One can also get an algebra for the commutator of the $p$'s among themselves. 
One finds (the operators referring all to the in- or all to the out-states)
\be
{}[p_\a(\ti\s),p_i(\ti\s')]&=&ip_\a(\ti\s')\,\pa_i\dts;\nn
{}[p_i(\ti\s),p_j(\ti\s')]&=&ip_i(\ti\s')\pa_j\dts +ip_j(\ti\s)\pa_i\dts,
\le{108}

Now we can also obtain an algebra that relates the in- and the out-operators. 
Using \eq{108}, we
get:
\be
{}[p_v^{\sm{in}}(\ti\s),p_i^{\sm{out}}(\ti\s')]&=&-iT\,\pa_iu(\ti\s')\,f^{-1}
\ts;\nn
{}[p_u^{\sm{out}}(\ti\s),p_i^{\sm{in}}(\ti\s')]&=&iT\,\pa_iv(\ti\s')\,f^{-1}\
ts;\nn
{}[p_i^{\sm{in}}(\ti\s),p_j^{\sm{out}}(\ti\s')]&=&-iT\,\pa_iv(\ti\s)\,\pa_ju(
\ti\s')\,
f^{-1}\ts\nn
&+&\frac{i}{T}\,p_v^{\sm{in}}(\ti\s)\,p_u^{\sm{out}}(\ti\s')\,\pa_jf\ts.
\le{110}
In reference \cite{g9607} it was not possible to find correct expressions for 
the commutators between in- and out-operators. The expected expression for 
the last of \eq{110} did not satisfy the Jacobi identity when combined with 
\eq{108}. One can check that the above expression does satisfy the Jacobi 
identity.

The algebra \eq{110} is very non-local and, furthermore, non-linear. It, 
however, can be significantly simplified by defining the total momentum
\be
P_\m=\int\dd^2\ti\s\,p_\m(\ti\s).
\le{111}
This leads to the following local expressions:
\be
{}[p_i^{\sm{in}}(\ti\s),P_j^{\sm{out}}]&=&i\pa_jp_i^{\sm{in}}(\ti\s);\nn
{}[p_i^{\sm{out}}(\ti\s),P_j^{\sm{in}}]&=&i\pa_jp_i^{\sm{out}}(\ti\s);\nn
{}[p_\a^{\sm{in}}(\ti\s),P_i^{\sm{out}}]&=&i\pa_ip_\a^{\sm{in}}(\ti\s);\nn
{}[p_\a^{\sm{out}}(\ti\s),P_i^{\sm{in}}]&=&i\pa_ip_\a^{\sm{out}}(\ti\s),
\le{112}
so the total transverse momentum generates translations.

One can check that the algebra between the transverse in-operators or the 
out-operators is similar to \eq{112}. However, we do not expect the 
theory to have
two different generators of transverse translations. So we expect
\be
\d P_{\sm{in}}^i=\d P^i_{\sm{out}}.
\le{113}
Integrating \eq{99} we indeed see that this is the case. The same holds for 
the 
lightcone
directions, as one sees from equation \eq{101}. Therefore, for the integrated 
momentum
operators we get the constraint
\be
\d P^\m_{\sm{in}}=\d P^\m_{\sm{out}}.
\le{114}
Recalling that these operators give the momentum transfer, this is nothing 
else than the expression of the conservation of momentum. As a constraint on 
Hilbert space, in our case it is also equivalent to the usual asymptotic 
completeness \cite{GaPa} of the in- and out-Hilbert spaces.

Equation \eq{114} implies that momentum is a globaly conserved quantity. But 
locally it is not conserved, as one can see from the individual local 
expressions. Only after integrating over $\ti\s$ the total momentum is 
conserved. This is also the usual expectation in field theory.

Recalling that we started off regarding the Minkowski plane as the 
near-horizon region of a Schwarzschild black hole, we have shown that one can 
go beyond the eikonal approximation and compute the momentum transfer, 
thereby respecting momentum conservation which is a minimal requirement for 
the unitarity of the S-matrix. The assumption that the S-matrix is unitary 
was the starting point of 't Hooft's considerations, as explained in the 
introduction. We now see that this assumption leads to a consistent algebra 
of momenta. In fact, it would be interesting to take the algebra 
\eq{108}-\eq{110} as the starting point of some field theory, the momentum 
being related to the stress-energy tensor in the usual way, and to study the 
Hilbert-space structure of this theory.

Since the results presented in this section are valid to the first 
non-trivial order in the eikonal approximation, it seems that the framework 
developed in \cite{VV} would be most appropriate to do an additional check of 
our results, and would perhaps provide some more conceptual insight in the 
near-eikonal regime of quantum gravity.

Some of the results in this section had already been found in \cite{g94} from 
general considerations. Here we learn that they straightforwardly follow when 
recoil effects are taken into account. Furthermore, we also get the 
additional equations \eq{110} and \eq{112}, which close the algebra. In the 
next section we perform another check of \eq{eomP}.

\section{Quantum gravity in 2+1 dimensions}\label{2+1}
When looking for a formulation of the S-matrix beyond the eikonal 
approximation, in four dimensions one encounters several problems 
\cite{g9607,SdHJHEP} that originate in the non-linearity of the equations. 
Indeed, as the dimension increases the equations become more and more 
non-linear \cite{SdHCQG}. However, when one reduces to 2+1 dimensions things 
simplify considerably as the algebra becomes linear.

In this section we compactify one of the space-time (and world-sheet) 
directions on a small circle of radius $R$ and assume the three remaining 
fields $X^\m$ to be independent of this internal dimension. We also assume 
that the momentum along this direction is zero and hence we only take the 
zero modes into account. For a complete theory one should of course also 
consider the excited modes. 

There are several ways to set up the theory. In references 
\cite{g9805,SdHJHEP} it was chosen to find the commutator for the $X$ fields 
from a covariant generalisation of the dimensionally-reduced system. 
Reference \cite{g9805} wrote the covariant formula only after deriving the 
commutators, whereas in \cite{SdHJHEP} the equation of motion for $X$ was 
first covariantly generalised, and from there the commutator was found. Both 
approaches gave the same result. In reference \cite{SdHJHEP} the commutator 
between the $X$'s and $P$'s was then found from the Jacobi identity, and it 
was checked that it agrees with the commutator one finds if one directly 
dimensionally reduces \eq{Amunu}. It was concluded that the covariant 
generalisation of the algebra is directly related to the transfer of 
momentum. This also served as a check of the four-dimensional algebra, which, 
as stressed in \cite{SdHJHEP}, is not free of problems.

Here we choose an alternative route, which, as we will see, is equivalent to 
that of \cite{SdHJHEP} and provides a nice check of our formulae in the 
previous section. We take the expression for the commutator between $X$ and 
$P$, equation \eq{Amunu}, as our starting point for the dimensional 
reduction. In four dimensions this expression comes from the transfer of 
momentum, \eq{eomP}. We then use the Jacobi identity to find the commutator 
between the $X$'s.

As said, both methods give the same results. The advantage of the latter 
method is that it only assumes transfer of momentum and not knowledge of the 
commutator between the $X$'s. Furthermore, in principle this method can be 
generalised to higher dimensions, where we can compute the transfer of 
momentum as in the previous sections, but we do not have a fully consistent 
equation of motion for $X$ for the reasons explained in \cite{SdHJHEP}. 
Solving the Jacobi identity should give the equation of motion for $X$. 
Nevertheless, we have not been able to find a solution to the Jacobi identity 
in four dimensions, although we do not see any reason why it should not have 
a solution.

We parametrise the compactified dimension by $\s_2=y$, $0\leq y\leq R_3$. We 
define $\s=\s_1$, $\pa={\pa\over\pa\s}$ and $\e_{\m\n\l}=\e_{\m\n\l y}$. 
Notice that the effective 2+1-dimensional Newton's constant is obtained from 
the 3+1-dimensional one by
\be
G_3={G_4\over R}.
\ee
Therefore we will find an effective coupling $T={RT_4}$. 

Since the momentum $p^\m(\s_1,\s_2)$ is a momentum density, we have to 
integrate over the internal direction to obtain the observable momentum from 
the three-dimensional point of view: $P^\m(\s_1) 
=\int\dd\s_2\,p^\m(\s_1,\s_2)$.

In 2+1 dimensions, \eq{Amunu} becomes
\be
{}[X^\m(\s),P^\n(\s')]=i\left(g^{\m\n}-\frac{1}{T}\,\e^{\m\n\l}\int\dd\s''\,P
_\l(\s'')\,\pa f(\s-\s'')\right)\d (\s-\s')
\le{2.21}
where the shift function is now given by $f(\s-\s') =\half|\s-\s'|$.

We can obtain the commutator between two $X$'s from the Jacobi identity. As 
remarked in \cite{SdHJHEP} and stressed in previous sections, it is better to 
consider its derivative, $\pa X^\m$, rather than $X$ itself, because the 
former satisfies a local algebra. So we work out the following relation:
\be
{}[{}[\pa X^\m(\s),P^\n(\s')],\pa X^\l(\s'')] +\mbox{cyclic}=0.
\ee
We get the following solution:
\be
[\pa X^\m(\s),\pa X^\n(\s')]= -{i\over T}\,\e^{\m\n\l}g_{\l\r}\,\pa 
X^\r(\s)\,\dts.
\le{2+1algebra}
This is the SO(2,1) algebra obtained in \cite{g9805,SdHJHEP}.

Following \cite{g9805}, the presence of the delta-function in \eq{2+1algebra} 
suggests to define the following integrated variables:
\be
x_A^\m=\int_A\dd\s\,\pa x^\m=x^\m(A_1)-x^\m(A_0),
\le{2.15b}
where $A$ is an interval $A=[A_0,A_1]$ along the line $\s$.

These variables have the nice property:
\be
[x^\m_A,x^\n_A]=-\frac{i}{T}\,\e^{\m\n\l}g_{\l\r}\,x^\r_A.
\le{2.15c}
As argued by 't Hooft, this gives rise to a time variable that is quantised 
in units of $t_{\sm{Pl}}/R$.

Another useful quantity is the total momentum flowing through $A$,
\be
p^\m_A\equiv\int_A\dd\s\,P^\m(\s).
\le{2.16}
The commutator then becomes
\be
[x^\m_A,p^\n_A]=iG^{\m\n},
\le{2.17}
with the ``generalised metric"
\be
G^{\m\n}=g^{\m\n}-\frac{1}{T}\,\e^{\m\n\l}g_{\l\r}\,p^\r_A.
\le{2.18}

The same results can be derived \cite{SdHJHEP} from a covariant 
generalisation of the three-dimensional equation of motion \eq{eom2}:
\be
\pa^2x^\m=\frac{1}{T}\,\e^{\m\n\l}g_{\l\r}\,\pa x^\r\,p_\n.
\le{2.11}

One can also work out the commutation relations in a way analogous to 
equations \eq{106}-\eq{107}, finding that $P$ again has an interpretation as 
the generator of translations:
\be
{}[u_A,p_x^A]&=&i(\pa u)_A\nn
{}[v_A,p_x^A]&=&i(\pa v)_A,
\le{2.20}
etc., in a way analogous to the 3+1-dimensional case.

It is particularly beautiful that the link between a covariant algebra, where 
all co-ordinates are treated on the same footing, and the inclusion of the 
transverse gravitational force, can be made so precise in 2+1 dimensions: 
including the transverse gravitational force leads to an algebra that is 
invariant under the full three-dimensional Lorentz group, and viceversa: 
writing the algebra in a manifestly SO(2,1) invariant form automatically 
accounts for transverse effects.

\section{Second quantisation of gravitationally interacting particles} 
\label{2ndquant}

Gravitational interactions at high energies lead to a non-commutative 
space-time. One can wonder what consequences this has for fields that live on 
this space-time. In reference \cite{VVexchange}, it was found that taking 
into account the back-reaction of particles on a black-hole horizon leads to 
quantised fields that satisfy a so-called exchange algebra. This exchange 
algebra exhibits great similarity with the Moyal product defined in 
non-commutative gauge theories.

The computation of \cite{VVexchange} uses 't Hooft's results to model a 
forming black hole with a horizon that fluctuates in time. The formation of 
the future horizon depends on the time of arrival of in-coming particles, and 
thus it matters whether we first add in-going particles and then measure the 
positions of out-going particles, or viceversa.

In this section we show that this effect is not at all an exclusive feature 
of time-dependent black holes (although black-holes are the natural scenario 
where these effects become important). Gravitationally interacting fields in 
Minkowski space already obey such an exchange algebra if they interact 
gravitationally. All the considerations in this section are independent of 
the dimension, except for the details of the eikonal approximation. This 
section is based on \cite{SdHHV}.

Consider two massless particles in Minkowski space. Particle 1 is ``hard" and 
carries a shock-wave with it, whereas particle 2 is ``soft" and so its 
back-reaction can be neglected. Particle 1 is a left-mover with momentum 
$k^-$ along $x^-$, and particle 2 is a right-mover with momentum $k^+$ along 
$x^+$. When particle 2 crosses the trajectory of particle 1 at $x^+=0$, it 
will get shifted:
\be
\d x^-=k^- f,
\le{shift2}
and the impact parameter is kept fixed. 

Next we consider quantised fields in this Minkowski background. For the 
moment we restrict ourselves to fields with no transverse momentum. These 
fields fall apart into a $+$ and a $-$ component:
\be
\phi(x^+,x^-)=\phi_+(x^+)+\phi_-(x^-).
\le{7.1}
Therefore, the Hilbert space decomposes into a left- and a right-moving part.

To have an S-matrix description, we must have some notion of asymptotic 
states. Because of \eq{7.1}, the Hilbert space of the in-states will fall 
apart into:
\be
|\mbox{in}\ket_-|\mbox{in}\ket_+,
\ee
and likewise for the out-states. The S-matrix will relate both sets of 
states. In a momentum representation, if there are for example $N$ in-going 
particles with momentum along the $x^-$-direction, we have a state
\be
|k^-_1,\cdots,k^-_{\tn{N}}\ket_{\tn{in},-}.
\le{7.2}
We now consider creation and annihilation operators of particles at $I_-$ and 
$I_+$. A creation operator $a_{+}^\dagger(k^+)$ that naturally acts on an 
in-state is defined by
\be
a_{+}^\dagger(k^+)|0\ket_{\tn{in},+}= |k^+\ket_{\tn{in},+},
\ee
and likewise for the $x^-$-direction (see Figure \ref{fig4}). We require 
these operators to satisfy the usual commutation rules
\be
[a_\a(k),a_\b^\dagger(k')]=\d(k-k')\,\d_{\a\b},
\le{7.3}
where the Greek indices stand for $+$ or $-$\footnote{This definition is 
slightly different from, but completely equivalent to, the usual Fock space. 
In usual Fock space, the states are characterised by the occupation numbers 
$|\{n_k\}\ket$. This is a more economic arrangement of the state \eq{7.2}, 
but it is not useful for our purposes.}. For the out-states we have a similar 
definition, and the corresponding operators will be called $b_\a(k)$.

We now consider the commutation rules between in- and out-operators. In the 
absence of any interactions, the S-matrix is simply unity and so the Hilbert 
spaces $|\mbox{in}\ket_+$ and $|\mbox{out}\ket_+$ are identified. $a_\a$ and 
$b_\b$ then satisfy
\be
[a_+(k),b_+^\dagger(k')]&=&\d(k-k')\nn
{}[a_-(k),b_-^\dagger(k')]&=&\d(k-k').
\le{7.4}
The $+$ and the $-$-operators mutually commute in this case.

\begin{center}
        \begin{figure}[h]\hspace{4cm}
        \psfig{figure=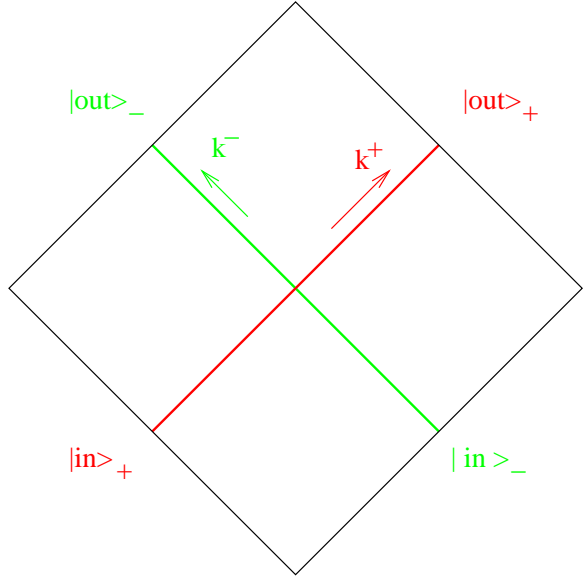}
        \caption{Asymptotic states in a two-particle collision}
	  \label{fig4}
        \end{figure}
\end{center}

We now include the gravitational interaction \eq{shift2}. We assume that the 
operators $a_-$ and $a_+$, and $b_-$ and $b_+$, will still 
commute\footnote{This is actually different from the philosophy advocated by 
't Hooft, who considered non-vanishing commutators for operators at spacelike 
separated distances, although still preserving causality. Notice, however, 
that even if two operators act at the same point of the light-cone 
$x^+$-$x^-$, they still can be separated by a spacelike distance since there 
is still a large transverse separation $\ti x-\ti y$. The extended nature of 
the shock-wave makes it impossible to avoid non-locality.}. Furthermore, 
since in the shock-wave approximation there are no self-interactions, 
\eq{7.4} still holds. When transforming a state $|\mbox{in}\ket_+$ into a 
state $|\mbox{out}\ket_+$, the S-matrix element is still trivial.

The interaction \eq{shift2} gives something non-trivial when one considers 
the commutators $[a_+,b_-]$ and $[a_-,b_+]$. In these cases, one has to take 
into account the shift \eq{shift2}.

The proposal is that in-going operators act also on the Hilbert space of 
out-going particles. When we add an in-going particle with momentum $k^-$, we 
are also shifting the trajectories of out-going particles with momentum along 
$x^+$ by the amount \eq{shift2}. So we define the operator $a_-(k)$ to act on 
out-states as follows:
\be
a_-^\dagger(k)|0\ket_{\tn{out},-} |k^+_1,\cdots,k^+_{\tn{N}}\ket_{\tn{out},+} 
&=&\exp\left[-i\sum_{i=1}^{\tn{N}}k^+_i k^-f(\ti x-\ti x^i)\right]\times\nn
&&\times |k^-\ket_{\tn{out},-} |k^+_1,\cdots,k^+_{\tn{N}}\ket_{\tn{out},+}.
\le{7.5}
So this operator translates out-coming particles by the corresponding shift. 
One can check that the states created by $a_-^\dagger$ form a natural set of 
states for the out-going Hilbert space. Indeed, solving the Klein-Gordon 
equation in a shock-wave geometry one finds that the complete set of 
wave-functions are not simply plane waves, but rather plane waves translated 
over the corresponding shift. Equation \eq{7.5} also defines the S-matrix.

Notice that the shift $f$ depends on the transverse position of each 
particle, but this has no meaning in a momentum representation, where we have 
taken $\ti k\approx0$, since in principle such a particle cannot be 
localised. However, one can neglect this effect as long as the transverse 
distances are large, so that quantum fluctuations are small. As soon as the 
transverse distance becomes small, one also has to take transverse momentum 
transfer into account, and the eikonal approximation \eq{7.5} is no longer 
valid.

We are now in a position to compute the difference between the products 
$a_-^\dagger b_+^\dagger$, $b_+^\dagger a_-^\dagger$:
\be
&b_+^\dagger(k)\,a_-^+(p)& 
|k^1,\cdots,k^{\tn{N}}\ket_{\tn{out},+}|p^1,\cdots,p^{\tn{M}} 
\ket_{\tn{out},-}= \nn
&&=\exp\left[-i\sum_{i=1}^{\tn{N}}k_i^+p^-f(\ti x-\ti x^i)\right] \times\nn
&&\times\,\,\,\,|k,k^1,\cdots,k^{\tn{N}}\ket_{\tn{out},+}|p,p^1,\cdots,p^{\tn
{M}} \ket_{\tn{out},-};\nn
&a_-^+(p)\,b_+^\dagger(k)& 
|k^1,\cdots,k^{\tn{N}}\ket_{\tn{out},+}|p^1,\cdots,p^{\tn{M}} 
\ket_{\tn{out},-}=\nn
&&=\exp\left[-i\sum_{i=1}^{\tn{N}}k_i^+p^-f(\ti x-\ti x^i)-ik^+p^-f\right] 
\times\nn
&&\times\,\,\,\,|k,k^1,\cdots,k^{\tn{N}}\ket_{\tn{out},+}|p,p^1,\cdots,p^{\tn
{M}} \ket_{\tn{out},-}.
\ee
Since the states considered here are arbitrary, we conclude that
\be
b_+^\dagger(k^+)a_-^\dagger(k^-)= 
\exp\left[-ik^+k^-f\right]a_-^\dagger(k^-)b_+^\dagger(k^+).
\le{7algebra}
Of course, the commutation rules for the annihilation operators can be 
obtained by replacing $k\rightarrow -k$.

Notice that the exchange factor \eq{7algebra} generates shifts both in the 
$+$ and in the $-$-direction, depending on the state it acts on. Therefore we 
require that $b_-$ and $a_+$ obey the same algebra:
\be
a_+^\dagger(k^+)b_-^\dagger(k^-)= \exp\left[-ik^+k^-f\right] b_-^\dagger(k^-) 
a_+^\dagger(k^+).
\ee

One can now define scalar fields $\phi(x^\pm)_{\tn{in},\pm}$, 
$\phi(x^\pm)_{\tn{out},\pm}$ in terms of these operators:
\be
\phi_{\tn{in},+}(x^+) &=&\int\dd k_+\,a_-(k_+)\,e^{ik_+x^+}\nn
\phi_{\tn{out},-}(x^-) &=&\int\dd k_-\,b_+(k_-)\,e^{ik_-x^-},
\le{7field}
and analogously for the other two fields. Notice that, since we integrate 
over positive and negative frequencies, these fields are automatically real 
and contain both creation and annihilation modes. As remarked before, one can 
also check that they satisfy the Klein-Gordon equation in the shock-wave 
geometry:
\be
\left[\pa_+\pa_--p^-f(\ti x-\ti x')\,\d(x^+)\pa_-^2\right]\phi(x)=0,
\ee
and we have neglected transverse derivatives which give factors quadratic in 
$\ti k$ and $\pa_{\ti x}f$, which are assumed to be small. In this 
approximation, the solution to this equation is:
\be
\phi(x)=\int\dd k_-\dd\ti k\,F(k_-,\ti k)\,\exp\left[ip^-k_-\th(x^+)f(\ti 
x)+i k_-x^-+ik_+x^++i\ti k\cdot\ti x\right],
\le{fullfield}
where $k_+=-\ti k^2/k_-$ and under the assumption that the main contribution 
to the integral comes from the region of small $\ti k$. The function $F$ is 
arbitrary, and has to be fixed by imposing some boundary conditions on the 
field and its derivative.

To simplify notation, we write \eq{7field} as
\be
\phi_{\tn{in}}(x^+) &=&\int\dd k_+\,a(k_+)\,e^{ik_+x^+}\nn
\phi_{\tn{out}}(x^-) &=&\int\dd k_-\,b(k_-)\,e^{ik_-x^-}.
\le{fields} 
Now these fields satisfy the following exchange algebra:
\be
\phi_{\tn{out}}(x^-) \phi_{\tn{in}}(x^+) 
=\exp\left[if\pa_+\pa_-\right]\phi_{\tn{in}}(x^+)\phi_{\tn{out}}(x^-),
\le{exchange}
which looks like the $M\rightarrow\infty$ limit of the algebra obtained in 
\cite{VVexchange}.

Ultimately we would like to consider not only zero modes but rather fields 
with transverse momentum. Indeed, the transverse distance has not properly 
been taken care of in \eq{exchange}. $f$ depends on the transverse separation 
of $\phi_{\tn{in}}$ and $\phi_{\tn{out}}$, so it is clear that the fields 
should depend on the transverse co-ordinates too. It is straightforward to 
include transverse momentum as long as we are in the eikonal regime. Consider 
fields like in \eq{fullfield},
\be
\phi_{\tn{in}}(x)=\int\dd k_+\dd\ti k\,a(k_+,\ti k) \,e^{ik_+x^++ik_-x^-+i\ti 
k\ti x},
\ee
where $k_-=-\ti k^2/k_+$. This expression is valid as long as we consider 
large transverse separations between the fields. One gets:
\be
\phi_{\tn{out}}(y)\phi_{\tn{in}}(x)=\exp\left[if^{\m\n}{\pa\over\pa 
x^\m}{\pa\over\pa y^\n} \right] \phi_{\tn{in}}(x)\phi_{\tn{out}}(y),
\le{exchange4}
where
$f^{\m\n}=\e^{\m\n}f(\ti x-\ti x')$, the indices running over the light-cone 
directions only. The epsilon-tensor is due to the minus sign coming from the 
antisymmetry under interchange of the in- and out-labels, and $x^+$ and $x^-$ 
in \eq{exchange4}. 

Notice that, by assumption, the main contribution to the integral over 
transverse momenta comes from the region of small $\ti k$. This would seem to 
imply that the effect of the shift of the in-field in the $x^+$-direction is 
negligible, since its momentum $k^-=-\ti k^2/k^+$ is small. However, this is 
not necessarily true as the factor appearing in the exponential is 
proportional to $-\ti k^2p^+f(\ti x-\ti y)/k^+$, so in 4-dimensional 
Minkowksi space, where $f\sim\log|\ti x-\ti y|$, this need not be small for 
large transverse separations and large momenta $p_+$. In other words, the 
eikonal approximation only requires the derivative of $f$ to be small, but 
the shift itself can be large in Planck units.

The above expression is suspiciously similar to the Moyal product that one 
gets in non-commutative field theory. The obvious guess is that this is 
related to our original commutator
\be
[x^\m,y^\n]=if^{\m\n}.
\le{xy}
Notice, however, that the situation here is slightly different from that in 
non-commutative field theory in that our back-ground is commutative now. 
Non-commuting particle co-ordinates have been replaced by non-commuting 
fields.

\section{Discussion and conclusions}\label{discconc}

The eikonal regime turns out to be a very interesting corner of the moduli 
space of quantum gravity. Things simplify so enormously in this regime that 
the theory becomes topological. Still it has non-trivial dynamics. Global 
variations of the fields correspond to massless particles in the bulk, 
similarly to the way massive particles in 2+1 dimensions correspond to 
topological defects. If the holographic principle is to be true, one should 
not be surprised by this conclusion but should rather wonder whether the same 
is true away from the eikonal regime.

It is also found that the theory is a non-commutative theory whose natural 
length scale is the Planck length. Furthermore, Heisenberg's relation is 
modified by a term proportional to Newton's constant. This has been proposed 
by other authors \cite{Maggiore,Kempf,max}, but in the context of collisions 
between particles at high energies it appears to be a simple consequence of 
the entanglement between the particles after interactions.

Attempts to construct the S-matrix for a two-particle collision at arbitrary 
angles and arbitrary momentum transfer have failed so far. Note that for this 
it is not at all necessary to have a two-particle solution of Einstein's 
equations as long as the rest mass of the particles is small, as one can 
always go to a frame where the momentum of one of the particles is small. We 
have performed a somewhat indirect analysis. The momentum transfer between 
the particles was computed, and from this it is easy to obtain the 
commutation rules between momenta. At every stage conservation of momentum 
was explicit. However, we were not able to find an algebra between the 
co-ordinates, although in principle this can be found by integrating the 
Jacobi identity. In 2+1 dimensions, this is easy to do and we get an SO(2,1) 
algebra between position operators, in agreement with earlier works 
\cite{g9805,SdHJHEP}. In the derivation given here it is clear that these 
complicated quantum mechanical effects are again rooted in the entanglement 
between the particles produced by the momentum transfer.

The precise analysis of the momentum transfer also gives interesting insights 
in the decoupling of the longitudinal and transverse degrees of freedom. 
Transverse momentum transfer is of the order $\d p_\perp\sim 
p_\parallel\,\ve$, whereas longitudinal momentum transfer is of the order $\d 
p_\parallel\sim p_\perp\,\ve$. Thus, as long as $p_\perp\ll p_\parallel$, 
transverse physics is frozen and the transverse modes can be treated 
classically, whereas the longitudinal modes are still rapidly fluctuating. In 
fact, we have explicitly seen that when this condition is no longer valid, 
the transverse modes start fluctuating and become quantum mechanical 
operators as well. 

A case of particular interest is AdS. Based on the AdS-shock-wave solution of 
by Horowitz and Itzhaki, we found that also in AdS interactions between 
massless particles are given in terms of shifts. In particular, for scalar 
fields the effect is a phase shift. In the next chapter we will find that, 
from the CFT point of view, the dual operator has a different expectation 
value inside the light-cone from its value outside.

Another interesting result concerns second quantisation of these 
gravitationally interacting particles. They satisfy an exchange algebra which 
is very similar to the Moyal product defined in non-commutative gauge 
theories, with the difference that in our case the $\theta$-parameter is a 
function of the transverse co-ordinates. The non-commutativity of the algebra 
is rooted in the non-commutativity of the first-quantised space-time. 
However, despite the similarity the situation is different from that in 
non-commutative gauge theories as our algebra is not derived from an action 
on a non-commutative space. In our case the co-ordinates commute. The 
non-commutativity arises when we include gravitational interactions. This is 
true both for the first and the second quantised system: in our case, it is 
always the matter fields of particles that are non-commuting. These are 
either co-ordinates of particles, $X^\m(\ti\s)$, or scalar fields, $\f(x)$. 
The underlying space-time ($\ti\s$ or $x$, respectively) is always 
commutative.

There are several interesting open questions which are left for future study.

One interesting problem is how to fully take into account the transverse 
effects in the non-commutative scalar field theory without having to restrict 
ourselves to the eikonal approximation. This could be done most easily in the 
2+1-dimensional context where we have the full commutator between the $x$'s, 
which satisfy the SO(2,1) algebra, and it may be easier to obtain the exact 
solutions to the Klein-Gordon equation.

Another crucial question in the context of holography is the interpretation 
of the commutators \eq{xy} and \eq{exchange4} in the context of the AdS/CFT 
duality. The computation of the trajectories in the AdS-shock-wave metric in 
Appendix \ref{appA1} reveals that once again the shift is proportional to the 
momentum, and so upon quantisation one expects co-ordinates to be 
non-commuting. However, one now has to take into account the additional 
problems with quantisation that arise in AdS. It is likely that the 
techniques developed in chapter \ref{reconstruction} can help us understand 
the meaning of the commutator \eq{xy} in terms of sources or operators 
related to the bulk fields $z^\m(s)$. Another, more straightforward approach, 
will be to directly study the algebra \eq{exchange4} from the point of view 
of the dual operators on the boundary. A previous step in this direction is 
taken in section \ref{sec3.7} of the next chapter.

As remarked by 't Hooft \cite{g9805}, the epsilon-tensor in \eq{2.15c} is 
directly related to the position of the observer with respect to a black hole 
horizon. In turn, the appearance of such an epsilon tensor can be traced back 
to the minus sign difference in \eq{02}, which gives rise to an epsilon 
tensor in the longitudinal space in formula \eq{comm0}. At the level of the 
S-matrix, this sign difference comes from the fact that particles with 
momentum $p_u$ are in-going, whereas those with momentum $p_v$ are out-going, 
as one easily sees from \eq{S-m}. Thus, this epsilon tensor is indeed 
connected with the distinction between in-going and out-going and thus with 
causality. It is at first somewhat surprising that the same epsilon tensor 
appears in \eq{EHred2} and \eq{comm-1}, but also here it has to do with the 
orientation with respect to the asymptotic boundary of the space-time, and 
thus again it is a global property closely related with causality.

Let us end with a somewhat speculative remark. Beyond the eikonal 
approximation, although there may still be some hidden redundancy in our 
formulae, we have seen that there are more than two fields $X^a$ whose 
variations contain physical information. In four dimensions, there are four 
such fields, $X^a$ and $Y^m$. One is therefore led to speculate that the 
path-integral approach in \cite{VV} at the next order in the eikonal 
approximation will still be topological, the physical fields now being the 
boundary values of $X^a$ and $Y^m$. Of course, at some point one expects to 
encounter the usual non-renormalisable infinities in quantum gravity, and at 
that point one may need to invoke string theory. 

\newpage

\chapter{Boundary Description of High-Energy Scattering in Curved 
Space-times}\label{GSM}

We show that for an eikonal limit of gravity in a space-time of any 
dimension with a non-vanishing cosmological constant, the Einstein -- 
Hilbert action reduces to a boundary action. This boundary action 
describes the interaction of shock-waves up to the point of evolution 
at which the forward light-cone of a collision meets the boundary of 
the space-time. The conclusions are quite general and in particular 
generalise the work of E. and H. Verlinde \cite{VVErice}. The role of the 
off-diagonal Einstein action in removing the bulk part of the action is 
emphasised. We discuss the sense in which our result is a particular
example of holography and also the relation of our solutions
in AdS to those of Horowitz and Itzhaki \cite{HI}. We also find a boundary 
action for the case of asymptotically de Sitter space. This is relevant to 
the discussions of holographic duals of de Sitter space in 
\cite{Bousso,HKS,FKMP}.

The contents of this chapter are based on \cite{GASdHMOL}.

\section{Introduction}\label{sec3.1}
Although one could claim that high-energy scattering in gravity should be 
treated in string theory the philosophy adopted in this chapter, based 
on the holographic principle is that such collisions should be treatable 
in the context of quantum gravity. The holographic principle is taken to 
be the guiding feature behind quantum gravity, rather than the string 
principle. As such it 
implies a reduction in the true number of quantum gravity degrees of 
freedom in line with the counting of degrees of freedom in string theory. 
Thus implementing correctly the holographic principle \cite{ghologr,
g9607,Susskind,AdSreview} in quantum gravity 
should result in a softening of amplitudes akin to that which occurs in 
string theory. From here on all discussions will take place in the context 
of gravity using the Einstein -- Hilbert action including cosmological 
constant apart from some string-theory related comments in the final
sections. 

The role of high energy scattering has been emphasized by 't Hooft in the 
context of the black hole evaporation process. As is well known, the
appearance of Hawking radiation can be attributed to the diverging red-shift
of outgoing wave packets when propagated back to the region close to
the horizon. Quantum gravitational effects are therefore expected to
play a fundamental role and their inclusion is expected to restore the
unitarity of the Hawking radiation. According to the picture of 
't Hooft these gravitational interactions close to the horizon can be 
effectively
described by shock wave configurations associated to the boosted
particles. They have non-trivial backreaction effects, bringing about
a shift in the geodesics of the outgoing particles and in the position
of the horizon, as we have seen in the previous chapters. These correlations 
should in principle reduce the enormous degeneracy of states at the horizon 
of the black hole that one naively calculates using quantum field theory in 
the curved space-time of the near-horizon geometry. In this picture
the horizon of the black hole becomes a sort of fluctuacting
membrane due to incoming and outgoing particles and information of the
bulk spacetime is  projected holographically onto this surface. 

In view of these developments it seems interesting to search for a 
more concrete relation between the general arguments of 't Hooft and 
Susskind and the AdS/CFT construction. In this chapter we discuss in 
general the eikonal limit of scattering in curved space-times
and find that under certain rather general assumptions about the 
relevant classical backgrounds, the dynamics of gravity is 
described by a theory that lives only on the boundary of the space-time. 
We also find that, from the bulk point of view, some of the classical 
solutions of this boundary theory describe shock-waves moving from the 
boundary to the bulk in Einstein spaces. 
On the way to finding classical backgrounds for our quantum theory we need 
the general solution of a two-dimensional gravity model analysed in 
\cite{BOL}.
Our solutions include the shock-wave solution constructed by Horowitz and 
Itzhaki \cite{HI} and this will be discussed in some detail in section 
\ref{sec3.7}.

We will also find a boundary description of scattering in the case of 
asymptotically de Sitter space-times. This boundary action is defined on the 
past and future space-like boundaries of the de Sitter space and may be 
important for discussions of causality and locality of holographic duals of 
de Sitter space \cite{Bousso,HKS,FKMP}. To our knowledge, this is the first 
explicit boundary descripition of the dynamics in de Sitter space.

This chapter is organised as follows. In section \ref{sec3.2} we will 
describe the setup in which our analysis takes place in particular reviewing 
the 
basic idea of \cite{VV,VVErice} in which a rescaling is made of the Einstein 
--
Hilbert action thus separating it into three pieces each scaling differently
in the eikonal limit. In section \ref{sec3.3} we discuss the solutions to the 
classical part of this action in various regimes. In section \ref{sec3.4} we 
introduce shock-wave configurations and then in section \ref{sec3.5} we show 
how the off-diagonal part of the Einstein equations will be implemented. In 
section \ref{sec3.6} we discuss the derivation and details of the resulting 
boundary action and in section \ref{sec3.7} we show how our analysis is 
related to and extends the construction of Horowitz and Itzhaki \cite{HI}. 
Finally in section \ref{sec3.8} we make some comments on our results and some 
other concluding remarks. 

\section{The setup}\label{sec3.2}

We consider high-energy scattering in spacetimes with a non vanishing 
cosmological constant $\L$. Our basic construction is a direct 
generalization of that used in \cite{VV,VVErice} and thus we will consider
an almost forward scattering situation. One introduces two scales, 
${\ell_{\parallel}}$ 
and ${\ell_{\perp}}$: the former
is the typical longitudinal wavelength of the particles while the  latter 
represents the 
impact parameter. Due to the presence of the cosmological constant we also
have an additional
scale ${\ell}$ -- the radius of curvature $\Lambda\sim\frac{1}{\ell^2}$.
For high-energy forward scattering
${\ell_{\parallel}}$ is typically of the order of the Planck length 
$\Pl$, ${\ell_{\perp}} \gg {\ell_{\parallel}}$. This set of 
length scales characterizes the eikonal limit of the scattering
process which for gravity is a linearized regime. We will also deal
with two different cases according to large or small values of the
cosmological constant present in the problem. In general we then 
find that for $\ell_{\perp}$ small on the cosmological scale the 
scattering takes place in the locally flat space-time. On the other
hand for impact parameters that are large on the cosmological scale, 
there are significant changes in the scattering process due to the
curvature. The final result is conceptually the same however as we
find that for shock-wave scattering the process can always be described by 
a lagrangian on the boundary at infinity of the space-time. 

Our general strategy will be to choose dimensionless 
co-ordinates by extracting the natural length scale in the corresponding 
directions and therefore we will consider the Einstein -- Hilbert 
action plus a non-vanishing cosmological constant and exterior 
curvature $K$,
\be
S={1\over\Pl^{d-2}}[\int_M\dd^dx\,\@{-G}\,(R-2\L) 
+\int_{\partial M}\dd^{d-1}x\@{\g}\,2K],
\ee
making a rescaling in the longitudinal $x^\a$  and transverse
co-ordinates $y^i$ according to the respective scales, as explained in 
section \ref{TFT} of the introduction. Under a rescaling of the metric, the 
action rescales as
\be
\e^{d-4}S_E = \left(\frac{S_0}{\epsilon^2}+\frac{S_1}{\epsilon} 
+ S_2\right)
\ee
where $\epsilon=\Pl/\ell_\perp\sim\ell_\parallel /\ell_\perp$ 
is a very small dimensionless
parameter. $S_2=S_\parallel$ therefore is the strongly coupled part of the
action while $S_0=S_\perp$ is the weakly coupled part. The former is
non pertubative while the latter is essentially classical. The role of 
$S_1$ will be discussed in the following but as is clear it also contributes
to the classical part of the action in the limit of small $\epsilon$.
Under the above rescaling the cosmological term scales as
${\ell^2_\perp\over\e^{d-4}}$ and thus becomes part of the
classical $S_\perp$ or the ``quantum'' $S_\parallel$ depending on the
size of $\ell_\perp$ in comparison to the cosmological scale, $\ell$.
 We will consider
both the case in which the cosmological constant is added to the
classical part of the action -- the ``strongly curved regime'' 
or the regime of 
large impact parameter -- and the case when 
the cosmological constant is included in the strongly coupled part 
of the action -- the ``flat regime'' or regime of small impact parameter. 

\subsection{Scaling and small fluctuations}

We will actually consider a metric that at leading order is block 
diagonal -- the blocks corresponding to the plane of the scattering 
and the plane transverse to the scattering. We will consider a rescaling
of the metric (equivalent but more convenient than that of the 
co-ordinates discussed above) such that
\be
G_{\mu\nu} = \left(\begin{array}{cc}
g_{\a\b} & h_{\a i}\\    
h_{i \a} & h_{ij}
\end{array}\right) \ra \left(\begin{array}{cc}
\ell_\parallel^2\,g_{\a\b} & \ell_\parallel\ell_\perp\, h_{\a i}\\    
\ell_\parallel\ell_\perp\, h_{i \a} & \ell_\perp^2\, h_{ij}
\end{array}\right).
\le{rescaling}
We use a notation where Greek indices label longitudinal variables, and Latin 
indices label transverse ones, $x^\m=(x^\a,y^i)$.

In addition to this rescaling of the energy scales, we will also make the 
assumption that the off-diagonal blocks of the metric are 
small. In the end then we will be making a double expansion of the 
action, in $\epsilon$ and in $h_{i\a}$. 

In the limit that $\epsilon\ra 0$ the leading terms in the action 
become classical
and thus we need to derive and examine first the equations of motion
arising from $S_0$ and $S_1$ given our choice of metric. 
$S_0$ always becomes a covariant $1+1$ dimensional 
action and has no terms linear
in the small off-diagonal part of the metric $h_{i\alpha}$.
$S_1$ starts at linear order in $h_{i\alpha}$ and the equation of
motion here comes from the variation with respect to $h_{i\alpha}$
imposing the vanishing of the off-diagonal block of the 
Ricci tensor $R_{i\a}$. The remaining part $S_2$ of the action
is the most interesting part as it is not removed in our
limit and basically describes the dynamics of the eikonal limit
of scattering at high-energy and large impact parameter. 
We will find that this action contains no bulk degrees of 
freedom and thus reduces to a boundary term.
The details of the scaling of the curvature components are in Appendix 
\ref{appB1}. At each order in $\e$ the action gets contributions from 
different orders in the expansion of the Ricci tensor. The leading order 
$\epsilon^{-2}$ term in the action has contributions from $R_{\a\b}$ at order 
$\epsilon^0$ and from $R_{ij}$ at order $\epsilon^{-2}$; the subleading order 
at $\epsilon^{-1}$ in the action comes solely from the leading term in 
$R_{i\a}$; while the final 
term at order $\epsilon^0$ has contributions from the remaining 
higher order terms in $R_{\a\b}$ and $R_{ij}$.
The resulting double expansion in $\epsilon$ and $h_{i\a}$ is:
\be
\epsilon^{d-4}\,S &=&
{1\over \epsilon^2}\int_{\cal M}\sqrt{-gh}\,\left(R_g + \frac{1}{4}
(h^{ik}h^{lm} - 
h^{il}h^{km})
\partial_\a h_{ik}\partial_\b h_{lm} g^{\a\b}\right)\nn
  &-& {2\over \epsilon} \int_{\cal M} \sqrt{-gh}\,h^{i\alpha} R_{i\alpha}\nn
  &+& \int_{\cal M}\sqrt{-gh}\left(R_h + \frac{1}{4}
(g^{\a\b}g^{\g\d} - g^{\a\g}g^{\b\d})
\partial_i g_{\a\b} \partial_j g_{\g\d}h^{ij}\right)\nn
&+& \int_{\partial{\cal M}} \@{\g}\,2K
 -2 \ell_\perp^2 \int_{\cal M}\sqrt{-gh}\,\Lambda.
\le{fullaction}

Considering a path integral for this action we see that the first two terms
become classical as $\epsilon\ra 0$. The cosmological constant can be moved
to different orders of $\epsilon$ depending on its scaling with
$\ell_\parallel$ or with $\ell_\perp$. 
Physically the mobility of the cosmological constant corresponds to the 
relationship between the scale of curvature of the space-time and the 
impact parameter of the scattering process under consideration. In the 
regime for which the curvature of the space-time does not really
enter into the dicussion we find that the analysis is similar to 
that in flat space, though with corrections to the boundary action coming
from the cosmological constant. In the other regime for which the 
impact parameter is larger than the radius of curvature, the space-time in 
the plane of scattering is curved and the analysis more subtle. The 
result again is that the scattering process can be described by a now
non-quadratic lagrangian that lives on the boundary of the 
space-time. 

The contribution of the exterior curvature will follow the usual 
construction of the Einstein -- Hilbert action. It will split under
rescaling to give contributions to the boundary
in such a way that these boundary terms 
have their usual effect. That is, at the leading ``classical'' orders
they will simply cancel boundary terms that come from integrating by 
parts when varying the action to get the equations of motion. The 
rescaling of the co-ordinates acts on the exterior curvature part
of the action in such a way that it only contributes to the action 
at order $\epsilon^{-2}$ and $\epsilon^{-1}$ and thus will not
provide any addition to our final boundary action which is at 
order $\epsilon^0$. The details of the 
scaling of the exterior curvature part of the action are given in 
Appendix \ref{appB2}.

The general setup that is obtained via this rescaling of the action
by the factor $\epsilon$ (which depends on the energy scales of the problem)
is one in which we have an energy dependent action. This means that 
we are not considering a high energy process in a theory that is already 
defined, but rather we are using the high energy ``eikonal'' limit to
define for us a new action that (hopefully) isolates the degrees of 
freedom that are important for the problem at hand. 
In particular, as we will see from the classical solutions that come
from the small $\epsilon$ limit, the space-time splits 
into a $2+(d-2)$ configuration
in which the two parts are coupled only through the constraint 
that the off-diagonal part of the curvature vanish. The interaction
between the two parts of the space-time - that transversal and that 
longitudinal -- is restricted by $R_{i\a} = 0$. 
Therefore in the case of large cosmological constant although one
may be tempted to interpret this as a limit of small AdS$_d$
it is not. It is more simply a case in which the separation of the 
shock-waves in the transverse part of the space-time is large, 
and the size of the AdS$_2$ in the longitudinal space corresponds to 
a large curvature. However this is not obviously the same
as a scattering in say the context of string theory in an AdS$_d$ with 
large curvature, though it does retain some of the important features.

In the next three sections we will discuss in turn each order in $\epsilon$ 
of 
this rescaled action.

\section{The solutions}\label{sec3.3}

The geometry in the longitudinal plane of the scattering is determined by the 
saddle point of $S_0$. This is the classical part of the action which is of 
order ${1\over\e^2}$:
\be
S_0[g,h]&=&\int_{\cal M}\@{gh}\,[R[g]-2\L + {1\over4}\, g^{\a\b}\pa_\a 
h_{ij}\pa_\b h_{kl}(h^{ik}h^{jl} -h^{ij}h^{kl})]\nn
&=&\int_{\cal M}\@{gh}\,[R[g]-2\L +{1\over4}h^{ij}\Box h_{ij} 
-\half(\na\log\@{h})^2].
\ee
The form of the action after the last equality sign is particularly useful to 
derive the equations of motion with respect to the transverse metric. We use 
an index-free notation where all derivatives are with respect to longitudinal 
variables. The action indeed does not contain transverse derivatives, and so 
the dynamics in the transverse space is trivial.

The equations of motion derived from this action are:
\be
g_{\a\b}[-\L +{1\over8}\,\Tr(\na h)^2 -{1\over8}\,(\Tr\na h)^2] 
-{1\over4}\Tr(\na h)_{\a\b}^2 +{1\over4}(\Tr\na h)^2_{\a\b}+&&\nn
+{1\over\@{h}}(\na_\a\na_\b -g_{\a\b}\Box)\@{h}&=&0\nn
\half h^{ij}[R[g]-2\L +{1\over4}h^{kl}\Box h_{kl}] -{1\over4}h^{ik}h^{jl}\Box 
h_{kl} +\half h^{ij}(\na\log\@{h})^2 \,+&& \nn
+\half\na\log\@{h}\na h^{ij} +{1\over4}\Box h^{ij} 
+{3\over4}h^{ij}\Box\log\@{h}&=&0.
\le{eomsaddle}
Note that taking the trace of the first equation we get:
\be
(\Box_g+2\L)\@{h}=0.
\ee
So far we made no assumptions concerning the particular form of the solution. 
However, to solve the equations of motion we need some specific ansatz that 
is suitable for describing a forward scattering situation and allows among 
the various possibilities for the presence of the cosmological constant. In 
particular, the form of the metric used in \cite{VV},
\be
h_{ij}=\ti h_{ij}(y),
\ee
does not allow for non-trivial solutions in all cases that we will study. We 
then need to assume that in general the transverse metric depends on the 
longitudinal co-ordinates through a warp factor
\be
h_{ij}(x^\mu)=e^{\chi(x,y)}\ti h_{ij}(y).
\le{transvmetric}
This is of course not the most general ansatz but it is general enough so as 
to give non-trivial solutions of the equations of motion. This ansatz can 
also be used to study radial scattering situations provided one chooses a 
time and a radial co-ordinate in the longitudinal directions. We will treat 
the $d=3$ case separately due to various inconvenient factors of $(d-3)$ in 
the following general analysis. 

Substituting the ansatz \eq{transvmetric} in the equations of motion 
\eq{eomsaddle} gives the same equations of motion that follow from the 
reduced action:
\be
S_0 = 
S_{\perp}=\int_M\@{-g\ti h}\,e^{(d-2)\chi\over 2}\left(R[g]-2\L-{(d-2)(d-3)
\over4}g^{\a\b}\pa_\a\chi\pa_\b\chi\right)
\le{Sperp0}
Making the following field redefinition
\be
\phi(x,y)= (\frac{d-3}{2(d-2)})^{1/2} \exp \left( (\frac{d-2}{4}) \chi(x,y) 
\right)
\ee
one gets
\be
S_{\perp}=-8\int_M\@{-g\ti h} \left(g^{\a\b}\pa_\a\phi\pa_\b
\phi-{d-2\over 4(d-3)}\phi^2(R[g]-2\L)\right).
\le{Sperp}
The eikonal limit restricts us to consider the extrema of 
\eq{Sperp}. It is interesting to note that with the assumption 
\eq{transvmetric} the problem is reduced to a general two-dimensional gravity
plus scalar field as studied in
\cite{BOL}. More properly, since the transverse fluctuations are
suppressed in the leading order (in $h_{i\a}$) term of the weakly coupled
action, its explicit expression will not contain, as shown by the
scaling arguments, transverse derivatives. 
Therefore the action still depends on all
four co-ordinates but the dependence on the transverse directions
is only parametric. 
The equations of motion for the metric and the scalar field $\phi$ are:
\be
\pa_\a\phi\pa_\b\phi-{1\over2}\,g_{\a\b}g^{\g\d}\pa_\g\phi\pa_\d
\phi={d-2\over4(d-3)}(\L\,g_{\a\b}\,\phi^2 + 
(g_{\a\b}\Box-\nabla_\a
\nabla_\b)\phi^2)
\le{eomI}
\be
\Box\phi+{(d-2)\over4(d-3)}(R[g]-2\L)\phi=0
\le{eomII}
As mentioned, these are the same equations of motion as \eq{eomsaddle} for 
our warped metric.

As proved in the paper \cite{BOL}, all classical solutions have a 
Killing vector that is perpendicular to the curves of constant scalar 
field. When $\L<0$, the solutions are static. Therefore we can in this case 
choose the longitudinal metric $g_{\a\b}$ to be of the form
\be
\dd s^2=-e(x)^2\dd t^2+g(x)^2\dd x^2,
\le{longmetric}
where also
\be
\phi = \phi(x),
\ee
in co-ordinates where $x^\a=(x,t)$. These solutions have a boundary at 
spacelike infinity. More properly in our case, as we will see below, $e$ and 
$g$ depend on the transverse co-ordinates too, since we are considering the 
two dimensional longitudinal manifold times the transverse space.

When $\L>0$, the solutions are time-dependent \cite{BOL}. They are de 
Sitter-type cosmological solutions with a spacelike boundary. These solutions 
can be obtained either by analytic continuation of the above solutions from 
negative to positive cosmological constant, or by choosing a time-dependent 
metric from the beginning:
\be
\dd s^2=-g(t)^2\dd t^2+e(t)^2\dd x^2,
\ee
where now
\be
\phi = \phi(t).
\ee
These are in fact the type of solutions analysed in \cite{BOL}.

\subsection{Large Curvature}

For the case of a negative cosmological constant, the general solution to 
the equations \eq{eomI}-\eq{eomII} can easily (details in Appendix 
\ref{appB3}) be found and is:
\be
\phi(r)&=&\psi(r)^\gamma\nn
e(r)&=&C\psi(r)^{\gamma\over 4Q}\dot\psi(r)
\le{solution}
where
\be
\psi(r) = Ae^{\@{\lambda\over 4Q\gamma}r} + Be^{-\@{\lambda\over 4Q\gamma}r},
\ee
\be
\dd r = g(x)\dd x,
\ee
and 
\be
\gamma = {4Q\over 1 + 8Q}
\ee
and where  $\l=-{(d-2)\over2(d-3)}\L$, 
$Q={(d-2)\over4(d-3)}$. 

Note that $A,B$ and $C$ are constant with respect to the 
longitudinal co-ordinates. However they can have an arbitrary 
dependence on the transverse co-ordinates $y^i$. Their precise form is 
fixed by imposing opportune boundary conditions depending on the spacetime 
under consideration.

It is also interesting to notice that if one takes either of 
$A$ or $B$ to zero, this two-dimensional metric has constant curvature
and is actually just the metric on AdS$_2$ -- the entire space-time metric
being AdS$_2$ times the $(d-2)$-dimensional  transverse geometry plus 
a warp factor. 

The fact that the transverse metric $\ti h_{ij}(y)$ is not determined by the 
equations of motion means that it is an arbitrary classical back-ground. This 
is also the case for the small curvature case to be considered in the next 
subsection.

\subsection{Small Curvature}

In the small curvature regime, the cosmological constant term belongs to the 
strongly coupled part of the action, as discussed. The classical action 
that we then have to consider is therefore \eq{Sperp0} with $\L=0$. 
However, putting the cosmological constant to zero in the solutions above 
is a singular limit. It is easy to directly solve for the metric in this 
case and one finds:
\be
\phi(r)&=&(Ar+B)^\gamma\nn
e(r)&=&C(Ar+B)^{\gamma\over 4Q},
\le{case1}

Again, $A,B$ and $C$ are allowed to depend on the transverse 
co-ordinates. The curvature is
\be
R[g]={16QA^2\over(1+8Q)^2(Ar+B)^2}.
\ee
Notice that it is always positive. In the limit $B\ra\infty$ we recover 
flat space, which was not a solution of the equations of motion in the 
strong curvature regime. In the region $r\ll B$, the space has locally 
positive 
constant curvature.

Note that there is also a degenerate flat space solution for which 
$\phi$ is constant and
\be
e = Ar + B.
\le{case2}

Even though these solutions could not be obtained directly from those 
with $\Lambda$ non-zero  by  setting $\Lambda$ to zero
they can be obtained as near-horizon limits of those geometries, and 
this exactly corresponds to first shifting the co-ordinate $r$ and then 
taking the limit of small cosmological constant.

The second case, \eq{case2}, is the near-horizon 
geometry for the solutions of
the previous section, for $B/A>0$. Indeed, a simple co-ordinate 
transformation brings the near-horizon metric into the 
form of the Rindler metric (see Appendix \ref{B3}). In the 
same way \eq{case1} is the near-horizon geometry in the case $B/A<0$, 
and again a simple co-ordinate transformation brings it into the form 
of a Rindler type metric with singular horizon (again see Appendix 
\ref{appB1}). 

\subsection{Three-dimensional space-time}

The above formulae are not directly applicable for $d=3$, although 
one can obtain the equations of motion by carefully setting $d=3$
in the above equations. The action in three dimensions is:
\be
S_{\perp}=\int\@{-g\ti h}\,\phi^2(R[g]-2\L),
\ee
where $\phi=e^{\chi/4}$. The equations of motion for the scalar 
field and the metric are \cite{DeJa}:
\be
\nabla_\a\nabla_\b\phi^2-g_{\a\b}\nabla^2\phi^2-g_{\a\b}\L\phi^2=0,
\ee
\be
R[g]=2\L,
\ee
and so the space-time always has constant curvature.
It is therefore not surprising that the only solution we find is AdS$_2$. 
These equations are totally symmetric 
under interchanges of $\phi$ and $e$, and under reflections 
$r\rightarrow-r$. Therefore, the general solution is
\be
\phi(r)&=&A\,e^{r/2\ell}\nn
e(r)&=&B\,e^{-r/\ell},
\ee
where $\ell$ is the AdS radius. This solution corresponds to pure AdS, 
as expected, with a scalar field $\phi$ that vanishes at the boundary 
and has a singularity at the horizon.

When $\L>0$, we obtain 3-dimensional de Sitter space. 

\section{Gravity at high energy and shock waves}\label{sec3.4}

This section is a necessary digression into the shock-wave solutions
to the classical part of our action. We need to 
understand the 
form of these shock-waves as they will motivate our final 
choice for the metric that we will use in the remaining non-classical
part of our action. The physics in the bulk that can be described 
classically via these shock-waves will then be the physics that is
encoded in the boundary action.

It turns out that scattering at Planckian energies is dominated by
the gravitational force. Therefore one should have a complete theory 
of quantum gravity to describe these
processes. However already in the eikonal regime that we are 
considering one can use
semiclassical methods to get useful information.

At leading order gravitational interactions can indeed be described  
by shock wave
configurations -- gravitational waves with a longitudinal impulsive 
profile. Essentially this is the gravitational field surrounding a 
particle whose mass is dominated by kinetic energy therefore representing 
a sort of massless regime of general relativity 
\cite{AiSe,ACV,SdHJHEP,gnp85,KO1,KO2,Penrose,g87}.

Explicit solutions in general spacetimes and their physical 
effects have been described by Dray and 't Hooft \cite{gnp85}, using 
the so called cut and paste method (see chapter \ref{HEscattering}). 

For example, the gravitational field of a massless particle in flat
spacetime can be described by a metric of the form
$$
\dd s^2=-\dd u\dd v -4p\ln(|x^i|^2)\,\delta(u)  \dd u^2 +\dd x^i\dd x_i,
$$
where $p$ is the momentum of the massless particle.
The physical effects of such configurations play a crucial role in 
't Hooft's description of the evaporation of a black hole. We refer to
\cite{g9607} for a detailed account and references.

Choosing $x^\mu=(x^+,x^-,x^i)$ and placing the massless particle at
$x^+=x^i=0$, a natural way to rewrite this metric is then 
$$
\dd s^2=\partial_\alpha X^- \dd x^\alpha \dd x^+ +  \dd x_i^2
$$
with
$$
X^-=x^-+p \,\theta(x^+)\ln (|x^i|^2)
$$
This means that if we want to describe the scattering of two high 
energy particles before a collision takes place we must use the 
generalised shockwave configuration with the metric 
in the longitudinal plane being of the form:
$$
\dd s^2=\partial_\alpha X^a \partial_\beta X^b \eta_{ab}\,\dd x^\a\dd x^\b,
$$
thus allowing a pair shock-waves of the above type in both $x^+$ and in 
$x^-$.
Here the SO(1,1) $X^a$ vectors can in principle depend on all space-time
co-ordinates. These are the configurations studied in \cite{VVErice} 
and below we will generalise this construction to include the presence of 
curvature in the longitudinal plane.

\section{The constraint and solution-ansatz}\label{sec3.5}

The second order in our expansion is quite simple. It is 
\be
-\frac{2}{\epsilon}\int\sqrt{-gh}h^{i\alpha}R_{i\alpha}
\le{constraint}
As this is order $\epsilon^{-1}$ we also need to implement the 
corresponding equation of motion (as we did for the leading order
in the previous section). In this order basically the equation of
motion appears as a constraint $R_{i\a} = 0$ on the general solutions. 

Before implementing this constraint we will go back to the construction
of \cite{VVErice} where it is shown how to change variables in 
a way that simplifies the following analysis.
The saddle-point of the transverse part of the
action $S_0$ gives the dominant vacuum
field configurations. In the absence of the cosmological constant
there was only
\be
R[g]&=&0\nn
h_{ij}&=&h_{ij}(y).
\le{cases}
As recounted in the previous section for massless shock-wave configurations 
we will choose a parametrization of the metric via diffeomorphisms
that represents these shock-waves,
\be
g_{\alpha\beta}= \partial_\alpha X^a \partial_\beta X^b 
\eta_{ab}
\ee
where the $X^a(x,y)$ are diffeomorphims which relate $g_{\alpha \beta}$
to the flat metric. Note that they are maps of the two dimensional $x^\alpha$
plane onto itself being however allowed to vary in the transverse
directions and therefore represent transverse co-ordinate dependent
displacements in the longitudinal co-ordinates. 
These $X^a$ fields have the appearance of diffeomorphisms in the world-volume 
of the two-dimensional sigma-model and as such would appear to not
introduce any new degrees of freedom. However, in the $d$-dimensional theory
this is no longer really true as we are not considering the full 
transformation 
of the higher dimensional metric under these transformations. Nevertheless, 
due to the constraint coming from the off-diagonal part of the 
Einstein action we will see that these fields do not contribute additional 
bulk degrees of freedom. 

An intermediate and useful step required to derive the boundary action 
and used to great effect in \cite{VV} is to express the
strongly coupled action in terms of fields $V_i^\alpha$ defined as
\be
\partial_i X^a =V_i^\alpha \partial_\alpha X^a.
\le{V}
These fields were introduced and motivated physically in terms of fluid 
velocity in \cite{VV}. In the gravitational setup presented here they 
can be thought of as zweibeins (see also \cite{Kallosh}) 
for a two-dimensional sigma-model describing
the embedding of the scattering plane into the transverse space.
They considerably simplify the action and help to conceptualize our
configuration from the sigma-model point of view.

This definition could also have been motivated by the
simple practical consideration that in order to rewrite the strongly
coupled action as a boundary action one needs to remove derivatives in 
the transverse directions to give one an action that is covariant
in the longitudinal directions. As a consequence one tries to express
every derivative in the transverse directions in terms of a derivative
in the longitudinal ones. This is precisely obtained utilizing this
definition of the $V_i^\alpha$ fields. In this way the indices labelling 
transverse directions act as an internal symmetry of the sigma-model
from the point of view of the longitudinal spacetime. This will 
become clear in the next section where we write the general explicit form 
for the boundary action for all $d \geq 3$ and in both the strong 
and weak curvature regimes.

This construction is basically identical for the more general metrics
considered here. 
As we have seen in Section 4, the conditions
\eq{cases} are too restrictive and one ends up in this general setup 
case with a
family of solutions specified by $g_{\alpha\beta}$ and
$\chi$. A natural generalization of  the above parametrization of
$g_{\alpha\beta}$ is then, 
\be
g_{\alpha\beta}=e^{\sigma(X)} \partial_\alpha X^a \partial_\beta X^b 
\eta_{ab}.
\le{anstzI}
thus allowing the presence of a warp factor in the $2+(d-2)$ decomposition
of the metric. 
In principle $\sigma$ may also have some explicit $y$ dependence, however 
this
would correspond to a more complicated sigma model than the one we
are presently considering. As stated several times, 
the introduction of $X$ is simply
a statement that the scattering configuration described by shock-waves 
is described simply via singular co-ordinate tranformations with 
support only along light-cones in the scattering plane and thus the classical 
solution $\sigma(x)$ after the shock wave ansatz becomes simply
$\sigma(X)$ and similarly $\chi(x)$ becomes $\chi(X)$. 
As in the previous case \cite{VV}, we define fields $V_i^\a$ by
\be
\pa_iX^a=V_i^\a\pa_\a X^a,
\le{Vfield}
which in turn gives when lowering the longitudinal index
\be
V_{i\a}=e^{\s(X)}\pa_iX^a\pa_\a X_a.
\ee
With the use of the $V_{i\a}$ the longitudinal metric 
changes under reparametrisations of the transverse co-ordinates 
according to,
\be
\pa_ig_{\a\b}=\nabla_\a V_{\b i}+\nabla_\b V_{\a i}.
\le{cmode}
Finally we also will have 
\be
h_{ij} = e^{\chi(X)} \tilde{h}_{ij}(y).
\le{anstzII}

Notice that this form of the solutions captures both the cases $\L<0$ 
and $\L>0$.

\section{The Effective Boundary Theory}\label{sec3.6}

We now examine how our classical solution -- ansatz leads us to the general
result that in this setup the transverse action $S_\parallel$ always
reduces to a boundary action. 

Making the substitutions  of our solution -- ansatz 
in the leading order ($\epsilon^0$) action,
\be
S_\parallel[g,h]&=&\int\@{-gh}\left[R[h] -{1\over4}\,h^{ij}\pa_ig_{\a\b}
\pa_jg_{\g\d}\,\e^{\a\g}\e^{\b\d}\right]\nn
&=&\int\@{-gh}\left[R[h] -\e_{\a\g}\e_{\b\d}\,h^{ij}\na^\a V^\b_i\na^\g 
V_j^\d +\half h^{ij}R_iR_j \right],
\ee
where
\be
R_i&=&\e^{\a\b}\na_\a V_{i\b}.
\ee

\subsection{Strong curvature regime}

Filling in the solutions of the classical equations of motion,
\be
g_{\a\b}&=&e^{\s(X)}\,\pa_\a X^a\pa_\b X^b\,\et_{ab}\nn
h_{ij}&=&e^{\chi(X)}\,\ti h_{ij},
\ee
we get
\be
S_\parallel&=&\int\@{-g\ti h}\,e^{{d-4\over2}\,\chi}\left[R[\ti h] -
\e_{\a\g}\e_{\b\d} \ti h^{ij}\na^\a V^\b_i\na^\g V_j^\d +\half R_i^2+
\right.\nn
&-&\left.(d-3)\Box_{\ti h}\chi -{1\over4}(d-3)(d-4)(\pa_i\chi)^2\right],
\ee
and from now on we raise and lower transverse indices by means of 
the rescaled metric $\ti h_{ij}$.

The action splits into a bulk and a boundary term:
\be
S_\parallel&=&S_{\sm{bulk}}+S_{\sm{bdry}}\nn
&=&\int_{\pa M}\dd x^\a\@{\ti h}\,e^{{d-4\over2}\,\chi} \left(R[\ti h]\,
e^\sigma \,X^0\pa_\a X^1 +e^\s\e_{ab}\,\frac{}{}
\pa_iX^a\times\right.\nn
&\times&\left.[\pa^i\pa_\a X^b +\half\na^\b\s(\pa_\a X^bV^i_\b -\pa_\b 
X^bV^i_\a)] -\half\,V_{i\a}R^i -{d-3\over2}\,\e_{\a\b}
V^{i\b}\pa_i\chi\right) +\nn
&+&\int_M\@{-g\ti h}\,e^{{d-4\over2}\,\chi}\,V^{i\a}R_{i\a},
\le{boundaryaction}
where by $X^a$ we mean the variation of $X^a$ around its infinite value. 
Filling in the constraint
\be
R_{i\a}&=&\half\,R[g]V_{i\a} +\half\,\e_{\a\b}\na^\b R_i +
\half\pa_\a\chi\na^\b V_{i\b} -{d-3\over2}\,\pa_i\pa_\a\chi +\nn
&+&{d-4\over4}\,R_i\,\na^\b\chi(\na_\a V_{i\b} + \na_\b V_{i\a}) =0,
\le{constrainteq}
it obviously reduces to a boundary action. Note that this action will 
generally consist of two disconnected pieces corresponding to the 
two boundaries of the longitudinal space-time. 

When $\L<0$, as we have been implicitly assuming in this section, the 
boundary is timelike. In the large curvature regime, the discussion for 
$\L>0$ is more intrincate. Formally, the above derived action is valid in 
de Sitter space as well, the boundary being now a spacelike boundary at 
the future and past infinities. This agrees with Bousso's considerations 
on de Sitter space in \cite{Bousso}. As in this case the boundary theory is 
defined on an Euclidean manifold $\pa M$, the physical interpretation in 
terms of causality and locality of a corresponding  holographic map is 
somewhat 
more mysterious than in the AdS case, as discussed in \cite{Bousso,HKS,FKMP}.

\subsection{Weak curvature regime}

In this regime there are  two types of solutions, curved (singular) and 
flat. The action is the same as in the strong curvature regime, apart
from an additional term proportional to the cosmological constant,
\be
S_\parallel[g,h]&=&\int\@{-gh}\left[R[h] -2\L -{1\over4}\,h^{ij}
\pa_ig_{\a\b}\pa_jg_{\g\d}\,\e^{\a\g}\e^{\b\d}\right]\nn
&=&\int\@{-gh}\left[R[h] -2\L-\e_{\a\g}\e_{\b\d}\,h^{ij}\na^\a 
V^\b_i\na^\g V_j^\d +\half h^{ij}R_iR_j \right].
\ee
Again filling in the general solutions we get:
\be
S_\parallel&=&S_{\sm{bulk}}+S_{\sm{bdry}}\nn
&=&\int_{\pa M}\dd x^\a\@{\ti h}\,\left((R[\ti h]-2 e^\chi\L)\,\frac{}{} 
\,X^0\pa_\a X^1 +e^{{d-4\over2}\,\chi+\s}\e_{ab}\,\pa_iX^a\times\right.\nn
&\times&\left.[\pa^i\pa_\a X^b +\half\na^\b\s(\pa_\a X^bV^i_\b -
\pa_\b X^bV^i_\a)] -e^{{d-4\over2}\,\chi}(\half\,V_{i\a}R^i +{d-3\over2}
\,\e_{\a\b}V^{i\b}\pa_i\chi)\right) +\nn
&+&\int_M\@{-g\ti h}\,e^{{d-4\over2}\,\chi}\,V^{i\a}R_{i\a}
\ee

The most interesting case is the flat-space solution, where the action 
is simply quadratic:
\be
S_\parallel&=& S_{\sm{bulk}}+S_{\sm{bdry}}\nn
&=&\int_{\pa M}\dd x^\a\@{\ti h}\,[\e_{ab}\,\pa_\a X^b \, (\half R[\ti h]-\L 
-\triangle_{\ti h})\, X^a -\half\,V_{i\a}R^i] \nn
&+&\int_M\@{-g\ti h}\,e^{{d-4\over2}\,\chi}\,V^{i\a}R_{i\a}
\ee
The constraint then reads
\be
R_{i\a}=\half\,\e_{\a\b}\na^\b R_i=0.
\ee
Note that unlike \cite{VVErice} this does not imply that $R_i = 0$ but
that $R_i$ is a function only of the transverse co-ordinates $R_i(y)$. In 
particular then even for the flat space we have found that the complete 
analysis of this limit actually implies that there can be an additional 
term in the boundary action. It would be interesting to understand the
physical meaning of this extra piece. 

The action can be rewritten as
\be
S_\parallel&=&S_{\sm{bdry}}\nn
&=&\int_{\pa M}\dd x^\a\@{\ti h}\,\left(\e_{ab}\,\pa_\a X^a (\triangle_{\ti 
h} 
+\L -\half R[\ti h])\,X^b + \half\partial_\a X^aR^i(y) \partial_i
X_a\right)
\le{smallcurv}
which will be convenient for the discussions in the next section. 
Needless to say that in this case the classical solutions are independent 
of the value of the cosmological constant, and therefore the action 
\eq{smallcurv} allows any value of $\L$. 
We thus find ourselves with a quadratic action like that of \cite{VV}.
Correspondingly there will be a way to quantize this action, write down
the S-matrix and to study the inevitable non-commutativity of the boundary 
co-ordinates. In the next section
we will consider the relationship between our construction and the 
curved space-time shock-wave scattering considered in particular in 
a paper by Horowitz and Itzhaki \cite{HI}. 

\section{Shock-waves from eikonal gravity and the AdS/CFT 
Correspondence}\label{sec3.7}

The boundary action found in the small curvature regime for $R_i(y) = 0 $  
is quadratic and therefore easy to deal with. 
In fact it is a straightforward generalisation 
of the boundary action found in \cite{VV,VVErice}. 

Let us briefly discuss its quantum mechanical properties when we couple it 
to point particles. In this regime, and restricting ourselves to the 
classical solutions of the equations of motion, the longitudinal space is 
basically flat. Therefore, the coupling to point particles in this case 
goes precisely along the lines of section 5.1 of \cite{VV}. For details about 
the stress-energy tensor of a pointlike particle we refer to Appendix 
\ref{appA1}.

Following \cite{VV}, we represent the stress-energy tensor in terms of a 
momentum flux $P_{a\a}$ as follows:
\be
T_{\a\b}=P_{a\a}\,\pa_\b X^a.
\ee
As shown in \cite{VV}, in a forward scattering situation where the 
stress-energy tensor is concentrated in the longitudinal plane, the matter 
part of the action depends only on the boundary values of $X^a$. Its 
variation equals:
\be
\d S_{\sm{matter}}=2\int_{\pa{\cal M}}\dd 
x^\a\@{h}\,\e_\a^{\,\,\,\b}\,P_{a\b}\d X^a.
\ee
Coupling this action to the gravitational part of the action, \eq{smallcurv}, 
gives the standard shift equation, see e.g. \eq{X}.

Quantisation is now straightforward and, as discussed in \cite{VV} and also 
in chapters \ref{Intro} and \ref{HEscattering} of this thesis, it leads to 
non-trivial commutators for the co-ordinates $X^a$:
\be
[X^a(y),X^b(y')]=i\e^{ab}f(y,y'),
\le{XaXb}
where $f$ now satisfies the Green's function equation
\be
(\triangle_{\ti h}+\L -\half\,R[\ti h])\,f(y,y')=\delta^{(d-2)}(y-y').
\le{greeneq}

As we have already discussed, we expect shock-waves to be described by our 
boundary action also. Let us first briefly discuss how shock-waves can be 
implemented in AdS \cite{HI}.

We write pure AdS in the following co-ordinates,
\be
\dd s^2={4\over(1-y^2/\ell^2)^2}\,\et_{\m\n}\dd y^\m\dd y^\n,
\ee
where $y^2=\et_{\m\n}y^\m y^\n$. The stress tensor of a massless particle 
is computed in Appendix \ref{appA1} and gives 
\be
T_{uu}=-p\,\d(u)\d(\r),
\ee
where $\r$ is the radial co-ordinate $\r=\sum_{i=1}^{d-2}y_i^2$.

Horowitz and Itzhaki found the following solution of Einstein's equations 
with a massless particle:
\be
\dd s^2={4\over(1-y^2/\ell^2)^2}\,\left(\et_{\m\n}\dd y^\m\dd y^\n +
8\pi\GN\,p \d(u)(1-\r^2/\ell^2)f(\r)\dd u^2\right)
\le{H-I}
provided
\be
\triangle_h f -4\,{d-2\over\ell^2}f=\d(\r).
\le{shiftads}
$\triangle_h$ is the Laplacian on the transverse hyperbolic space,
\be
\dd s^2={\dd\r^2+\r^2\dd\O^2_{d-3}\over(1-\r^2/\ell^2)^2},
\ee
and therefore \eq{shiftads} takes the form
\be
f''+{d-3+(d-5)\rho^2/\ell^2\over\rho(1-\rho^2/\ell^2)}f'
-{4(d-2)\over\ell^2(1-\rho^2/\ell^2)^2}\,f&=&\d(\rho).
\ee

The solutions to \eq{shiftads} are given by:
\be
{\mbox{AdS}_3:}\,\,\,\,\,\,\,\,\,\,\, f(\r)&=&{\ell\over2}
(C+\th(\r))\,\sinh\log\left({\ell+\r\over\ell-\r}\right) +
\ell D\cosh\log\left({\ell+\r\over\ell-\r}\right)\nn
{\mbox{AdS}_4:}\,\,\,\,\,\,\,\,\,\,\, f(\r)&=&C\,{1+\r^2/\ell^2\over1-
\r^2/\ell^2}\,\log(\r/D)+{2C\over1-\r^2/\ell^2}\nn
{\mbox{AdS}_5:}\,\,\,\,\,\,\,\,\,\,\, f(\r)&=&{C\over1-\r^2/\ell^2}
\left({1\over\r}+{6\r\over\ell^2} +{\r^3\over\ell^4}\right) +
{D\over\ell}\,{1+\r^2/\ell^2\over1-\r^2/\ell^2},
\le{fsol}
$D$ is an arbitrary constant to be determined by boundary conditions. 
$C$ is a constant of order 1 that can be computed either by explicit 
computation or by matching with the Minkowski solutions.
The shift function $f$ of course behaves like the solutions 
for shock-wave in Minkowski space $f\sim{1\over|x|^{d-4}}$ 
in the limit when the AdS radius divided by 
the impact parameter goes to infinity, $\ell/\r\rightarrow\infty$.
In fact the metric \eq{H-I} was derived by boosting a black hole to the 
speed of 
light while sending its mass to zero and keeping its energy fixed. 

Notice that for an Einstein space with negative curvature and curvature 
radius $\L=-{1\over2\ell^2}\,(d-1)(d-2) $, the above general equation for the 
shift function derived via our boundary action method \eq{greeneq} 
reduces to the condition \eq{shiftads} 
found by Horowitz and Itzhaki precisely when the transverse space is
Euclidean AdS$_{d-2}$:
\be
\dd s^2=4\,{\dd\r^2+\r^2\dd\O_{d-3}^2\over(1-{\r^2\over\ell^2})^2}.
\ee
In this case, the transverse curvature is
\be
R[\ti h]=-{1\over\ell^2}\,(d-2)(d-3).
\ee

The class of solutions to \eq{greeneq} is 
however much larger than only shock-waves in pure AdS. It allows for values 
of the transverse curvature that are postitive, negative or zero, and the 
cosmological constant is also allowed to be positive. In the limit 
$\L\rightarrow0$, all our results of course 
agree with the results found in \cite{gnp85}.

It is not surprising that we find an approximate shock-wave from our boundary 
action only in the small curvature regime. These shock-waves have a smooth 
limit as $\Lambda\rightarrow0$ which of course could not happen in 
the large curvature regime. Note that, as for the Dray-'t Hooft Ansatz 
\eq{8.0}, the transverse metric $\ti h_{ij}(y)$ is not determined by 
Einstein's equations and is to be treated as a classical back-ground.

Horowitz and Itzhaki have argued that the CFT duals of shock-waves are 
``light-cone states'' -- states with their energy-momentum tensor 
localised on the boundary light-cone. It is tempting to argue that our 
boundary description should somehow be related to these light-cone states. 
Indeed, we have shown that our boundary theory describes bulk shock-waves 
in an approximate fashion. Hence one is 
led to speculate that our boundary action is somehow related to some sort 
of eikonal limit of a boundary CFT perturbed by the addition of light-cone 
states. Notice, however, that it is not at all clear how to prove such a 
relation.  In particular it is not clear how light-cone states 
should be precisely described in field theory, although some attempts have 
been made in \cite{PST}. Related discussions can be found in 
\cite{PolS-m,SussS-m,Steve} and, recently, in \cite{GiLi}. Furthermore
if quantum gravity has a boundary description at all energies we have 
taken the eikonal limit of it, thereby explicitly breaking covariance of 
the boundary theory. An interesting question is whether it is possible to 
do an eikonal approximation in a covariant way, or whether it is possible 
to restore covariance afterwards, as discussed in the previous chapter. See 
also \cite{g9607,g9805,SdHJHEP}. In particular, as we saw before, in the 
simplified 2+1 -- dimensional setup, restoring Lorentz covariance is 
tantamount to going beyond the extreme eikonal regime.

It would be extremely interesting if we could find an analog of \eq{XaXb} 
in the context of the AdS/CFT correspondence. 
This would amount to identifying 
the operators $X^a$ in the CFT and to interpreting them from the bulk point 
of 
view. Based on previous considerations by 't Hooft and a computation of 
the trajectories of massless particles outlined in Appendix 
\ref{appA1}, they 
are expected to correspond to the positions of colliding particles, however
a careful analysis is required. This could most easily
be done using the techniques in \cite{KSS1} where boundary sources and 
operators are related to the coefficients of the perturbative expansion of 
bulk fields.

\subsection{Boundary description of scalar fields}

So far we have discussed single particles in an AdS background and 
interactions between quantum mechanical particles by means of shock-waves. 
One would however be ultimately interested 
in considering second quantised fields that interact gravitationally. 
In flat space, 
computing an S-matrix and extracting from it the amplitude for scattering 
between massless particles is a relatively straightforward task even if 
the interactions are gravitational \cite{g87,g9607}. In AdS, however, 
things are much more complicated due to the presence of the timelike 
boundary and the impossibility to separate wavepackets. These problems 
can be sidestepped by imposing appropriate boundary conditions on the 
fields and ensuring that the S-matrix is unitary \cite{AvIsSt}. However, 
this is not possible for all the modes, and in the context of the AdS/CFT 
correspondence we are interested in considering both normalisable and 
non-normalisable modes. For other discussions of the AdS S-matrix, see 
\cite{PolS-m,SussS-m,Steve}. In this chapter we will not consider this issue, 
but rather concentrate on the CFT duals of scalar fields with generic 
boundary conditions.

As a first step towards considering the full quantum mechanics of scalar 
fields interacting gravitationally in AdS, we consider scalar fields on an 
AdS-shock-wave background. We concentrate on conformally coupled scalar 
fields. These have the nice property that their equation of motion is 
invariant under Weyl rescalings, up to a certain weight. The Klein-Gordon 
equation for these fields is
\be
\left(\Box_G -{d-2\over4(d-1)}\,R[G]\right)\f(y)=0,
\le{37}
and so in the AdS-shock-wave background they have mass $m^2=-
{d^2-1\over4\ell^2}$. In this section and in the following chapters, $d$ will 
denote the dimension of the boundary of the $(d{+}1)$-dimensional bulk AdS.

We perform a conformal transformation by which we remove the double pole 
of the metric:
\be
\dd s^2=G_{\m\n}\,\dd y^\m\dd y^\n ={1\over\O(y)^2}\,\bar 
G_{\m\n}\dd y^\m\dd y^\n,
\ee
$G$ being the AdS-shockwave metric \eq{H-I}. The Klein-Gordon equation 
transforms into
\be
\stackrel{-}{\Box}\! \bar\phi(y)=0,
\le{38}
calculated in the metric $\bar G_{\m\n}$, and 
\be
\bar\f(y)=\O^{1-d\over2}\,\,\f(y).
\ee
There is no curvature term in \eq{38} because $R[\bar G]=0$. For the 
metric $\bar g_{\m\n}$, the Laplacian factorises into a flat piece plus 
a shock-wave part,
\be
\stackrel{-}{\Box}\!\bar\f(y)=\et^{\m\n}\pa_\m\pa_\n\bar\f(y) -p\,\d(u)
\,F(\rho)\,\pa_v^2\bar\f(y)=0.
\le{40}
Equation \eq{40} is difficult to solve in general due to the transverse 
derivatives\footnote{For exact solutions, see \cite{Garriga}.} but it can be 
readily solved in the eikonal approximation. A simple plane-wave solution is 
given by 
\be
\phi(y)&=&\O(y)^{{d-1\over2}}\,\exp \left[ikv+ ikp\,\th(u)\,F(\rho)\right],
\le{44}
as one would expect from a computation of trajectories: the only effect 
of the shock-wave is a shift of the wave function over a distance given 
by the shift. The full solution gives, in the eikonal approximation,
\be
\phi(y)=\O(y)^{d-1\over2}\int\dd^dk\, a(k)\, e^{ipk\,\th(u)F(\rho) +
ik_\m y^\m} +\mbox{c.c.},
\le{45}
where $k_\m^2=0$. Note that this sense of eikonal approximation is 
the same as in previous sections -- all transverse derivatives
are set to zero. 

To interpret this classical field from the CFT point of view, it is 
easiest to go to Poincare co-ordinates where the boundary is at $r=0$. 
The above field then has the following expansion \cite{KSS1}
\be\label{expansion}
\phi(r,x)=r^{d-1\over2}\phi_0(x)+\cdots
\ee
as it approaches the boundary. This is the expected behaviour for a field 
of mass $m^2=-{d^2-1\over4\ell^2}$. As explained in the introduction, for 
a field of such mass the value of the field at infinity can have an 
interpretation either as the expectation value of the operator dual to the 
field or as a source. The coefficient at order $r^{d+1\over2}$ is then 
interpreted as the source or as the dual operator respectively.

Let us consider the $\D_+$-theory, where $\f_0$ corresponds to an operator 
of dimension $\D={d-1\over2}$,
\be
\bra O(x)\ket=-\f_0(x).
\ee
The expression for $\f_0$ can be obtained from \eq{45}. Notice that
as $r\rightarrow0$ the step function approaches 
$\th(u)\rightarrow\th({t^2-\vec{x}^2\over t})$. This means that the 
operator $O(x)$ has different expectation values on either side of the 
light-cone, $|t|>|\vec{x}|$ and $|t|<|\vec{x}|$, and furthermore there 
is a reflection as $t\rightarrow-t$. The operator acquires a certain 
``dressing" inside the light-cone. In the $\D_-$-theory, where $\f_0$ is 
interpreted as a source for $O(x)$, we see that the effect of the 
shock-wave is to introduce an explicit time-dependence in the source. 

As pointed out in \cite{HI} shock-waves in AdS correspond 
to states with a stress-energy tensor concentrated on the light cone. 
We have found that when we also turn on a source for an operator 
of dimension $\D={d-1\over2}$ in the back-ground of these light-cone 
states, the operator aquires different values on either side of the 
light-cone. Elaborating this a little bit further along the lines of 
\cite{BKLT} let us add that there is a map between the creation 
and the annihilation 
operators of the field $\phi$ and the composite operators in terms of
which $O(x)$ is 
expanded. This however assumes a well-defined field theory for the 
scalar field $\phi$ in AdS, which we certainly have not constructed here 
(see however \cite {AvIsSt,SussS-m,PolS-m,BKL}). One has to find a complete 
set of operators 
that generate the Hilbert space of the boundary theory and that have a 
well-defined inner product. This imposes additional conditions on the 
solutions \eq{45} for them to be normalisable, like the quantisation of 
the frequencies. It would be most interesting to work out all these 
details, and to have an explicit field theory realisation of these phenomena.

The next step would be to consider gravitationally interacting fields in 
this AdS back-ground. In reference \cite{VVexchange} it was shown that 
fields interacting by means of shock-waves on a black hole horizon satisfy 
an exchange algebra, of the form:
\be
\phi_{\tn{out}}(y)\phi_{\tn{in}}(x)=\exp\left[if^{ab}(x-y){\pa\over\pa x^a}
{\pa\over\pa y^b} \right] \phi_{\tn{in}}(x)\phi_{\tn{out}}(y),
\le{exchange3}
where $f^{ab}(x-y)=\e^{ab}f(x-y)$ depends on the transverse distance between 
the points $x$ and $y$. Here the non-commutativity of the fields was 
ascribed to the fluctuations of the horizon due to in-coming and out-going 
shock-waves. In chapter \ref{HEscattering}, an alternative derivation of this 
exchange algebra has been given for Minkowski space. The derivation does not 
use 
the presence of a horizon, but only the fact that creation operators create 
particles that carry shock-waves with them and thus produce shifts on the 
back-ground 
space-time. This is closely related to the form \eq{45} of the solutions of
the Klein-Gordon equation, which up to a conformal factor is the same in 
AdS and in Minkowski space. Therefore it seems reasonable to expect that 
a similar kind of non-commutative behaviour is to be found in AdS.
It would be interesting to interpret this in terms of operators in the CFT. 
Note however that when performing such a derivation one can no longer 
ignore the problem of correct quantisation of fields in AdS.

It seems likely that yet another way to derive the algebra \eq{exchange3} 
is by coupling our boundary action not to point particles but to scalar 
fields whose energy-momentum tensor is concentrated mainly in the 
longitudinal space. 

When considering point particle fields in AdS and comparing them to 
quantities in the CFT, the UV/IR duality plays a crucial role 
\cite{BKLT,SuWi,PePo}. Bulk translations correspond to boundary rescalings. 
This will be particularly important if one develops an S-matrix theory, like 
we do in the following section for asymptotically flat spaces. In the usual 
Poincare co-ordinates of AdS, it is easy to see that a constant rescaling of 
the bulk co-ordinate $r\rightarrow e^\l r$ together with a Weyl rescaling of 
the boundary metric is a symmetry of the metric. In the commonly used AdS 
co-ordinates where the double pole is only in the warp factor,
\be
\dd s^2=\dd y^2+e^{2y}g_{ij}\,\dd x^i\dd x^j,
\ee
a translation $y\rightarrow y+\l$ induces a Weyl rescaling on the boundary, 
$g_{ij}\rightarrow e^{2\l}g_{ij}$. This symmetry is of course not a symmetry 
of the full theory due to infrared divergences, as we will study in more 
detail in chapter \ref{reconstruction}.

\section{Comments and Conclusions}\label{sec3.8}

Our analysis is a semi-classical analysis in the sense that we have 
setup a path-integral involving $S_\parallel$ that in addition involves only 
the fluctuations with insertion of fields all taking place on the boundary. 
Thus we have actually constructed a general proof of a particular form of 
holography -- that corresponding to interactions of massless particles
via gravitational shock-wave dynamics encoded in a theory of 
fluctuations on the boundary.

We would like to point out that our derivation requires no specific gauge 
choice. This agrees with \cite{VVErice}. However we do impose the requirement 
on our metric that it is of an approximately $2+(d-2)$ block-diagonal form 
with small off-diagonal
components $h_{i\a}$. We have seen that in the course of our construction
it was indeed very important to retain the small off-diagonal 
$h_{i\a}$ as it was precisely due to this that the constraint $R_{i\a}$ was
derived and which was of importance to remove all bulk terms in the 
theory. 

The fact that in this eikonal limit the theory becomes holographic in the
sense described above is due not {\it only} to the fact that we treat
essentially as a classical background the transverse metric but also
to the crucial fact that one has an additional constraint to
impose. As already remarked this constraint arises from the
linear fluctuations in $h_{i\a}$ at order $\epsilon^{-1}$ in the rescaled
action and is therefore associated to small off-diagonal pieces of the
metric. In the end there is therefore no complete decoupling of transverse
and longitudinal components of the metric as they are 
tied together by the non-trivial constraint $R_{i\a} = 0$ \eq{constrainteq}. 

The off-diagonal constraint essentially restricts the variations
of our solutions in the transverse directions. This is where the
dependence on the transverse direction is really taken into account. 
If follows that                            
to effectively obtain a boundary theory one simply imposes this
constraint on the transverse dynamics. We could rephrase the state of
affairs by saying that Einstein gravity in the eikonal is a topological
theory on a two-dimensional manifold embedded in $d$ dimensions, 
provided some constraints are
imposed on the ``lapse function'' $V_{i\a}$ which allows one to move from one 
plane
to another by means of transverse deformations.

We also recall that as stressed by 't Hooft 
the gravitational interactions close
to the horizon of a black hole or more generally at high energies
are precisely described by shock wave configurations associated to
boosted particles. They have non-trivial back-reaction effects, bringing 
about a shift in the geodesics of the outgoing particles which induces a form 
of non-commutativity at the quantum level. This has 
been observed in the analysis of \cite{VVErice}  
and similarly occurs here in the particular subset of cases considered for 
which the boundary action is quadratic. 

A particularly interesting result is the boundary action that we found for 
the asymptotically de Sitter case. This action is defined on a space-like 
surface which is the past or the future boundary of de Sitter space. This 
suggests that observables in a holographic description of de Sitter space can 
somehow be defined as correlation functions of a theory living on a 
space-like surface \cite{Bousso}. On the other hand, in \cite{HKS} it is 
argued that such a prescription may have little physical meaning as there is 
no physical observer that can collect together the data of measurements on 
such a surface. We regard the solution of this conundrum as an important 
challenge for future research.

Throughout this chapter we have worked with the Einstein-Hilbert action
without including higher curvature corrections. However
our method is perfectly 
applicable for these higher order terms also, and in fact considering them 
is important when the energy is increased above $1/\Pl$. 
When embedding our theory 
in a specific string theory, one also has to include additional matter 
fields. Notice, however, that for the case $d=5$ it should be straightforward 
to embed our results in string theory by condidering backgrounds with a 
constant dilaton and a covariantly constant self-dual 5-form compactified
for example on an $S^5$. This is left for future research.

\newpage

\chapter{Holographic Reconstruction of Space-time in the AdS/CFT 
Correspondence}
\label{reconstruction}

The contents of this chapter are based on \cite{KSS1}. For a review and 
explicit examples see also \cite{Kstrings00}, and for a short account of the 
AdS/CFT correspondence see section \ref{AdS/CFT} of the introduction and also 
reference \cite{AdSreview}. 

We develop a systematic method for renormalising the AdS/CFT prescription for 
computing correlation functions. This involves regularising the bulk on-shell 
supergravity action in a covariant way, computing all divergences, adding 
counter-terms to cancel them and then removing the regulator. We explicitly 
work out the case of pure gravity up to six dimensions and of gravity coupled 
to scalars, but the techniques can be easily applied for other matter fields. 
The method can also be viewed as providing a holographic reconstruction of 
the bulk space-time metric and of bulk fields on this space-time, out of 
conformal field theory data. Knowing which sources are turned on is 
sufficient in order to obtain an asymptotic expansion of the bulk metric and 
of bulk fields near the boundary to high enough order so that all infrared 
divergences of the on-shell action are obtained. To continue the holographic 
reconstruction of the bulk fields one needs new CFT data: the expectation 
value of the dual operator. In particular, in order to obtain the bulk metric 
one needs to know the expectation value of stress-energy tensor of the 
boundary theory. We provide completely explicit formulae for the holographic 
stress-energy tensors up to six dimensions. We show that both the 
gravitational and matter conformal anomalies of the boundary theory are 
correctly reproduced. We also obtain the conformal transformation properties 
of the boundary stress-energy tensors.

\section{Introduction and summary of the results}

Holography states that a $(d{+}1)$-dimensional gravitational 
theory\footnote{In this and the next chapter, we use the convention the 
boundary is $d$-dimensional whereas the bulk is $(d{+}1)$-dimensional} 
(referred to as the bulk theory) should have a 
description in terms of a $d$-dimensional
field theory (referred to as the boundary theory)
with one degree of freedom per Planck area
\cite{ghologr,Susskind}. The arguments leading to the 
holographic principle use rather generic properties
of gravitational physics, indicating that holography
should be a feature of any quantum theory of gravity.
Nevertheless it has been proved a difficult task 
to find examples where holography is realised,
let alone to develop a precise dictionary 
between bulk and boundary physics. The AdS/CFT 
correspondence \cite{Malda} provides such a realisation \cite{Wit,SuWi}
with a rather precise computational framework \cite{Gubs,Wit}.
It is, therefore, desirable to sharpen the existing 
dictionary between bulk/boundary physics as much as possible. 
In particular, one of the issues one would like to
understand is how space-time is built 
holographically out of field theory data.

The prescription of \cite{Gubs,Wit} gives a concrete 
proposal for a holographic computation of physical
observables. In particular, the partition function
of string theory compactified on AdS spaces 
with prescribed boundary conditions for the 
bulk fields is equal to the generating functional 
of conformal field theory correlation 
functions, the boundary value of fields being now 
interpreted as sources for operators of the dual conformal 
field theory (CFT). 
String theory on anti-de Sitter (AdS) spaces is still incompletely
understood. At low energies, however, the theory 
becomes a gauged supergravity with an AdS ground
state coupled to Kaluza-Klein (KK) modes. On the 
field theory side, this corresponds to the large $N$
and strong 't Hooft coupling regime of the 
CFT. So in the AdS/CFT context the question is how
one can reconstruct the bulk space-time out of
CFT data. One can also pose the converse 
question: given a bulk space-time, what 
properties of the dual CFT can one read off?

The prescription of \cite{Gubs,Wit} equates the 
on-shell value of the supergravity action 
with the generating functional of connected graphs
of composite operators, see \eq{correspondence}-\eq{correspondence2}. Both 
sides of this correspondence, however, suffer from infinities
---infrared divergences on the supergravity side
and ultraviolet divergences on the CFT side.
Thus, the prescription of \cite{Gubs,Wit}
should more properly be viewed as an equality between
bare quantities. One needs to renormalise the theory to obtain a 
correspondence between 
finite quantities. It is one of the aims of this 
chapter to present a systematic way of performing 
such a renormalisation.
 
The CFT data\footnote{We assume that the 
CFT we are discussing has an AdS dual.
Our results only depend on the space-time 
dimension and apply to all cases where the AdS/CFT duality 
is applicable, so we shall not specify any particular
CFT model.} that we will use are: which 
operators are turned on, and what is their vacuum 
expectation value. Since the boundary metric (or, more properly,
the boundary conformal structure) couples to the boundary 
stress-energy tensor, the reconstruction of the 
bulk metric to leading order involves a detailed
knowledge of the way the energy-momentum  tensor 
is encoded holographically. 
There is by now an extended literature on the 
study of the stress-energy tensor in the context of the 
AdS/CFT correspondence starting from \cite{BK,Myers}.
We will build on these and other related works \cite{EJM,Mann,KLS}. 
Our starting point will be the calculation of the 
infrared divergences of the on-shell gravitational action
\cite{HS}. Minimally subtracting the divergences 
by adding counter-terms \cite{HS} leads straightforwardly to the 
results in \cite{BK,EJM,KLS}. After the subtractions 
have been made one can remove the (infrared) regulator
and obtain a completely explicit formula for
the expectation value of the dual 
stress-energy tensor in terms of the gravitational solution.

We will mostly concentrate on the gravitational sector,
i.e. on the reconstruction of the bulk metric, 
but we will also discuss the coupling to scalars. 
Our approach will be to build perturbatively an Einstein manifold 
of constant negative curvature (which we will sometimes 
refer to as an asymptotically 
AdS space) as well as a solution to the scalar field equations
on this manifold out of CFT data. The CFT data we start from 
is what sources are turned on. We will include 
a source for the dual stress-energy tensor as well 
as sources for scalar composite operators. 
This means that in the bulk we need to solve the 
gravitational equations coupled to scalars 
given a conformal structure at infinity and 
appropriate Dirichlet boundary conditions for the 
scalars. It is well-known that if one considers the standard Euclidean AdS 
(i.e., with isometry $SO(1,d+1)$), the scalar field
equation with Dirichlet boundary conditions 
has a unique solution. In the Lorentzian case,
because of the existence of normalisable modes, the 
solution ceases to be unique. Likewise,
the Dirichlet boundary condition problem 
for (Euclidean) gravity has a unique 
smooth solution (up to diffeomorphisms) in the case the bulk manifold is 
topologically a ball and the boundary conformal 
structure is sufficiently close to the standard one \cite{GrahamLee}.
However, given a boundary topology there may be 
more than one Einstein manifold with this boundary. 
For example, if the boundary has the topology
of $S^1 \times S^{d-1}$, there are two possible bulk manifolds
\cite{PageH,Wit}:
one which is obtained from standard AdS by global identifications
and is topologically $S^1 \times R^d$,
and another, the Schwarzschild-AdS black hole, 
which is topologically $R^2 \times S^{d-1}$.

We will make no assumption
on the global structure of the space nor its
signature. The CFT should provide additional 
data in order to retrieve this information.
Indeed, we will see that only the information 
about the sources leaves undetermined the 
part of the solution which is sensitive to 
global issues and/or the signature of space-time.
To determine that part one needs new CFT data.
To leading order these are  
the expectation values of the CFT operators.

In particular, in the case of pure gravity, we find that 
generically a boundary conformal structure 
is not sufficient in order to 
obtain the bulk metric. One needs more CFT data.
To leading order one needs to specify  
the expectation value of the boundary stress-energy tensor.
Since the gravitational field equation is a second
order differential equation, one may expect that these data are sufficient 
in order to specify the full solution. However, higher point functions of the 
stress-energy tensor may be necessary if higher derivatives corrections
such as $R^2$-terms are included in the action.
We emphasise that we make no assumption about the regularity 
of the solution. Under additional assumptions 
the metric may be determined by fewer data. 
For example, as we mentioned above, under certain 
assumptions on the topology and the 
boundary conformal structure one obtains 
a unique smooth solution \cite{GrahamLee}.
Another example is the case when
one restricts oneself to conformally
flat bulk metrics. Then a conformally flat boundary metric 
does yield a unique bulk metric, up to diffeomorphisms and 
global identifications \cite{KS}.

Turning things around, given a specific solution,
we present formulae for the expectation values 
of the dual CFT operators. In particular, in the 
case the operator is the stress-energy tensor,
our formulae have a ``dual'' meaning \cite{BK}:
both as the expectation value of the 
stress-energy tensor of the dual CFT 
and as the quasi-local stress-energy tensor 
of Brown and York \cite{BrownYork}. We provide very explicit
formulae for the stress-energy tensor associated with
any solution of Einstein's equations with 
negative constant curvature. 

Let us summarise these results for space-time 
dimension up to six\footnote{In this chapter the dimension we refer to is the 
dimension of the boundary. So, $d=6$ corresponds to asymptotically AdS$_7$.}. 
The first step is to 
rewrite the solution in the Graham-Fefferman
co-ordinate system \cite{FeffermanGraham}
\be \label{GrFe}
\dd s^2=G_{\m \n} \dd x^\m dx^\n = {l^2 \over r^2}\left(\dd r^2 + 
g_{ij}(x,r) \dd x^i \dd x^j\right),
\eea
where 
\be
g(x,r)=g_{(0)} + r^2 g_{(2)}+ \cdots + r^d g_{(d)} + h_{(d)} r^{d} \log r^2 +
{\cal O}(r^{d+1}).
\eea
The logarithmic term appears only in even dimensions. $l$ is a parameter of 
dimension of length related to the cosmological constant as 
$\L=-{d(d-1) \over 2 l^2}$. Any asymptotically AdS metric can be brought in 
the form (\ref{GrFe}) near the boundary (\cite{GrahamLee}, 
see also \cite{GrahamWitten,Graham}). Once this co-ordinate 
system has been reached, the expectation value of the boundary stress-energy 
tensor reads
\be \label{tx}
\<T_{ij}\>={d l^{d-1} \over 16 \p \GN}\, g_{(d)ij} + X_{ij}[g_{(n)}],
\eea
where $X_{ij}[g_{(n)}]$ is a function of $g_{(n)}$ with $n<d$.
Its exact form depends on the space-time dimension and it reflects the
conformal anomalies of the boundary conformal field theory.
In odd (boundary) dimensions, where there are no gravitational conformal
anomalies, $X_{ij}$ is equal to zero. The expression for $X_{ij}[g_{(n)}]$
for $d=2,4,6$ can be read off from the formulae that will be given in 
(\ref{T2}), (\ref{T4}) and (\ref{T6}), respectively. The universal part of 
(\ref{tx}) (i.e. with $X_{ij}$ omitted)  
was obtained previously in \cite{Myers}. 
Actually, to obtain the dual stress-energy tensor it is 
sufficient to only know $g_{(0)}$ and $g_{(d)}$ as $g_{(n)}$ with 
$n<d$ are uniquely determined from $g_{(0)}$, as we will see. 
The coefficient $h_{(d)}$ of the logarithmic term 
in the case of even $d$ is also directly related to the 
conformal anomaly: it is proportional to the metric 
variation of the conformal anomaly.
 
It was pointed out in \cite{BK} that this prescription for 
calculating the boundary stress-energy tensor  
provides also a novel way, free of divergences\footnote{
We emphasise, however, that one 
has to subtract the logarithmic divergences in even dimensions
in order for the stress-energy tensor to be finite.},
of computing the gravitational quasi-local 
stress-energy tensor of Brown and York \cite{BrownYork}.
Conformal anomalies reflect infrared divergences 
in the gravitational sector \cite{HS}. 
Because of these divergences
one cannot maintain the full group of isometries even asymptotically.
In particular, the isometries of AdS that rescale the radial 
co-ordinate (these correspond to dilations in the CFT) 
are broken by infrared divergences. 
Because of this fact, 
bulk solutions that are related by diffeomorphisms that
yield a conformal transformation in the boundary do not 
necessarily have the same mass. Assigning zero mass 
to the space-time with boundary $R^d$,
one obtains that, due to the conformal anomaly,
the solution with boundary $R \times S^{d-1}$ has 
non-zero mass. This parallels exactly the discussion 
in field theory. In that case, starting from the 
CFT on $R^d$ with vanishing expectation value
of the stress-energy tensor, one obtains 
the Casimir energy of the CFT on $R \times S^{d-1}$
by a conformal transformation \cite{CC}. The
agreement between the gravitational ground-state energy 
and the Casimir energy of the CFT is a direct consequence
of the fact that the conformal anomaly 
computed by weakly coupled gauge theory and by 
supergravity agree \cite{HS}. It should be noted that, 
as emphasised in \cite{BK}, agreement between gravity/field
theory for the ground state energy is achieved only after
all ambiguities are fixed in the same manner on both
sides.
 
A conformal transformation in the boundary theory is realised in the bulk as 
a 
special diffeomorphism that preserves the form of the co-ordinate system 
(\ref{GrFe}) \cite{ISTY}. Using these diffeomorphisms one can easily study 
how the (quantum, i.e., with the effects of the conformal anomaly taken into 
account) stress-energy tensor transforms under conformal transformations.
Our results, when restricted to the cases studied in the literature 
\cite{CC},
are in agreement with them. We note that the present determination is 
considerably easier than the one in \cite{CC}.

Let us briefly discuss in more detail how conformal invariance is broken. As 
is well-known \cite{Malda}, the bulk metric does not quite induce a metric on 
the boundary, but only a conformal class of metrics. Since the metric has a 
double pole on the boundary \cite{FeffermanGraham}, one can define a metric 
by extracting this pole. That is, pick a positive function $r$ with a single 
zero at the boundary. The induced boundary metric is then given by 
$g_{(0)}=r^2G|_{\pa M}$ where $\pa M$ is the boundary of the manifold $M$. 
However, there is an obvious arbitrariness in this definition in that any 
other function $r'=e^wr$ with a single zero gives an equally valid boundary 
metric. Therefore, the metric on the boundary is defined up to a conformal 
transformation. This already indicates that the holographic dual should be a 
conformal theory, and is very similar to how in the eikonal regime of quantum 
gravity bulk time translations give rise to Lorentz boosts of the boundary 
theory.

On the other hand, infrared divergences break the symmetries of the bulk. To 
renormalise the theory we introduce a cut-off on the radial variable at 
$r=\e$. One can then renormalise the action by adding covariant counter-terms 
which are evaluated at the cut-off $r=\e$. When sending the cut-off to zero, 
the action should be finite. However, the presence of a logarithmic 
divergence gives rise to an anomaly when we perform a conformal 
transformation on $g_{(0)}$, $g_{(0)}'=e^{2\s}g_{(0)}$. This is a special 
kind of bulk diffeomorphism and so one would naively expect it to be a 
symmetry of the action. But it transforms as \cite{HS,Kstrings00}:
\be
S_{\sm{ren}}[e^{2\s}g_{(0)}]=S_{\sm{ren}}[g_{(0)}] +{\cal A}[g_{(0)},\s],
\ee
where the anomaly ${\cal A}$ is a conformally invariant functional of the 
metric \cite{HS} and it precisely corresponds to the conformal anomalies 
found on the gauge theory side.

The fact that infrared divergences break bulk diffeormorphisms means that 
only diffeomorphisms that do not induce a Weyl rescaling on the boundary are 
true symmetries of the theory. This implies that bulk solutions which are 
related by a diffeomorphism may have different dual stress-tensors when the 
diffeomorphism induces a conformal transformation on the boundary. 

The discussion is qualitatively the same when one 
adds matter to the system. We discuss 
scalar fields but the discussion generalises straightforwardly
to other kinds of matter. We study both the case when the 
gravitational background is fixed and the case when gravity is dynamical. 

Let us summarise the results for the case of scalar fields
in a fixed gravitational background (given by a 
metric of the form (\ref{GrFe})). We look for 
solutions of massive scalar fields with mass
$m^2=(\D-d) \D$
that near the boundary have the form (in the co-ordinate system (\ref{GrFe}))
\be \label{frexp}
\F(x,r)&=&r^{d-\D}\left(\f_{(0)} + r^2 \f_{(2)} + \cdots +
r^{2\D-d} \f_{(2\D-d)}+\right.\nn
&&\,\,\,\,\,\,\,\,\,\,\,\,\,\,\,\,\left.+\, r^{2\D-d} \log r^2 
\psi_{(2\D-d)}\right)
+{\cal O}(r^{\D+1}).
\eea
The logarithmic terms appears only when $2\D-d$ is an integer
and we only consider this case in this chapter.
We find that $\f_{(n)}$, with $n<2\D-d$, and $\psi_{(2\D-d)}$
are uniquely determined from the scalar field equation.
This information is sufficient for a complete 
determination of the infrared divergences of the 
on-shell bulk action. In particular, 
the logarithmic term $\psi_{(2\D-d)}$ in (\ref{frexp})
is directly related to matter conformal anomalies.
These conformal anomalies were shown not to renormalise
in \cite{PeSk}. We indeed find exact agreement with the computation 
in \cite{PeSk}. Adding counter-terms to cancel 
the infrared divergences we obtain the renormalised on-shell
action. We stress that even in the case of a free 
massive scalar field in a fixed AdS background 
one needs counter-terms in order for the on-shell action
to be finite (see (\ref{finiteact})).
The coefficient $\f_{(2\D-d)}$ is left undetermined
by the field equations. It is determined, however, by the 
expectation value of the dual operator $O(x)$. Differentiating the 
renormalised on-shell action one finds (up to terms 
contributing contact terms in the 2-point function)
\be
\< O (x)\> = (2 \D- d) \f_{(2\D-d)}(x).
\eea
This relation, with the precise proportionality coefficient,
has first been derived in \cite{KleWit}. The value of the proportionality 
coefficient is crucial in order to obtain the correct 
normalisation  of the 2-point function in standard
AdS background \cite{FMMR}.

In the case when the bulk geometry is dynamical we find that, 
for scalars that correspond to irrelevant operators, our
perturbative treatment is consistent only if one considers
single insertions of the irrelevant operator, i.e. the source
is treated as an infinitesimal parameter, in agreement with the discussion 
in \cite{Wit}. For scalars that correspond to marginal and
relevant operators one can compute perturbatively the back-reaction
of the scalars to the gravitational background. One can then 
regularise and renormalise as in the discussion of pure 
gravity or scalars in a fixed background. For illustrative
purposes we analyse a simple example.

This chapter is organised as follows. In the next section we discuss 
the Dirichlet problem for AdS gravity and we obtain an asymptotic 
solution for a given boundary metric (up to six dimensions).
In section \ref{holstente} we use these solutions to obtain 
the infrared divergences of the on-shell gravitational action.
After renormalising the on-shell action by adding counter-terms,
we compute the holographic stress-energy tensor. Section \ref{cotrpr} is 
devoted
to the study of the conformal transformation properties of the 
boundary stress-energy tensor. In section \ref{matter} we extend the analysis
of sections \ref{Dirichletgrav} and \ref{holstente} to include matter. 
In appendices \ref{EinSol} and \ref{as-sc}
we give  the explicit form of the solutions discussed 
in section \ref{Dirichletgrav} and section \ref{matter}. Appendix 
\ref{div-ind} contains the explicit form of the counter-terms discussed 
in section \ref{holstente}. In appendix \ref{h-a} we present a proof that the 
coefficient of the logarithmic term in the metric (present in even 
boundary dimensions) is proportional to the metric variation of the conformal 
anomaly.

\section{Dirichlet boundary problem for AdS gravity}\label{Dirichletgrav}

The Einstein-Hilbert action for a theory on a manifold $M$ 
with boundary $\pa M$ is given by\footnote{
Our curvature conventions are as follows:
$R_{ijk}{}^l=\pa_i \G_{jk}{}^l + \G_{ip}{}^l \G_{jk}{}^p - (i
\leftrightarrow j)$ and $R_{ij}=R_{ikj}{}^k$. We these conventions
the curvature of AdS comes out positive, but we will 
still use the terminology ``space of constant negative
curvature''. Notice also that we take
$\int \dd^{d+1} x = \int \dd^d x \int_0^\infty \dd r$
and the boundary is at $r=0$ (in the co-ordinate system (\ref{GrFe})). 
The minus sign in front of the trace of the second 
fundamental form is correlated with the choice of having $r=0$ in 
the lower end of the radial integration.} 
\be \label{action}
S_{\sm{gr}}[G]={1 \over 16 \p \GN}[\int_{M}\dd^{d+1}x\, 
\sqrt{G}\, (R[G] + 2 \L) 
- \int_{\pa M} \dd^d x\, \sqrt{\g}\, 2 K],
\eea
where $K$ is the trace of the second fundamental form (see \eq{fundform}) and
$\g$ is the induced metric on the boundary. The 
boundary term is necessary in order to get an action which 
only depends on first derivatives of the metric \cite{GibbonsHawking},
and it guarantees that the variational 
problem with Dirichlet boundary conditions is well-defined.

According to the prescription of \cite{Gubs,Wit}, the conformal 
field theory effective action is given by evaluating the 
on-shell action functional. The field specifying the 
boundary conditions for the metric is regarded as a source 
for the boundary operator. We therefore need to obtain solutions
to Einstein's equations,
\be \label{feq1}
R_{\m \n} - \half R G_{\m \n} = \L G_{\m \n},
\eea
subject to appropriate Dirichlet boundary conditions.

As explained above, metrics $G_{\m \n}$ that satisfy (\ref{feq1}) have a 
second order pole at infinity. Therefore, they do not induce a 
metric at infinity but only a conformal class. This is achieved by 
introducing a defining function $r$, i.e. a positive function in the 
interior of the manifold $M$ that  has a single zero and non-vanishing 
derivative at the boundary. Then one obtains a metric at the boundary by 
$g_{(0)}= r^2 G|_{\pa M}$ {}\footnote{Throughout this chapter the metric 
$g_{(0)}$
is assumed to be non-degenerate. For studies of the AdS/CFT 
correspondence in cases where $g_{(0)}$ is degenerate 
we refer to \cite{BPSV,marika}.}. 

We are interested in solving (\ref{feq1}) given a conformal structure
at infinity. This can be achieved by working in the co-ordinate 
system (\ref{GrFe}) introduced by Feffermam and Graham 
\cite{FeffermanGraham}.
The metric in (\ref{GrFe}) 
contains only even powers of $r$ up to the order we are interested in
\cite{FeffermanGraham} (see also \cite{GrahamWitten,Graham}).
For this reason, it is convenient to use the variable $\r=r^2$ \cite{HS},
{}\footnote{Greek indices, $\m,\n,..$ are used for $d+1$-dimensional indices, 
Latin ones, $i,j,..$ for $d$-dimensional ones. 
To distinguish the curvatures of the various metrics introduced in 
(\ref{coord}) we will often use the notation $R_{ij}[g]$ to 
indicate that this is the Ricci tensor of the metric $g$, etc.} 
\bea \label{coord}
&&ds^2=G_{\m \n} dx^\m dx^\n = l^2 \left({d\r^2 \over 4 \r^2} + 
{1 \over \r} g_{ij}(x,\r) dx^i dx^j \right), \nonu
&&g(x,\r)=g_{(0)} + \cdots + \r^{d/2} g_{(d)} + h_{(d)} \r^{d/2} \log \r + 
...,
\eea
where the logarithmic piece appears only for even $d$.
The sub-index in the metric expansion (and in all other 
expansions that appear in this chapter) indicates the number
of derivatives involved in that term, i.e. $g_{(2)}$ contains 
two derivatives, $g_{(4)}$ four derivatives, etc., as one can see from the 
explicit expressions given in appendix \ref{EinSol}. It follows
that the perturbative expansion in $\rho$ is also 
a low energy expansion. We set $l=1$ from now on. One can easily 
reinstate the factors of $l$ by dimensional analysis.

One can check that the curvature of $G$ satisfies
\be \label{ads}
R_{\k \l \m \n}[G] = (G_{\k \m} G_{\l \n} - G_{\k \n} G_{\l \m}) + {\cal 
O}(\r).
\eea
In this sense the metric is asymptotically anti-de Sitter. The 
Dirichlet problem for Einstein metrics satisfying (\ref{ads}) 
exactly (i.e. not only to leading order in $\rho$) was solved in \cite{KS}.
 
In the co-ordinate system (\ref{coord}), Einstein's equations read \cite{HS}
\bea
\rho \,[2 g^{\prime\prime} - 2 g^\prime g^{-1} g^\prime + \Tr\,
(g^{-1} g^\prime)\, g^\prime] + {\rm Ric} (g) - (d - 2)\,
g^\prime - \Tr \,(g^{-1} g^\prime)\, g & = & 0, \cr
\nabla_i\, \Tr \,(g^{-1} g^\prime) - \nabla^j g_{ij}^\prime  & = & 0, \cr
\Tr \,(g^{-1} g^{\prime\prime}) - \frac{1}{2} \Tr \,(g^{-1} g^\prime
g^{-1}
g^\prime) & = & 0 , \label{eqn}
\eea
where differentiation with respect to $\rho$ is denoted with a prime,
$\nabla_i$ is the covariant derivative constructed from the metric
$g$, and ${\rm Ric} (g)$ is the Ricci tensor of $g$.

These equations are solved order by order in $\r$. This is achieved
by differentiating the equations with respect to $\r$ and then setting 
$\r=0$. For even $d$, this process would have broken down at order $d/2$
if the logarithm was not introduced in (\ref{coord}). $h_{(d)}$
is traceless, $\Tr\, \gi h_{(d)}=0$,
and covariantly conserved, $\nabla^i h_{(d)ij}=0$.
We show in appendix \ref{h-a} that $h_{(d)}$
is proportional to the metric variation of the 
corresponding conformal anomaly, i.e. it is proportional to the 
stress-energy tensor of the theory with action 
the conformal anomaly. In any dimension, only the trace of 
$g_{(d)}$ and its covariant divergence are determined.
Here is where extra data from the CFT are needed:
as we shall see, the undetermined part is specified 
by the expectation value of the dual stress-energy tensor.

We collect in appendix \ref{EinSol} the results for 
$g_{(n)}$, $h_{(d)}$ as well as the results for the trace and 
divergence $g_{(d)}$. In dimension $d$ the latter are 
the only constraints that equations (\ref{eqn}) 
yield for $g_{(d)}$. From this information
we can parametrise the indeterminacy by finding the most general
$g_{(d)}$ that has the determined trace and divergence. 

In $d=2$ and $d=4$ the equation for the coefficient 
$g_{(d)}$ has the form of a conservation law
\be \label{gA}
\nabla^ig_{(d)ij}=\nabla^iA_{(d)ij}~~,\qquad d=2, 4,
\eea
where $A_{(d)ij}$ is a symmetric tensor explicitly constructed from
the coefficients $g_{(n)},~n<d$. 
The precise form of the tensor $A_{(d)ij}$ is given
in appendix \ref{EinSol} (eq.(\ref{Ad})). 
The integration of this equation obviously involves an
``integration constant'' $t_{ij}(x)$, a symmetric covariantly conserved
tensor the precise form of which cannot be determined from 
Einstein's equations.

In two dimensions, we get \cite{KS} (see also \cite{bautier})
\be \label{g2}
g_{(2) ij}= \half (R\, g_{(0) ij} + t_{ij}),
\eea
where the symmetric tensor $t_{ij}$ should satisfy
\be \label{t2}
\nabla^i t_{ij}=0, \qquad \Tr\, t = - R.
\eea

In four dimensions we obtain\footnote{From now on we will suppress 
factors of $g_{(0)}$. For instance,
$\Tr\, g_{(2)} g_{(4)}= \Tr\, [g_{(0)}^{-1}g_{(2)}g_{(0)}^{-1} g_{(4)}]$. 
Unless we explicitly mention the contrary, indices will be raised and
lowered with the metric $g_{(0)}$, and all contractions will 
be made with this metric.}
\be \label{g4}
g_{(4)ij}={1 \over 8} g_{(0)ij} \,[(\Tr\, g_{(2)})^2-\Tr\, g_{(2)}^2] + 
\half (g_{(2)}^2)_{ij} - {1 \over 4} g_{(2)ij}\, \Tr\, g_{(2)}
+t_{ij}.
\eea
The tensor $t_{ij}$ satisfies
\be \label{t4}
\nabla^i t_{ij}=0, \qquad \Tr\, t 
= -{1 \over 4} [(\Tr\, g_{(2)})^2 - \Tr\, g_{(2)}^2].
\eea 

In six dimensions  the equation determining the 
coefficient $g_{(6)}$ is more subtle than the one in (\ref{gA}).
It is given by
\be \label{dg6}
\nabla^ig_{(6)ij}=\nabla^i A_{(6)ij}
+{1\over 6}\Tr (g_{(4)} \nabla_jg_{(2)}),
\eea
where the tensor $A_{(6)ij}$ is given in (\ref{Ad}). It 
contains a part which is antisymmetric in the indices $i$ and $j$. 
Since $g_{(6)ij}$ is by definition a symmetric tensor
the integration of equation (\ref{dg6}) is not straightforward. 
Moreover, it is not obvious that
the last term in (\ref{dg6}) takes a form of divergence of some local tensor. 
Nevertheless, this is indeed the case 
as we now show. Let us define the tensor $S_{ij}$, 
\begin{eqnarray} 
\label{Sij}
&&S_{ij}=\nabla^2C_{ij}-2R^{k \ l}_{\ i \ j} C_{kl}
+4(g_{(2)}g_{(4)}-g_{(4)}g_{(2)})_{ij}
+{1\over 10}(\nabla_i\nabla_jB
-g_{(0)ij}\nabla^2 B) \nonumber \\
&&\hspace{1cm}
+{2\over 5}g_{(2)ij}B+g_{(0)ij}(-{2\over 3}\Tr \, g_{(2)}^3
-{4\over 15}(\Tr \,g_{(2)})^3+
{3\over 5}\Tr \, g_{(2)}\Tr \, g^2_{(2)})~~,
\end{eqnarray}
where 
$$
C_{ij}=(g_{(4)}-{1\over 2}g^2_{(2)}+{1\over 4}g_{(2)}\Tr \,g_{(2)})_{ij}+
{1\over 8}g_{(0)ij}B~~,~~
B=\Tr \, g^2_2-(\Tr \, g_2)^2~~.
$$
The tensor $S_{ij}$ is a local function of the Riemann tensor. Its 
divergence and trace read
\be \label{dtS}
\nabla^iS_{ij}=-4\Tr (g_{(4)} \nabla_j g_{(2)})~~,~~
\Tr S=-8\Tr(g_{(2)}g_{(4)})~~.
\eea
With the help of the tensor $S_{ij}$ the equation (\ref{dg6}) 
can be integrated in a way similar to the $d=2,4$ cases. One obtains
\be \label{g6}
g_{(6)ij}=A_{(6)ij}-{1\over 24}S_{ij} +t_{ij}~~.
\eea
Notice that tensor $S_{ij}$ contains an antisymmetric part which 
cancels the antisymmetric part of
the tensor $A_{(6)ij}$ 
so that $g_{(6)ij}$ and $t_{ij}$ are symmetric tensors, as they should.
The symmetric tensor $t_{ij}$ satisfies 
\be \label{t6}
\nabla^it_{ij}=0~~,~~\Tr\, t =-{1\over 3}[{1\over 8}(\Tr g_{(2)})^3
-{3\over 8}\Tr g_{(2)}\Tr g^2_{(2)}
+{1\over 2}\Tr g^3_{(2)}-\Tr g_{(2)} g_{(4)}]~~.
\eea

Notice that in all three cases, $d=2,4,6$, the trace of $t_{ij}$ is 
proportional to the holographic conformal anomaly.
As we will see in the next section, the symmetric tensors $t_{ij}$ 
are directly related to the expectation value of the boundary 
stress-energy tensor.

When $d$ is odd the only constraint on the coefficient $g_{(d)ij}(x)$ is 
that it is conserved and traceless
\be
\nabla^ig_{(d)ij}=0~~, \qquad \Tr \, g_{(d)}=0~~.
\label{odd}
\eea
So that we may identify
\be
g_{(d)ij}=t_{ij}~~.
\eea

\section{The holographic stress-energy tensor}\label{holstente}

We have seen in the previous section that given a conformal
structure at infinity we can determine an asymptotic expansion
of the metric up to order $\r^{d/2}$. We will now show that this term 
is determined by the expectation value of the dual stress-energy tensor. 

According to the AdS/CFT prescription, the expectation value
of the boundary stress-energy tensor is determined by 
functionally differentiating the on-shell gravitational 
action with respect to the boundary metric. 
The on-shell gravitational action, however, diverges. 
To regulate the theory we restrict
the bulk integral to the region $\r\geq\e$
and we evaluate the boundary term at $\r=\e$.
The regulated action is given by
\bea \label{regaction}
S_{\sm{gr,reg}}&=&{1 \over 16 \p \GN}\left[\int_{\r\geq\e} 
\dd^{d+1}x\, \sqrt{G} \,(R[G] + 2 \L) 
- \int_{\r=\e} \dd^d x \sqrt{\g}\, 2 K\right]= \\
&=&{1 \over 16 \p \GN} \int \dd^d x \left[ 
\int_\epsilon \dd\rho\,{d\over\rho^{d/2+1}}\,\sqrt{\det g(x,\rho )}\right.\nn
&+&\left. {1\over \rho^{d/2}}
(-2 d \sqrt{\det g(x,\rho )} +4 \rho\partial_\rho \sqrt{\det g (x,\rho 
)})|_{\rho=\epsilon}\right], 
\nonumber
\eea
Evaluating (\ref{regaction}) for the solution we obtained in the 
previous section we find that the divergences 
appears as $1/\e^k$ poles plus a logarithmic divergence \cite{HS},
\be \label{regaction1}
S_{\sm{gr,reg}} &=& {l \over 16 \pi \GN} \int \dd^d x \sqrt{\det g_{(0)}} 
\left( 
\epsilon^{-d/2} a_{(0)} + \epsilon^{-d/2+1} a_{(2)} + \ldots 
+ \epsilon^{-1} a_{(d - 2)}\right.\nn
&-&\left.{\over} \log \epsilon\, a_{(d)} \right) + {\cal O}(\e^0),
\eea
where the coefficients $a_{(n)}$ are local covariant expressions
of the metric $g_{(0)}$ and its curvature tensor. We give the
explicit expressions, up to the order we are interested in, in appendix 
\ref{appC}.

We now obtain the renormalised action by subtracting  
the divergent terms, and then removing the regulator,
\be \label{renaction}
S_{\sm{gr,ren}}[g_{(0)}]&=&\lim_{\e \to 0}{1 \over 16 \p \GN}
[S_{\sm{gr,reg}} -\int \dd^d x \sqrt{\det g_{(0)}} \left( 
\epsilon^{-d/2} a_{(0)} + \epsilon^{-d/2+1} a_{(2)} + \ldots\right.\nn
&+&\left.{\over} \epsilon^{-1} a_{(d - 2)} - \log \epsilon\, a_{(d)} 
\right)].
\eea
The expectation value of the stress-energy tensor of the 
dual theory is given by
\be \label{tij1}
\<T_{ij}\> = {2 \over \sqrt{\det \g_{(0)}}} 
{\pa S_{\sm{gr,ren}} \over \pa g_{(0)}^{ij}}
=\lim_{\e \to 0} 
{2 \over \sqrt{\det g(x, \e)}} {\pa S_{\sm{gr,ren}} \over \pa g^{ij}(x,\e)}  
=\lim_{\e \to 0}\left( {1 \over \e^{d/2-1}}\, T_{ij}[\g]\right),
\eea
where $T_{ij}[\g]$ is the stress-energy tensor of the theory 
at $\r=\e$ described by the action in (\ref{renaction}) but before the
limit $\e \to 0$ is taken ($\g_{ij}=1/\e\, g_{ij}(x,\e)$ is the 
induced metric at $\r=\e$). 
Notice that the asymptotic expansion of the metric only 
allows for the determination of the divergences of the 
on-shell action. We can still obtain, however, a formula 
for $\<T_{ij}\>$ in terms of $g_{(n)}$
since, as (\ref{tij1}) shows, we only need to know
the first $\e^{d/2-1}$ orders in the expansion of $T_{ij}[\g]$.
 
The stress-energy tensor $T_{ij}[\g]$ contains two contributions,
\be \label{tij2}
T_{ij}[\g]=T^{\sm{\sm{reg}}}_{ij}+T^{\sm{\sm{ct}}}_{ij},
\eea
$T^{\sm{\sm{reg}}}_{ij}$ comes from the 
regulated action in (\ref{regaction}) and $T^{\sm{\sm{ct}}}_{ij}$ 
is due to the counter-terms. The first contribution is equal to 
\be \label{regtij} 
T_{ij}^{\sm{reg}}[\g]&=&-{1 \over 8 \p \GN} (K_{ij} - K \g_{ij})\nn
&=&-{1 \over 8 \p \GN}\,(-\pa_\e g_{ij}(x,\e) + g_{ij}(x,\e)\, 
\Tr [g^{-1}(x,\e) \pa_\e g(x,\e)]\nn
&+&{1-d \over \e} g_{ij}(x,\e)).
\eea
The contribution due to counter-terms can be obtained from 
the results in appendix \ref{div-ind}. It is given by
\bea \label{counterT}
T^{\sm{ct}}_{ij}&=&-{1 \over 8 \p \GN} \left( (d-1) \g_{ij} + {1 \over (d-2)}
(R_{ij} - \half R \g_{ij})+ \right.\nonu 
&&\left.-{1 \over (d-4) (d-2)^2}[-\nabla^2 R_{ij} + 2R_{ikjl} R^{kl} 
+{d-2 \over 2 (d-1)} \nabla_i \nabla_j R - {d \over 2 (d-1)} R R_{ij} \right.
\nonu
&&\left.- \half \g_{ij} (R_{kl} R^{kl} - {d \over 4 (d-1)} R^2 
- {1 \over d-1} \nabla^2 R)] - T^a_{ij} \log \e \right),
\eea
where $T^a_{ij}$ is the stress-energy tensor of the action 
$\int \dd^d x\, \sqrt{\det \g}\, a_{(d)}$. As is shown in Appendix \ref{h-a},
$T^{a}_{ij}$ is proportional to the tensor $h_{(d)ij}$ appearing
in the expansion (\ref{coord}).

The stress tensor $T_{ij}[g_{(0)}]$ is 
covariantly conserved with respect to the metric $g_{(0)ij}$. 
To see this, notice that each of $T^{\sm{reg}}_{ij}$
and $T_{ij}^{\sm{ct}}$ is separately covariantly conserved 
with respect to the induced metric $\g_{ij}$ at $\r=\e$:
for $T^{\sm{reg}}_{ij}$ one can check this by using the 
second equation in (\ref{eqn}), for $T_{ij}^{\sm{ct}}$
this follows from the fact that it was obtained by varying 
a local covariant counter-term. 
Since all divergences cancel in (\ref{tij1}),
we obtain that the finite part in (\ref{tij1}) 
is conserved with respect to the metric $g_{(0)ij}$.

We are now ready to calculate $T_{ij}$.
By construction (and we will verify this below) the 
divergent pieces cancel between $T^{\sm{reg}}$ and $T^{\sm{ct}}$.

\subsection{$d=2$}

In two dimensions we obtain
\be
\<T_{ij}\>={l \over 16 \p \GN}\, t_{ij},
\eea
where we have used (\ref{g2}) and (\ref{t2})
and the fact that $T^a_{ij}=0$ since $\int R$ is a topological 
invariant (and reinstated the factor of $l$).
As promised, $t_{ij}$ is directly related to the boundary stress-energy 
tensor. Taking the trace we obtain
\be
\<T^i_i\> = -{c \over 24 \p}\, R,
\eea
where $c=3l/2\GN$, which is the correct conformal anomaly 
\cite{BrownHenneaux}. 

Using our results, one can immediately obtain the stress-energy tensor 
of the boundary theory associated with a given solution $G$
of the three dimensional Einstein equations: one needs
to write the metric 
in the co-ordinate system (\ref{coord}) and then use the formula
\be \label{T2}
\<T_{ij}\>={2 l \over 16 \p \GN}\, (g_{(2)ij} - g_{(0)ij}\,\Tr\, g_{(2)}).
\eea
{}From the gravitational point of view  this is the quasi-local stress 
energy tensor associated with the solution $G$.

\subsection{$d=4$}

To obtain $T_{ij}$ we first need to rewrite the expressions in 
$T^{\sm{ct}}$ in terms of $\g_{(0)}$.
This can be done with the help of the relation
\be
R_{ij}[\g] &=& R_{ij}[\g_{(0)}] + {1 \over 4}\, \e \left(2 R_{ik} R^k{}_{j}
-2 R_{ikjl} R^{kl} -{1 \over 3} \nabla_i \nabla_j R + \nabla^2 R_{ij}
-{1 \over 6} \nabla^2 R g_{(0) ij}\right)\nn
&+& {\cal O}(\e^2).
\eea

After some algebra one obtains,
\bea
\<T_{ij}[\gzero]\>&=&-{1 \over 8 \p \GN} \lim_{\e \to 0}
\left[{1 \over \e} (-g_{(2)ij} + g_{(0)ij} \Tr\, g_{(2)}
+ \half R_{ij} - {1 \over 4} g_{(0)ij} R)
\right. \nonu
&&\left. +\log \e\, (-2 h_{(4)ij} - T^a_{ij}) -2 g_{(4)ij} - h_{(4)ij} - 
g_{(2)ij} \Tr\, g_{(2)} - \half g_{(0)ij} 
\Tr\, g_{(2)}^2 \right. \nonu
&&\left.+{1 \over 8}( R_{ik} R^k{}_{j}
-2 R_{ikjl} R^{kl} -{1 \over 3} \nabla_i \nabla_j R + \nabla^2 R_{ij}
-{1 \over 6} \nabla^2 R g_{(0) ij}) \right. \nonu
&&\left.-{1 \over 4} g_{(2)ij} R + {1 \over 8} g_{(0)ij}
(R_{kl} R^{kl} -{1 \over 6} R^2) \right].
\eea
Using the explicit expression for $g_{(2)}$ and $h_{(4)}$ given 
in (\ref{gexp}) and (\ref{h4}) one finds that both the 
$1/\e$ pole and the logarithmic divergence cancel.
Notice that had we not subtracted the logarithmic divergence from 
the action, the resulting stress-energy tensor would
have been singular in the limit $\e \to 0$. 

Using (\ref{g4}) and (\ref{t4}) and after some algebra we obtain
\be
\<T_{ij}\>=-{1 \over 8 \p \GN} [-2 t_{ij} -3 h_{(4)}].
\eea
Taking the trace we get
\be
\<T^i_i\>={1 \over 16 \p \GN} (-2 a_{(4)}),
\eea
which is the correct conformal anomaly \cite{HS}.

Notice that since $h_{(4)ij}=-\half T^a_{ij}$ the contribution 
in the boundary stress-energy tensor proportional to $h_{(4)ij}$
is scheme-dependent. Adding a local finite counter-term proportional
to the trace anomaly will change the coefficient of this term. 
One may remove this contribution from the boundary stress
energy tensor by a choice of scheme.

Finally, one can 
obtain the energy-momentum  tensor of the boundary theory
for a given solution $G$ of the five dimensional Einstein
equations with negative cosmological constant. It is given by 
\be \label{T4}
\<T_{ij}\>={4  \over 16 \p \GN} [g_{(4)ij}
-{1 \over 8} g_{(0)ij} [(\Tr\, g_{(2)})^2-\Tr\, g_{(2)}^2] -
\half (g_{(2)}^2)_{ij} + {1 \over 4} g_{(2)ij} \Tr\, g_{(2)}],
\eea
where we have omitted the scheme-dependent $h_{(4)}$-terms. From 
the gravitational point of view  this is the quasi-local stress 
energy tensor associated with the solution $G$.

\subsection{$d=6$}

The calculation of the boundary stress tensor in the $d=6$ case
goes along the same lines as in $d=2$ and $d=4$ cases although
it is technically involved.  Up to a local traceless covariantly conserved 
term
(proportional to $h_{(6)}$) the result is 
\be \label{T6}
\<T_{ij}\>={3  \over 8 \p \GN}\, 
(g_{(6) ij}-A_{(6)ij}+{1\over 24}S_{ij})~~.
\eea
where $A_{(6)ij}$ is given in (\ref{Ad}) and $S_{ij}$ 
in (\ref{Sij}).
It is covariantly conserved and has the correct trace
\be \label{T6trace}
\<T^i_i\>={1 \over 8\p \GN}(-a_{(6)})~~,
\eea
reproducing correctly the conformal anomaly in six dimensions \cite{HS}.

Given an asymptotically AdS solution in six dimensions equation
(\ref{T6}) yields the quasi-local stress-energy tensor associated
with it.

\subsection{$d=2k+1$}

In this case one can check that the counter-terms only 
cancel infinities. Evaluating the finite part we get 
\be \label{Todd}
\<T_{ij}\>={d  \over 16 \p \GN}\, g_{(d)ij},
\eea
where $g_{(d)ij}$ can be identified with a traceless 
covariantly conserved tensor $t_{ij}$.
In odd boundary dimensions there are no gravitational conformal
anomalies, and indeed (\ref{Todd}) is traceless. 
As in all previous cases, one can also read (\ref{Todd})
as giving the quasi-local stress-energy tensor associated 
with a given solution of Einstein's equations.

\subsection{Conformally flat bulk metrics}

In this subsection we discuss a special case
where the bulk metric can be determined to all orders 
given only a boundary metric. It was shown in \cite{KS} that,
given a conformally flat boundary metric, 
equations (\ref{eqn}) can be integrated to all orders
if the bulk Weyl tensor vanishes\footnote{
In \cite{KS} it was proven that if the bulk metric satisfies 
Einstein's equations and it has a vanishing
Weyl tensor, then the corresponding  boundary
metric has to be conformally flat. The converse is not necessarily true: 
one can have Einstein metrics with non-vanishing Weyl tensor
which induce a conformally flat metric in the boundary.}.
We show that the extra condition in the bulk metric
singles out a specific vacuum of the CFT.

The solution obtained in \cite{KS} is given by
\be\label{KS}
g(x,\rho )=g_{(0)}(x)+g_{(2)}(x)\rho+g_{(4)}(x)\rho^2~~,~~g_{(4)}
={1\over 4}(g_{(2)})^2,
\eea
where $g_{(2)}$ is given in (\ref{gexp}) (we consider $d>2$), 
and all other coefficients $g_{(n)}$, $n>4$ vanish.
Since $g_{(4)}$ and $g_{(6)}$ are now known, one can 
obtain a local formula for the dual stress-energy tensor 
in terms of the curvature by using (\ref{g4}) and (\ref{g6}).

In $d=4$, using (\ref{g4}) and $g_{(4)}={1\over 4}(g_{(2)})^2$, one obtains
\be\label{CF1}
t_{ij}=t^{\sm{cf}}_{ij}\equiv 
-{1\over 4}(g_{(2)})^2_{ij}+{1\over 4}g_{(2)ij}\Tr \, g_{(2)}-
{1\over 8}g_{(0)ij}[(\Tr \, g_{(2)})^2-\Tr \, g^2_{(2)}]~~.
\eea
It is easy to check that trace of $t^{\sm{cf}}_{ij}$ reproduces (\ref{t4}).
Furthermore, by virtue of the Bianchi identities, one can show that 
$t^{\sm{cf}}_{ij}$ is covariantly conserved. It is well-known that the 
stress-energy tensor of a quantum field theory on a conformally flat 
space-time is a local function of the curvature tensor (see for example the 
book by Birrell and Davies, \cite{Birrell-Davies}). Our equation (\ref{CF1}) 
reproduces the corresponding expression given in \cite{Birrell-Davies}.

In $d=6$, using (\ref{g6}) and $g_{(6)}=0$ we find
\bea \label{CF2}
&&t_{ij}=t^{\sm{cf}}_{ij}\equiv 
[ {1\over 4}g^3_{(2)}-{1\over 4}g^2_{(2)}\Tr \, g_{(2)}
+{1\over 8}g_{(2)}(\Tr g_{(2)})^2-{1\over 8}g_{(2)}\Tr \, g_{(2)} \nonu
&&\hspace{0.5cm}
+g_{(0)}({1\over 8}\Tr \, g_{(2)}\Tr \, g^2_{(2)}
-{1\over 12}\Tr \, g^3_{(2)}-{1\over 24}
(\Tr \, g_{(2)})^3)]_{ij}~~.
\eea
One can verify that the trace of $t^{\sm{cf}}_{ij}$ reproduces (\ref{t6}) 
(taking into account that $g_{(4)}={1\over 4} g^2_{(2)}$ 
and that $t^{\sm{cf}}_{ij}$ is covariantly conserved (by virtue of the 
Bianchi identities)).

Following the analysis in the previous subsections
we obtain
\be
\<T_{ij}\>={d  \over 16 \p \GN}\, t^{\sm{cf}}_{ij}.
\eea
So, we explicitly see that the global condition we imposed on the bulk 
metric implies that we have picked a particular vacuum in the 
conformal field theory. 

Note that the tensors $t^{\sm{cf}}_{ij}$ in (\ref{CF1}), (\ref{CF2})
are local polynomial functions of the Ricci scalar and the 
Ricci tensor (but not of the Riemann tensor) of the metric $g_{(0)ij}$.
It is perhaps an expected but still a surprising result that in
conformally flat backgrounds the anomalous stress tensor 
is a local function of the curvature.

\section{Conformal transformation properties of the stress-energy 
tensor}\label{cotrpr}

In this section we discuss the conformal transformation 
properties of the stress-energy tensor. These can be 
obtained by noting \cite{ISTY} that conformal transformations
in the boundary originate from specific diffeomorphisms that 
preserve the form of the metric (\ref{coord}).
Under these diffeomorphisms $g_{ij}(x,\r)$ transforms infinitesimally 
as \cite{ISTY}
\be
\delta g_{ij}(x, \r) = 2 \s (1 - \r \pa_\r)\, g_{ij}(x, \r) +
\nabla_i a_j(x,\r) + \nabla_j a_i(x,\r),
\eea
where $a_j(x,\r)$ is obtained from the equation
\be
a^i(x,\r)=\half \int_0^\r \dd\r' g^{ij}(x, \r') \pa_j \s(x).
\eea
This can be integrated perturbatively in $\r$,
\be
a^i(x,\r)=\sum_{k=1} a_{(k)}^i \r^k.
\eea
We will need the first two terms in this expansion,
\be \label{acoe}
a_{(1)}^i= \half \pa^i \s, \qquad
a_{(2)}^i=-{1 \over 4} g_{(2)}^{ij} \pa_j \s.
\eea

We can now obtain the way the $g_{(n)}$'s transform
under conformal transformations \cite{ISTY}
\bea \label{conftra}
\delta g_{(0)ij} &=& 2 \s g_{(0)ij}, \nonu
\delta g_{(2)ij} &=& \nabla_i a_{(1) j} + \nabla_j a_{(1) i}, \nonu
\delta g_{(3)ij} &=& - \s g_{(3)ij}, \nonu
\delta g_{(4)ij} &=&-2 \s (g_{(4)} + h_{(4)}) 
+a^k_{(1)} \nabla_k g_{(2) ij}
+\nabla_i a_{(2) j}  + \nabla_j a_{(2) i}\nn
&+&g_{(2) ik} \nabla_j a_{(1)}^k 
+g_{(2) jk} \nabla_i a_{(1)}^k, \nonu
\delta g_{(5)ij} &=& - 3 \s g_{(3)ij},
\eea
where the term $h_{(4)}$ in $g_{(4)}$
is only present when $d=4$.
One can check from the explicit expressions for 
$g_{(2)}$ and $g_{(4)}$ in (\ref{gexp}) that 
they indeed transform as (\ref{conftra}).
An alternative way to derive the transformation rules
above is to start from (\ref{gexp}) and perform a
conformal variation.
In \cite{ISTY} the variations (\ref{conftra}) were
integrated leading to (\ref{gexp}) up to 
conformally invariant terms. 

Equipped with these results and the explicit form of the 
energy-momentum tensors, we can now easily calculate 
how the quantum stress-energy tensor transforms under conformal
transformations. We use the term ``quantum stress-energy 
tensor'' because it incorporates the conformal anomaly.
In the literature such transformation rules were obtained \cite{CC} by first 
integrating the conformal anomaly to an effective 
action. This effective action is a functional of the 
initial metric $g$ and of the conformal factor $\s$. It can be
shown that the difference between the stress-energy tensor
of the theory on the manifold with metric $g e^{2 \s}$ and the one
on the manifold with metric $g$ is given by  
the stress-energy tensor derived by varying the effective action 
with respect to $g$. 

In any dimension 
the stress-energy tensor transforms {\em classically} under 
conformal transformations as
\be \label{clasTr}
\delta \<T_{\m \n}\> = -(d-2)\, \s\, \<T_{\m \n}\>.
\eea
This transformation law is modified by the quantum conformal
anomaly. In odd dimensions, where there is no
conformal anomaly, the classical transformation rule 
(\ref{clasTr}) holds also at the quantum level.
Indeed, for odd $d$, and by using (\ref{Todd}) and (\ref{conftra}),
one easily verifies that the holographic 
stress-energy tensor transforms correctly.
 
In even dimensions, the transformation (\ref{clasTr}) is modified.
In $d=2$, it is well-known that one gets an extra contribution
proportional to the central charge.
Indeed, using (\ref{T2}) and the formulae above we obtain
\be
\delta \<T_{ij}\>
={l \over 8 \p \GN}\, (\nabla_i \nabla_j \s - g_{(0)ij} \nabla^2 \s) 
={c \over 12}\, (\nabla_i \nabla_j \s - g_{(0)ij} \nabla^2 \s),
\eea
which is the correct transformation rule.

In $d=4$ we obtain,
\bea 
\delta \<T_{ij}\>&=&- 2 \s \<T_{ij}\> \nn
&+&{1 \over 4 \p \GN} \left(-2 \s h_{(4)}+ {1 \over 4} \nabla^k \s [\nabla_k 
R_{ij} 
- \half (\nabla_i R_{jk} + \nabla_j R_{ik}) - {1 \over 6} \nabla_k R 
g_{(0)ij}]
\right. \nonu 
&& \left.+{1 \over 48}(\nabla_i \s \nabla_j R + \nabla_i \s \nabla_j R) 
 +{1 \over 12} R (\nabla_i \nabla_j \s - g_{(0)ij} \nabla^2 \s) 
\right. \nonu 
&& \left.+{1 \over 8} [R_{ij} \nabla^2 \s 
- (R_{ik} \nabla^k \nabla_j \s + R_{jk} \nabla^k \nabla_i \s)
+ g_{(0)ij} R_{kl} \nabla^k \nabla^l \s] \right).
\eea
The only other result known to us is the result in \cite{CC},
where they computed the finite conformal transformation of the
stress-energy tensor but for a conformally flat metric $g_{(0)}$.
For conformally flat backgrounds, $h_{(4)}$ vanishes because 
it is the metric variation of a topological invariant. 
The terms proportional to a single derivative of $\s$  
vanish by virtue of Bianchi identities and the fact that the Weyl tensor 
vanishes for conformally flat metrics. 
The remaining terms, which only contain second derivatives 
of $\s$, can be shown to coincide with the
infinitesimal version of (4.23) in \cite{CC}.
 
One can obtain the conformal transformation
of the stress-energy tensor in $d=6$ in a similar fashion
but we shall not present this result here.
 
\section{Matter} \label{matter}

In the previous sections we examined how space-time is 
reconstructed (to leading order) holographically out of CFT data. In this 
section we wish to examine how field theory describing 
matter on this space-time is encoded in the CFT. 
We will discuss scalar fields but the techniques are readily applicable
to other kinds of matter. 

The method we will use is the same as in the case of 
pure gravity, i.e. we will start by specifying the 
sources that are turned on, find how far we can go 
with only this information and then input more CFT data.
We will find the same pattern: knowledge of the sources
allows only for determination of the divergent part of the 
action. The leading  finite part (which depends on global issues
and/or the signature of space-time) is determined by the 
expectation value of the dual operator. We would like 
to stress that in the approach we follow, i.e.
regularise, subtract all infinities by adding counter-terms and finally 
remove the regulator to obtain the renormalised action,
all normalisations of the physical correlation functions
are fixed and are consistent with Ward identities. 

Other papers that discuss similar
issues include \cite{AreVol,Nojod,NiTa,marika2}.

In order to couple gravity to matter, one has to solve the coupled system of 
Einstein's equations and the matter field equations. This is non-trivial, as 
in general it is hard to solve them exactly. In particular, it is not enough 
to have a solution $G_{\m\n}$ of Einstein's equations given some matter 
fields, denoted collectively by $\F(x)$, which enter Einstein's equations 
through the stress-energy tensor $T_{\m\n}$. One also has to ensure that the 
fields $\F(x)$ remain a solution of the matter field equations for the metric 
$G_{\m\n}$ with back-reaction. A simple example where this is the case is the 
shock-wave solution discussed in chapter \ref{HEscattering}. This solution is 
an exact solution of Einstein's equations with stress-energy tensor 
$T_{vv}=-p\,\d(v)\,\d^{(d-2)}(x)$. Now a straightforward analysis of the 
geodesics in the shock-wave metric, \eq{7}, shows that the null geodesic 
$v=0$, $x^i=0$ remains a null geodesic in the shock-wave metric. The reason 
is that the shock-wave metric still has an isometry along the $u$-direction. 
So, the stress-energy tensor does not change and the shock-wave solution 
solves both the Einstein and the matter field equations exactly: there is no 
gravitational self-interaction.

In general, however, it is hard to find exact solutions and one takes a 
perturbative approach, assuming that the matter content perturbs the 
space-time only slightly. So, as long as the geometry is not too violently 
modified, one can set up a perturbative expansion where the expansion 
parameter is the Planck length divided by the typical length scale set by 
matter. So, one usually neglects the second-order back-reaction which is 
produced by the changes in the matter field equations induced by the first 
order back-reaction. This is the approach we will pursue here.

In addition, since we look for perturbative solutions of Einstein's equations 
near the boundary, also the matter system should have perturbative solutions 
near the boundary. In other words, we need a perturbative expansion of the 
stress-energy tensor in $r$. The existence of perturbative solutions of 
Einstein's equations sets constraints on the allowed behaviour of the 
stress-energy tensor near the boundary. For the scalar fields of mass $m$ 
that we will study in the next section, this implies $m^2\leq0$. With these 
constraints, it is easy to check that the leading behaviour of the 
stress-energy tensor does not change when we take into account the 
back-reaction on the metric. This is because, from the CFT point of view, 
turning on a source $\f_{(0)}$ or giving a non-vanishing expectation value to 
the operator $O(x)$ of dimension $\D$ to which the source couples only 
changes the expectation value of the stress-energy tensor, but not the metric 
$g_{(0)}$. In other words, we still have a genuine Dirichlet problem in the 
bulk and only normalisable solutions change. It is possible to find the 
general expansion of the stress-energy tensor in $r$ up to the desired order 
and including an arbitrary number of back-reaction steps, but in most cases 
the second-order back-reaction effects do not affect the bulk metric to the 
order we are interested in.

\subsection{Coupling gravity to matter}

In this section we make some preliminary remarks concerning the existence of 
solutions of Einstein's equations coupled to matter. We do this very 
generally, without assuming any specific matter model.

The local analysis in the previous sections revealed that undeterminacies in 
the bulk metric in asymptotically AdS spaces are directly related to 
information about expectation values of operators in the CFT. For future 
reference, let us write the three components of Einstein's equations as 
follows:
\be 
E_{ij}&=&0\nn
E_{ri}&=&0\nn
E_{rr}&=&0
\le{einstein}
where
\be
E_{\m\n}=R_{\m\n}-\half\,G_{\m\n}\,R-\L\,G_{\m\n}+8\p\GN T_{\m\n}.
\ee
The three components of \eq{einstein} are of course the three components of 
\eq{eqn} coupled to arbitrary matter.

Now an essential fact in our analysis of the previous sections was that the 
$(ij)$-component of Einstein's equations \eq{einstein} left undetermined the 
metric coefficient $g_{(d)}$. Its trace was determined by the third of 
\eq{einstein}, and the second of \eq{einstein} gave additional information 
about the traceless part of $g_{(d)}$. This seems to be at odds with the fact 
that Einstein's equations have some degeneracy related to co-ordinate 
invariance, and the first and third of \eq{einstein} together with the 
Bianchi identities are generally sufficient to solve the second one. We will 
see that this general expectation is only true up to ``integration 
constants". It is interesting to study this in some detail, as the 
information missing from the second of \eq{einstein} was exactly the 
information about the dual stress-energy tensor. Indeed, as we shall now see, 
one can prove that under certain constraints the first and third of 
\eq{einstein} are enough to satisfy the second of \eq{einstein} {\it up to a 
certain coefficient}. This coefficient is exactly the one that specifies the 
dual stress-energy tensor. The same is true for the third of \eq{einstein}: a 
certain integration constant has to be set to zero, and this in turn gives 
the right value for the conformal anomaly. Our only restrictions are that 
\eq{einstein} has perturbative solutions in $r$, and that we work with the 
lowest-order supergravity action without $\a'$-corrections.

In this section we work in the $r$-co-ordinate system \eq{GrFe}. It is 
convenient to first work out the Ricci tensor in \eq{einstein}:
\be
R_{ij}&=&R_{ij}(g) +{d\over r^2}\,g_{ij} -{d-1 \over2r} \,g'_{ij} 
+{1\over2}\,g''_{ij} -{1\over2}(g'g^{-1}g')_{ij}\nn
&+&{1\over4}\,g_{ij}' \Tr(g^{-1}g') -{1\over2r}\,g_{ij}\,\Tr(g^{-1}g')\nn
R_{ir}&=&{1\over2}(g^{-1})^{jk}\left(\nabla_ig'_{jk} 
-\nabla_kg'_{ij}\right)\nn
R_{rr}&=&{d\over r^2}-{1\over2r}\,\Tr(g^{-1}g')+{1\over2}\,\Tr(g^{-1}g'') 
-{1\over4}\,\Tr(g^{-1}g')^2.
\le{ricci}
We see from \eq{ricci} and \eq{einstein} that, for the existence of 
perturbative solutions, the stress-energy tensor is not allowed to diverge 
worse than $1/r$. Thus, we consider the perturbative expansion: 
$T_{\m\n}={1\over r}\,T_{(-1)\m\n} +T_{(0)\m\n}+\dots$ For $T_{ir}$ we have 
the stronger requirement $T_{(-1)ir}=0$. In later sections we will make some 
comments on stress-energy tensors that have a more violent decay near the 
boundary. The stress-energy tensor can also contain logarithmic 
contributions, but usually these appear at higher order and we will not 
consider them here.

In the co-ordinate system \eq{GrFe}, the Bianchi identities take the 
following form:
\bea
[(d-1)-{r\over2}\,\Tr(g^{-1}g')]E_{ir} -rE_{ir}'&=&r\nabla^kE_{ik}\\ 
\label{bianchi1}
[(d-2)-{r\over2}\,\Tr(g^{-1}g')]rE_{rr} -r^2E_{rr}'&=&r\Tr(g^{-1}E) 
-{r^2\over2}\,\Tr(g^{-1}g'g^{-1}E)+\nn
&&+\,r^2\nabla^kE_{rk},\label{bianchi2}
\eea
Substituting our ansatz for the metric, \eq{coord}, for the first Bianchi 
identity at lowest order we get:
\be
(d-1)E_{ir}|_{r=0}&=&E_{(-1)ij}.
\le{33c2}
Now if the first Einstein equation is satisfied at lowest order, 
$E_{(-1)ij}=0$, then so is the second, $E_{(0)ir}=0$.

Now we can use induction to see whether, if $E_{ij}=0$ to all orders, 
$E_{ir}=0$ is true to all orders as well. We take successive derivatives of 
\eq{bianchi1}, which at order $n$ gives the expression:
\be
\sum_{k=0}^{n+1}a^{n+1}_k(r)E_{ir}^{(k)}|_{r=0}&=&0,
\le{33c3}
$a$ being some coefficient with the property $a^{n+1}_{n+1}(r=0)=0$. The 
vanishing of \eq{33c3} would be enough to ensure $E_{ir}=0$ at each order. 
However, if some $a^{n+1}_n$ vanishes, the equation cannot be solved and so 
at that order we may need to introduce logarithmic terms. This happens 
exactly for $n+1=d$. So, the perturbative analysis reveals that $E_{ij}=0$ 
ensures $E_{ir}=0$ only up to order $d-1$. Let us analyse this in some more 
detail.

Assuming $E_{ij}=0$ to all orders, \eq{bianchi1} reduces to
\be
[(d-1)-{r\over2}\,\Tr(g^{-1}g')]E_{ir}-rE_{ir}'&=&0.
\le{33c4}
This we can integrate exactly, getting
\be
E_{ir}&=&c_i\,r^{d-1}e^{-H(r)},
\le{33c5}
where $H(r)={1\over2}\int\dd r\Tr(g^{-1}g')$ and therefore it has the same 
regular power expansion as $g$. We thus see that, in general, we need to 
impose the additional constraint $c_i=0$ for $E_{ir}=0$ to be true. This is 
equivalent to setting
\be
c_i&=&E_{ir}^{(d-1)}|_{r=0}=0.
\le{33c6}
The fact that \eq{33c6} is met at order $d-1$ is not accidental. This is 
exactly the same behaviour we encountered in the vacuum case. So, it is true 
that the first of Einstein's equations together with the Bianchi equation 
imply the second Einstein equation, only if \eq{33c6} is satisfied. The 
latter condition in turn implies that the $d$-th derivative of $g$ is not 
specified by the first Einstein equations and has to be imposed additionally. 
Thus, the second of Einstein's equations gives us information about the 
traceless part of $g_{(d)}$.

The same analysis can be done for the second Bianchi 
identity \eq{bianchi2}. We get
\be
[(d-2)-{r\over2}\,\Tr(g^{-1}g')]rE_{rr}-r^2E_{rr}'&=&r\Tr(g^{-1}E) 
-{r^2\over2}\,\Tr(g^{-1}g'g^{-1}E)+\nn
&+&r^2\nabla^kE_{rk}.
\le{33c7}
Now, assuming $E_{ij}=0$ and $c_i=0$ implies $E_{ir}=0$ by the previous 
argument, and this gives an equation for $E_{rr}$ with the following exact 
solution
\be
E_{rr}&=&D\,r^{d-2}e^{-H(r)}.
\le{33c8}
Thus, we also have to impose $D=0$, which gives the condition
\be
D&=&E_{rr}^{(d-2)}|_{r=0}=0.
\le{33c9}
This ensures $E_{rr}=0$ to all orders and imposes a further constraint on the 
$d$-th derivative of $g$, which now has to satisfy \eq{33c6} 
and \eq{33c9}: in the vacuum case, the latter condition determines the trace 
of $g_{(d)}$.

To summarise, we have found that $E_{ij}=0$ and the Bianchi identities are 
not enough to have a solution of Einstein's equations. One needs to set to 
zero two additional integration constants, and these determine (part of) the 
coefficient $g_{(d)}$ in the expansion of the metric. Notice, however, that 
setting these integration constants to zero only ensures the existence of a 
solution of Einstein's equations, but does not necessarily specify all the 
coefficients of the metric uniquely. In fact, as we saw in the previous 
sections, the traceless part of $g_{(d)}$ is still undetermined.

It of course remains to be shown that the first of \eq{einstein} indeed has 
solutions to all orders given an arbitrary boundary condition $g_{(0)}$. For 
maximally symmetric spaces this was done in \cite{KS}.

\subsection{Dirichlet boundary problem for scalar fields in a fixed 
background}

In this section we consider scalars on a fixed
gravitational background. This is taken to be of the 
generic form (\ref{coord}). In most of the literature
the fixed metric was taken to be that of standard AdS,
but with not much more effort one can consider
the general case. 

The action for a massive scalar is given by
\be \label{mataction}
S_{\tnnn{M}}=\half \int \dd^{d+1}x \,\sqrt{G}\,
\left( G^{\m \n} \pa_\m \F \pa_\n \F
+ m^2 \F^2 \right),
\eea
where $G_{\m \n}$ has an expansion of the form
(\ref{coord}). 

We take the scalar field $\F$ to have an expansion 
of the form
\be \label{F}
\F(x,\r)=\r^{(d-\D)/2}\, \f(x,\r), \qquad 
\f(x,\r)=\f_{(0)} + \f_{(2)} \r + ... ~,
\eea
where $\D$ is the conformal dimension of the dual operator. 
We take the dimension $\D$ to be quantised as 
$\D={d \over 2} + k, k=0,1,..$. This is often the case
for operators of protected dimension. For the case of
scalars that correspond to operators 
of dimensions  ${d \over 2}-1 \leq \D < {d \over 2}$
we refer to \cite{KleWit}.
Inserting (\ref{F}) in the field equation, 
\be
(-\Box_G + m^2) \F =0,
\eea
where $\Box_G \F = {1 \over \sqrt{G}} \pa_\m (\sqrt{G} G^{\m \n} \pa_\n \F)$, 
we obtain that the mass $m^2$ 
and the conformal dimension $\D$ are related as $m^2=(\D-d)\D$, as explained 
in the introduction, see \eq{mass}. $\f$ satisfies
\be \label{phieq}
[-(d-\D) \pa_\r \log g\, \f + 2 (2 \D -d -2) \pa_\r \f 
-\Box_g \f] + \r [-2 \pa_\r \log g\, \pa_\r \f - 4 \pa^2_\r \f]=0.
\eea
Given $\f_{(0)}$ one can determine recursively $\phi_{(n)}, n>0$.
This is achieved by differentiating (\ref{phieq}) and setting 
$\r$ equal to zero.
We give the result for the first couple of orders in appendix \ref{as-sc}.
This process breaks down 
at order $\D-d/2$ (provided this is an integer, which we assume throughout
this section)
because the coefficient of $\phi_{(2 \D -d)}$ (the 
field to be determined) becomes zero. This is exactly 
analogous to the situation encountered for even $d$
in the gravitational sector. Exactly the same way as there, we introduce 
at this order a logarithmic term, i.e. the expansion of $\F$  now reads,
\be \label{fexp}
\F = \r^{(d-\D)/2}\, (\f_{(0)} + \r \f_{(2)} + ...) 
+ \r^{\D/2}\, (\phi_{(2 \D -d)} + \log \r\, \psi_{(2 \D -d)} + ...).
\eea
The equation (\ref{phieq}) now determines all terms up to 
$\phi_{(2 \D -d -2)}$, the coefficient of the logarithmic term 
$\psi_{(2 \D -d)}$, but leaves undetermined
$\phi_{(2 \D -d)}$. This is analogous to the situation 
discussed in section \ref{Dirichletgrav}, where the term $g_{(d)}$ was 
undetermined. 
It is well known \cite{BKL,BKLT,KleWit} that precisely at order 
$\r^{\D/2}$ one finds the expectation value of the dual
operator. We will review this argument below, and also 
derive the exact proportionality coefficient. Our 
result is in agreement with \cite{KleWit}.

We proceed to regularise and then renormalise the theory.
We regulate by integrating in the bulk from $\r \geq \e$,{}\footnote{
This regularisation for scalar fields in a fixed AdS background
was considered in \cite{Mvis,FMMR}. In these papers the 
divergences were computed in momentum space, but no counter-terms
were added to cancel them. Addition 
of boundary counter-terms to cancel infinities for scalar
fields was considered in \cite{Gordon}, and more recently 
in \cite{KleWit}.}
\bea \label{matreg}
S_{\tnnn{M}\sm{,reg}}&=&\half \int_{\r \geq \e} \dd^{d+1} x\, \sqrt{G}
\left( G^{\m \n} \pa_\m \F \pa_\n \F
+ m^2 \F^2 \right) \nonu
&=&-\int_{\r=\e} \dd^d x\, 
\sqrt{g(x,\e)} \e^{-\D+d/2}\, [\half\, (d-\D) \f^2(x,\e)
+ \e\, \f(x,\e) \pa_\e \f(x,\e)] \\
&=& \int \dd^d x\, \sqrt{g_{(0)}}\,
[\e^{-\D+d/2} a^{\tnnn{M}}_{(0)} + \e^{-\D+d/2+1} a^{\tnnn{M}}_{(2)}
+ ... + \e\,  a^{\tnnn{M}}_{(2 \D -d +2)}\nn
&-& \log \e\, a_{(2 \D -d)}] 
+ {\cal O}(\e^0). \nonumber
\eea
Clearly, with $\D-d/2$ a positive integer there is a finite number 
of divergent terms. The logarithmic divergence appears exactly  
when $\D=d/2+k, k=0,1,..$, 
in agreement with the analysis in \cite{PeSk}, 
and is directly related to the logarithmic term in (\ref{fexp}).
The first few of the power law divergences read
\be
a^{\tnnn{M}}_{(0)}=-\half (d-\D) \f_{(0)}^2, \qquad
a^{\tnnn{M}}_{(2)}=-{1 \over 4} \Tr\, g_{(2)}\, \f_{(0)}^2 + (d-\D +1)\, 
\f_{(0)} \f_{(2)}.
\eea
Given a field of specific dimension it is straightforward to 
compute all divergent terms.

We now proceed to obtain the renormalised action
by adding counter-terms to cancel the infinities,
\be
S_{\tnnn{M}\sm{,ren}}&=&\lim_{\e \to 0} [S_{\tnnn{M}\sm{,reg}} - \int \dd^d 
x\, \sqrt{g_{(0)}}\,
[\e^{-\D+d/2} a^{\tnnn{M}}_{(0)} + \e^{-\D+d/2+1} a^{\tnnn{M}}_{(2)}
+ ... + \e\, a^{\tnnn{M}}_{(2 \D -d +2)}\nn
&-& \log \e\,  a_{(2 \D -d)}].
\eea
Exactly as in the case of pure gravity, and since the 
regulated theory lives at $\r=\e$, one needs to rewrite the
counter-terms in terms of the field living at $\r=\e$, i.e. 
the induced metric $\g_{ij}(x, \e)$ and the field $\F(x,\e)$,
or equivalently $g_{ij}(x,\e)$ and $\f(x,\e)$.
This is straightforward but somewhat tedious:
one needs to invert the relation between $\f$ and $\f_{(0)}$ 
and between $g_{ij}$ and $g_{(0) ij}$ to sufficiently high order.
This then allows to express all $\f_{(n)}$, and therefore
all $a_{(n)}^{\tnnn{M}}$, in terms of $\f(x,\e)$ and $g_{ij}(x,\e)$ 
(the $\f_{(n)}$'s are determined in terms  of $\f_{(0)}$
and $g_{(0)}$ by solving (\ref{phieq}) iteratively).
Explicitly, the first two orders read
\bea \label{finiteact}
S_{\tnnn{M}\sm{,ren}}&=&\lim_{\e \to 0} \left[\half \int_{\r \geq \e} 
\dd^{d+1} x \sqrt{G}
\left( G^{\m \n} \pa_\m \F \pa_\n \F + m^2 \F^2 \right) \right. \\
&&\left.+ \int_{\r=\e}\, \sqrt{\g}\, [{(d-\D) \over 2} \F^2(x, \e) +
{1 \over 2(2 \D -d -2)}\, (\F(x,\e) \Box_\g \F(x,\e)\right.\nn
&+&\left. {d-\D \over 2 (d-1)} R[\g] \F^2(x,\e)) + ...] \right]. \nonumber 
\eea
The addition of the first counter-term was discussed in \cite{KleWit}.
The action (\ref{finiteact}) with only 
the counter-terms written explicitly is finite 
for fields of $\D < d/2 +2$. As remarked above, it is
straightforward to  obtain all counter-terms needed in order 
to make the action finite for any field of any mass.  
These counter-terms contain also logarithmic subtractions
that lead to the conformal anomalies discussed in \cite{PeSk}.
For instance, if $\D=\half d +1$, the coefficient 
$[2 (2 \D -d -2)]^{-1}$ in (\ref{finiteact}) is replaced 
by $-{1 \over 4} \log \e$.
An alternative way to derive the counter-terms is to demand that 
the expectation value $\<O\>$ is finite. This holds in the case of pure 
gravity too, i.e. the counter-terms can also be derived by requiring
finiteness of $\<T_{\m \n}\>$ \cite{BK}.

The expectation value of the dual operator is given by
\be \label{oexp}
\< O (x) \> = - {1 \over \sqrt{\det g_{(0)}}}
{\delta S_{\tnnn{M}\sm{,ren}} \over \delta \f_{(0)}} =
- \lim_{\e \to 0} {1 \over \sqrt{\det g(x,\e)}}
{\delta S_{\tnnn{M}\sm{,ren}} \over \delta \f(x,\e)}. 
\eea
Exactly as in the case of pure gravity, the expectation 
value receives a contribution both from the regulated part 
and from the counter-terms. We obtain,
\be \label{oexp1} 
\< O (x) \> = (2 \D - d)\, \f_{(2 \D -d)} 
+ F(\f_{(n)}, \psi_{(2 \D -d)}, g_{(m)}), \qquad n<2 \D-d,
\eea
where we used that $\f_{(2 \D -d)}$ is linear in $\f_{(0)}$
(notice that the action (\ref{mataction}) does not include 
interactions). $F(\f_{(n)},\psi_{(2 \D -d)}, g_{(m)})$ 
is a local function of $\f_{(n)}$ with
$n<2\D{-}d$, $\psi_{(2 \D -d)}$ and $g_{(m)}$. 
These terms are related to 
contact terms in correlation functions of $O$ with itself and
with the stress-energy tensor. Its exact form is straightforward 
but somewhat tedious to obtain (just use (\ref{finiteact}) and
(\ref{oexp})). 

As we have promised, we have shown that the 
coefficient $\f_{(2 \D -d)}$ is related with the 
expectation value of the dual CFT operator. 
In the case that the background geometry is the standard 
Euclidean AdS one can readily 
obtain $\f_{(2 \D -d)}$ from the unique 
solution of the scalar field equation 
with given Dirichlet boundary conditions.
One finds that  $\f_{(2 \D -d)}$ is proportional to 
(an integral involving) $\f_{(0)}$. Therefore, $\f_{(2 \D -d)}$
carries information about the 2-point 
function. The factor $(\D - d/2)$ is crucial
in order for the 2-point function to be 
normalised correctly \cite{FMMR}. We refer to \cite{KleWit}
for a detailed discussion of this point.

We finish this section by calculating the conformal 
anomaly associated with the scalar fields and in the 
case the background is (locally) standard AdS (i.e. $g_{(n)}=0$, 
for $0<n<d$). Equation (\ref{phieq}) simplifies and can be 
easily solved. One gets 
\bea
\f_{(2n)}&=&{1 \over 2n(2 \D -d -2n)}\, \Box_0 \f_{(2n-2)}, \nonu
\psi_{(2\D-d)}&=&-{1 \over 2 (2 \D- d)}\, \Box_0 \f_{(2\D-d-2)}=
-{1 \over 2^{2k} \G(k) \G(k+1)}\, (\Box_0)^k \f_{(0)} \label{psian},
\eea
where $k=\D-{d \over 2}$ and $\Box_0$ is the Laplacian 
of $g_{(0)}$. The regularised action written in terms
of the fields at $\r=\e$ contains the following explicit 
logarithmic divergence:
\be
S_{\tnnn{M}\sm{,reg}}=-\int_{\r=\e} \dd^dx \,\sqrt{\g}\, 
[\log \e\, 
(\D - {d \over 2})\, 
\f(x,\e)\, \psi_{(2\D-d)}(x,\e)+\cdots]\, ,
\eea
where the dots indicate power law divergent and finite terms,
$\psi_{(2\D-d)}(x,\e)$ is given by (\ref{psian}) with 
$g_{(0)}$ replaced by $\g$ and $\f_{(0)}$ by $\f(x,\e)$.
Using the same argument as in \cite{HS} we obtain 
the matter conformal anomaly,
\be
\ca_{\tnnn{M}}=\half \left({1 \over 2^{2k-2} (\G(k))^2} \right) \f_{(0)} 
(\Box_0)^k \f_{(0)}.
\eea
This agrees exactly with the anomaly calculated in \cite{PeSk}
(compare with formulae (10), (37) in \cite{PeSk}).

\subsection{Scalars coupled to gravity} \label{back}

In the previous section we ignored the back-reaction 
of the scalars to the bulk geometry. The purpose of
this section is to discuss this issue. The action 
is now the sum of (\ref{action}) and (\ref{mataction}),
\be \label{totaction}
S=S_{\sm{gr}}+S_{\tnnn{M}}.
\eea
The gravitational field equation in the presence of matter reads
\be
R_{\m \n} - \half (R + 2 \L) G_{\m \n}= - 8 \p \GN T_{\m \n}.
\eea
In the co-ordinate system (\ref{coord}) and with the 
scalar field having the expansion in (\ref{fexp}),
these equations read
\bea \label{eqnmatter}
\rho \,[2 g^{\prime\prime}_{ij} - 2 (g^\prime g^{-1} g^\prime)_{ij} + \Tr\,
(g^{-1} g^\prime)\, g^\prime_{ij} \,] &+& R_{ij} (g) - (d - 2)\,
g^\prime_{ij} - \Tr \,(g^{-1} g^\prime)\, g_{ij} =
\\  \hspace{3cm}
&=&- 8 \p \GN\, \r^{d-\D-1}\left[{(\D-d) \D \over d-1}\, \f^2\, g_{ij} 
+ \r\, \pa_i \f \pa_j \f\right], \nonu  
\nabla_i \Tr \,(g^{-1} g^\prime) - \nabla^j g_{ij}^\prime &=&
-16 \p \GN\, \r^{d-\D-1} \left[{d-\D \over 2}\, \f \pa_i \f 
+ \r\, \pa_\r \f \pa_i \f\right], \nonu
\Tr \,(g^{-1} g^{\prime\prime}) - \frac{1}{2} \Tr\, (g^{-1} g^\prime
g^{-1} g^\prime) &=& - 16 \p \GN\, \r^{d-\D-2}
\left[{d(\D-d)(\D-d+1) \over 4 (d-1)}\, \f^2 \right.\nonu
&+&\left. (d-\D)\, \r\, \f \pa_\r \f 
+ \r^2\, (\pa_\r \f)^2{\over}\right], \nonumber 
\eea

If $\D>d$, the right-hand side diverges near the boundary
whereas the left-hand side is finite. Operators with dimension
$\D>d$ are irrelevant operators. Correlation functions of these operators
have a very complicated singularity structure at coincident points.
As remarked in \cite{Wit}, one can avoid such problems by considering the 
sources to be infinitesimal and to have disjoint support, so that these 
operators are never at coincident points.  
Requiring that the equations in (\ref{eqnmatter}) are satisfied to leading 
order in $\r$ yields 
\be
\f_{(0)}^2=0, 
\eea
which is indeed the prescription advocated in \cite{Wit}. 
 
If $\D \leq d$, which means that we deal with marginal or relevant 
operators, one can perturbatively calculate the back-reaction of the 
scalars to the bulk metric. At which order the leading back-reaction
appears depends on the mass of the field. For fields that 
correspond to operators of dimension $\D=d-k$ the leading 
back-reaction appears at order $\r^k$, except when $k=0$
(marginal operators), where the leading 
back-reaction is at order $\r$. 

Let us see how conformal anomalies arise in this
context. The logarithmic divergences are coming from 
the regulated on-shell value of the bulk integral in 
(\ref{totaction}). The latter reads
\bea \label{totreg}
S_{\sm{reg}}(\mbox{bulk}) &=& \int_{\r \geq \e} \dd \r\, \dd^d x\, \sqrt{G}\, 
[{d \over 8 \p \GN} - {m^2 \over d-1}\, \Phi^2] \nonu
&=&\int_{\r \geq \e} \dd \r\, \dd^d x\, {1\over \r}\, \sqrt{g(x,\r)}\,
[{d\,\r^{-d/2}  \over 16 \p \GN}\, - {m^2\,\r^{-k} \over 2(d-1)}\, 
\f^2(x,\r)\,],
\eea
where $k=\D-d/2$. We see that gravitational conformal anomalies
are expected when $d$ is even and matter conformal anomalies
when $k$ is a positive integer, as it should.

In the presence of sources the expectation value 
of the boundary stress-energy tensor is not conserved but 
rather it satisfies a Ward identity that relates its covariant divergence
to the expectation value of the operators that couple to the 
sources. To see this consider the generating functional
\be
Z_{\tnnn{CFT}}[g_{(0)}, \phi_{(0)}]=\big< 
\exp \int \dd^d x\, \sqrt{g_{(0)}}\,[\half\, g_{(0)}^{ij} T_{ij} - \f_{(0)} 
O] \big>.
\eea
Invariance under infinitesimal diffeomorphisms,
\be 
\delta g_{(0)ij} = \nabla_i \xi_j + \nabla_j \xi_i, 
\eea
yields the Ward identity,
\be \label{WI}
\nabla^j \< T_{ij} \> = \< O \>\, \pa_i \f_{(0)}.
\eea 
As we have remarked before, $\<T_{ij}\>$ has a dual meaning\,\cite{BK},
both as the expectation value of the dual stress-energy tensor 
and as the quasi-local stress-energy tensor of Brown and York.
The Ward identity (\ref{WI}) has a natural explanation from
the latter point in view as well. According to \cite{BrownYork}
the quasi-local stress-energy tensor is not conserved in the 
presence of matter but it satisfies
\be \label{BY}
\nabla^j \<T_{ij}\> = - \t_{i\r},
\eea
where $\t_{i\r}$  expresses the flow of matter energy-momentum  through 
the boundary. Evidently, (\ref{WI}) is of the form (\ref{BY}).

Solving the coupled system of equations (\ref{eqnmatter}) and (\ref{phieq})
is straightforward but somewhat tedious. 
The details differ from case to case. 
For illustrative purposes we present a sample calculation:
we consider the case of two-dimensional massless scalar field
($d=\D=2, k=1$). 

The equations to be solved are (\ref{phieq}) and (\ref{eqnmatter})
with $d=\D=2$ and the
expansion of the metric and the scalar field are given 
by (\ref{coord}) and (\ref{fexp}) (again with $d=\D=2$), respectively.
Equation (\ref{phieq}) determines $\psi_{(2)}$,
\be \label{psi}
\psi_{(2)}=-{1 \over 4} \Box_0 \f_{(0)}.
\eea
Equations (\ref{eqnmatter}) determine $h_{(2)}$, the trace of the 
$g_{(2)}$ and provide a relation  
between the divergence of $g_{(2)}$ and $\f_{(2)}$,
\bea \label{d2ex}
&&h_{(2)}=- 4 \p \GN \left(\pa_i \f_{(0)}\pa_j \f_{(0)}
-\half\, g_{(0)ij}\, (\pa \f_{(0)})^2 \right), \nonu
&&\Tr\, g_{(2)} = \half\, R + 4 \p \GN\, (\pa \f_{(0)})^2, \nonu
&&\nabla^i g_{(2)ij}=\pa_i \Tr\, g_{(2)} + 16 \p \GN\, \f_{(2)} \pa_i 
\f_{(0)}.
\eea
Notice that $g_{(2)}$ and $\f_{(2)}$ are still undetermined
and are related to the expectation values of the dual operators
(\ref{tij1}) and (\ref{oexp1}), respectively.
Notice that $h_{(2)}$ is equal to the stress-energy tensor of a 
massless two-dimensional scalar. 

Going back to (\ref{totreg}), we see that the second term 
drops out (since $m^2=0$) and one can use the result already 
obtained in the gravitational sector,
\be \label{ga}
\ca&=&{1 \over 16 \p \GN} (-2 a_{(2)})= {1 \over 16 \p \GN} (- 2 \Tr\, 
g_{(2)})\nn
&=& - {1 \over 16 \p \GN} R + \half \f_{(0)} \Box_0 \f_{(0)} 
- {1\over 2}\,\nabla_i(\f_{(0)}\nabla^i \f_{(0)}),
\eea
which is the correct conformal anomaly \cite{HS,PeSk} (the last term can be 
removed by adding a covariant counter-term).

The renormalised boundary stress tensor reads
\begin{equation}
\<T_{ij}(x)\>={1\over 8\pi \GN}
\left(g_{(2) ij}+h_{(2) ij}-g_{(0) ij} \Tr\, g_{(2)} \right)(x).
\label{bT}
\end{equation}
Its trace gives correctly the conformal anomaly (\ref{ga}). 
On the other hand, taking the covariant derivative of (\ref{bT}) we get
\begin{eqnarray}
\nabla^j \<T_{ij}\>=\<O(x)\>\, \partial_i \phi_0(x)~~,\nonumber \\
\<O(x)\>=2(\phi_2(x)+\psi_2(x)).
\label{Ward1}
\end{eqnarray}
in agreement with equations (\ref{WI}) and (\ref{oexp1}).

\subsection{Pointlike particles}

The method developed in the previous subsections is quite generic and can be 
applied to other matter fields. Although we have not worked out all the 
details, in this section we give a further example for illustrative purposes: 
we consider pointlike particles. This is in our opinion a very important 
example for our understanding of holography in the AdS/CFT correspondence, 
and we hope to report the full details elsewhere. Indeed, one can do 
interesting gedanken experiments with point particles and black holes in AdS 
\cite{PST,SuTo,LoTh} to test the causality and locality properties of the 
boundary theory.

So we couple the Einstein-Hilbert action to the action for a pointlike 
particle. One then needs to solve Einstein's equations coupled to the 
geodesic equation and the constraint
\be\label{massive}
G_{\m\n}(z)\dot{z}^\m\dot{z}^\n=-\varepsilon
\ee
($\varepsilon=1$ for massive particles and $\ve=0$ for massless particles). 
In the massive case, we get the following stress-energy tensor:
\be\label{stress}
T^{\m\n}(x)={m\over\@{|G(x)|}}\int\dd t\,\delta^{(d+1)}(x-z(t))\, 
\dot{z}^\m\dot{z}^\n.
\ee
In the massless case, the stress-energy tensor is given by \eq{10}. We have 
also analysed the tachyonic case, but we will not present the results here.

We are interested in computing the back-reaction effects of the particle on 
the 
metric near the boundary. This will allow us to compute the expectation value 
of 
the stress-energy tensor of the dual theory \cite{KSS1}, which will depend in 
a crucial manner on the boundary conditions on the position and the speed of 
the particle. Therefore we are interested in the asymptotic behaviour of the 
stress-energy tensor as $r\rightarrow0$. This is given by the part of the 
trajectory satisfying $r(t)\rightarrow0$. Hence, the problem of finding the 
asymptotics of the stress-energy tensor translates itself into finding the 
region of the trajectory $\gamma$ near the boundary. For the massless 
particle 
and the tachyon it is known that they can travel from the boundary to the 
bulk 
and viceversa, so we expect that there are values of $t$ corresponding to 
$r=0$.
However, the particle with positive mass squared never reaches the boundary, 
and 
so we expect it not to contribute to the stress-energy tensor at $r=0$. As we 
will see, this turns out to be true also for Einstein spaces with arbitrary 
boundary metric.

The strategy will be the following. To identify the region of $t$ for which 
$r(t)\rightarrow0$, we solve the geodesic equation perturbatively in $r$ and 
find the solutions $r(t)$ and $z^i(t)$ perturbatively in some function of 
$t$. If there are such solutions, the perturbative expansion makes sense; if 
there are not, the geodesic equation cannot be solved perturbatively near the 
boundary.\\
\\
{\bf The massless particle}\\
\\
To lowest order, the geodesic equations for massless particles are solved by
\bea
r(t)&=&{1\over c(t-d)}\nonu
z^i(t)&=&z^i_0+r(t)v^i,
\eea
where $v^i$ is now a timelike vector, $g_{ij}v^iv^j=-1$, defined in general 
by $v^i(r)\equiv {\ddd z^i\over\ddd r}$. In this case, the 
stress-energy tensor can be cast in the form
\be
T_{\m\n}={pc\ell^2\over\@{g}}\left({r\over\ell}\right)^{d+3}\, v_\m v_\n\, 
\delta^{(d)}(x-z(r)),
\le{AdSstress}
where $v^\m(r)$ is defined by $v^\m(r)=(1,v^i(r))$. It is null in the 
space-time metric and satisfies
\bea
\partial_\m v^\m&=&0\nonu
v^\m\partial_\m\delta^{(d)}(x-z(r))&=&0.
\eea
All components of the above stress-energy tensor are proportional to 
$r^{d-1}$ 
in leading order in $r$ as $r\rightarrow0$. Therefore, it will contribute to 
$g_{(d)}$ but not to $h_{(d)}$, just as in the tachyonic case. It is now also 
straightforward to compute the dual stress-energy tensor. This will have an 
interesting behaviour \cite{HI}: to start with, unless one chooses very 
special boundary conditions, the effective Hamiltonian will be time-dependent 
due to the covariance of our formulae in the boundary co-ordinates. Notice 
that to get agreement with the results in \cite{HI}, where the stress-energy 
is centred on the light-cone, one may need to first perform a co-ordinate 
transformation. As mentioned in the previous sections, such a co-ordinate 
transformation changes the value of the stress-energy tensor if it induces a 
boundary Weyl rescaling.\\
\\
{\bf The massive particle}\\
\\
The bulk trajectory of a particle with positive mass squared is given by
\be \label{trajectory}
r(t)={r_0\over|\cos(t/\ell+c)|},
\ee
and, like in the tachyonic case, $r_0$ and $c$ are to be determined by the 
boundary conditions only. In this case, however, we see that $r(t)$ can never 
be 
zero unless $r_0=0$, in which case the particle stays forever at the boundary 
and never reaches the bulk. Therefore, a perturbative solution of the 
geodesic 
equation in powers of $r$ does not make sense in this case, as the world-line 
of 
the particle actually never reaches the boundary. Therefore, one can only 
hope 
to solve the geodesic equation for simple exact solutions of the vacuum 
Einstein equations. For example, it is an elementary exercise to solve for 
the case of a flat boundary, the trajectory being given by 
(\ref{trajectory}). In that case, one finds an expression for the 
stress-energy tensor analogous to that for the tachyon, but now involving a 
step function $\theta(r-r_0)$, hence with support only on the region $r>r_0$.

In this case, the particle contributes only a finite piece to the action. 

\section{Conclusions}\label{sec4.6}

Most of the discussions in the literature
on the AdS/CFT correspondence are concerned with obtaining
conformal field theory correlation functions using 
supergravity. Here we started investigating the 
converse question: how can one obtain information 
about the bulk theory from CFT correlation functions?
How does one decode the hologram? 

Answering these questions in all generality, but within the 
context of the AdS/CFT duality, entails developing 
a precise dictionary between bulk and boundary 
physics. A prescription for relating bulk/boundary 
observables is already available \cite{Gubs,Wit},
and one would expect that it would allow us to
reconstruct the bulk space-time from the boundary CFT.
The prescription of \cite{Gubs,Wit}, however, relates infinite quantities.
One of the main results presented here is the systematic
development of a renormalised version of this prescription.
Equipped with it, and with no other 
assumption (except that the CFT has an AdS dual),
we then proceeded to reconstruct the bulk 
space-time metric and bulk scalar fields to the
first non-trivial order.

Our approach to the problem is to start from the boundary 
and try to build iteratively bulk solutions. Within 
this approach, the pattern we find is the following: \\
\newline
$\bullet$ Sources in the CFT determine an asymptotic expansion
of the corresponding bulk field near the boundary to high enough order 
so  that {\em all infrared divergences} of the bulk on-shell
action can be computed. This then allows to obtain a 
renormalised on-shell action by adding boundary counter-terms 
to cancel the infrared divergences. 
\\
\newline
$\bullet$ Bulk solutions can be extended one order 
further by using the 1-point function of the corresponding dual CFT 
operator.
\\

In the case the bulk field is the metric, our results show
that a conformal structure at infinity is not in general
sufficient in order to obtain a bulk metric. The first 
additional information one needs is the expectation 
value of the boundary stress-energy tensor.

As a by-product, we have obtained ready-to-use formulae
for the Brown-York quasi-local stress-energy tensor
for arbitrary solution of Einstein's equations with 
negative cosmological constant up to six dimensions. The six-dimensional 
result is particularly interesting because,
via AdS/CFT, it provides new information about the 
still mysterious $(2,0)$ theory. Furthermore, we 
have obtained the conformal transformation properties
of the  stress-energy tensors. These transformation 
rules incorporate the trace anomaly and provide 
a generalisation to $d>2$ of the well-known Schwartzian 
derivative contribution in the conformal 
transformation rule of the stress-energy tensor in $d=2$.

Our discussion extends straightforwardly to the case of different matter. We 
expect that in all cases obstructions in extending the solution to the deep 
interior region will be resolved by additional CFT data. An interesting case 
to study in this framework is point particles. Reconstructing the  trajectory 
of the bulk point particle out of CFT data will present a model of how 
holography works with time dependent processes. Furthermore, following 
\cite{HI}, one could study the interplay between causality and holography. 
Another extension is to study renormalisation group flows using the present 
formalism. This amounts
to extending the discussion in section \ref{Dirichletgrav} by adding a 
potential for the scalars. Another application of our results is in the 
context 
of Randall-Sundrum (RS) scenarios \cite{RS}. 
Incorporating such a scenario in string theory, 
in the case the bulk space is AdS,
may yield a connection with the AdS/CFT duality \cite{herman,Wrs}. 
As advocated in \cite{Wrs}, 
one may view the RS scenario as $4d$ gravity 
coupled to a cut-off CFT. The regulated theory in 
our discussion provides a dual description of a
cut-off CFT. In this context, the 
counter-terms are re-interpreted as providing 
the action for the bulk modes localised 
on the brane \cite{KS,Gub,GKR}. We see, for instance,
that the counter-terms in (\ref{finiteact}) can be 
re-interpreted as an action for a bulk scalar mode localised on the 
brane (see, e.g., \cite{DK}). This is the subject of study in the next 
chapter.
 
\chapter{Warped Compactifications and the Holographic Stress Tensor} 
\label{warped}

The contents of this chapter are based on \cite{KSS2}. We study gravitational 
aspects of brane-world scenarios. We show that the bulk Einstein equations 
together with the junction  condition imply that the induced metric on the 
brane satisfies the full non-linear Einstein equations with a specific 
effective stress-energy tensor. This result holds 
for any value of the bulk cosmological constant.
The analysis is done by either placing the brane close to infinity
or by considering the local geometry near the brane. 
In the case that the bulk space-time is asymptotically AdS, we show that the 
effective stress-energy tensor is equal to the sum
of the stress-energy tensor of matter localised on the 
brane and of the holographic stress-energy tensor appearing in 
the AdS/CFT duality. In addition, there are specific
higher-curvature corrections to Einstein's equations. 
We analyse in detail the case of asymptotically flat space-time. 
We obtain asymptotic solutions of Einstein's
equations and show that the effective Newton's constant on the brane
depends on the position of the brane. 
 
\section{Warped Compactifications and AdS/CFT holography}\label{sec5.1}

The previous chapters dealt mainly with holography from the point of view of 
a bulk observer. We used the existence of a holographic dual to find how 
information about the boundary is encoded in the bulk. Now we change 
perspective and ask ourselves where the boundary observer finds the 
information about the bulk geometry and fields. We do this in the context of 
warped compactifications, where the boundary observer lives on a 
brane\footnote{The sense in which ``holography" is used here differs from the 
original sense. Here the boundary theory is a gravitational theory, and there 
is not a duality between bulk and boundary, but rather an embedding of the 
boundary in the bulk.}. We find that the information about the bulk, and in 
particular global information that is not captured by the local analysis, is 
encoded in the stress tensor on the brane.

In the AdS/CFT correspondence, the supergravity partition function is 
related to the generating functional of conformal field
theory (CFT) correlation functions as
\be\label{partf}
Z[\F]=\int_\f D\F\,\exp(iS[\F])=W_{\tnnn{CFT}}[\f],
\ee
where $\F$ denotes collectively all fields and $\f$ is a field parametrising
the boundary condition of $\F$ at infinity. In the conformal field theory
the boundary fields $\f$ are interpreted as sources for CFT operators.
In particular, the metric at infinity, $g_{(0)}$, is considered as the 
source for the stress-energy tensor of the dual CFT. The relation 
(\ref{partf})
suffers from divergences and has to be regularised and renormalised.

On the CFT side, there are UV divergences when operators come to 
coincident points. These correspond to IR divergences on the gravitational
side. To regulate the gravitational theory one may cut-off the 
asymptotically AdS space-time at some radius $\r=\e$ near the boundary.
One can then compute all infrared divergences.
The renormalised theory is obtained by adding counter-terms to 
cancel the infinities and then removing the cut-off.

One may, however,  wish to consider situations where the infrared cut-off 
is kept finite instead of being sent to zero. This is the case in warped 
compactifications, where the AdS space-time is cut-off by the presence of 
a brane. In this case, (\ref{partf}) does not have any infrared divergences 
and so one does not need to add counter-terms. 

In the cut-off space-time, the induced metric at the boundary $\g$
corresponds to a normalisable mode and so one should integrate over it:
\be\label{1}
\int D\g_\e\int_{\g_\e}DG\exp(iS[G])=\int D\g_{\e}\,
W_{\tnnn{CFT}}[\g, \e],
\ee
Under these circumstances, gravity becomes 
dynamical on the brane, and the brane theory is a CFT coupled to 
dynamical gravity.

Consider a space-time $M$ with a boundary $\pa M$. 
The action in (\ref{1}) is given by\footnote{
Our curvature conventions are as follows:
$R_{ijk}{}^l=\pa_i \G_{jk}{}^l + \G_{ip}{}^l \G_{jk}{}^p - (i
\leftrightarrow j)$ and $R_{ij}=R_{ikj}{}^k$. With these conventions
the curvature of AdS comes out positive, but we will 
still use the terminology ``space of constant negative
curvature''. Notice also that we take
$\int \dd^{d+1} x = \int \dd^d x \int_0^\infty \dd \r$
and the boundary is at $\r=0$.
The minus sign in front of the trace of the second 
fundamental form is correlated with the choice of having $\r=0$ in 
the lower end of the radial integration.}
\bea
S[\F, G]&=&{1 \over 16 \p G_{d+1}}[\int_{M}\dd^{d+1}x\, 
\sqrt{G}\, (R[G] + 2 \L) 
- \int_{\pa M} \dd^d x\, \sqrt{\g}\, 2 K ] \nonu
&&+\int_{M}\dd^{d+1}x\,\sqrt{G}\,  {\cal L}^{\sm{bulk}}  
+\int_{\pa M} \dd^d x\, \sqrt{\g}\, {\cal L}^{\sm{bdry}}
\eea
where ${\cal L}^{\sm{bulk}}$ denotes the Lagrangian for bulk matter
and ${\cal L}^{\sm{bdry}}$ the Lagrangian for matter
living on the boundary. Einstein's equations read\footnote{
The different signs in the right hand side of these two equations
is related to our conventions discussed in the previous 
footnote.}:
\bea\label{eom}
R_{\m\n}[G]-{1\over2}\,(R[G]+2\L)\,G_{\m\n} &=&
-8\p G_{d+1}\,T^{\sm{bulk}}_{\m\n}[G] \\
K_{ij}[\g]-\g_{ij}\,K[\g]&=&8\p G_{d+1}\,T_{ij}^{\sm{bdry}}[\g]. 
\label{junction}
\eea
These equations describe the case the bulk space-time ends on the 
brane. This is in fact half of the Randall-Sundrum (RS) space-time \cite{RS}.
In the RS scenario one glues on the other side of the brane an 
identical space-time. Then the substitution
$$
K_{ij} \to \lim_{\delta \to 0}[K_{ij}(\r=\e+\delta) 
- K_{ij}(\r=\e-\delta)],
$$
in (\ref{junction})
yields the junction condition (see, for example, \cite{jct} for a
derivation). $\r=\e$ is the position of the brane.
In the RS context,
$K_{ij}(\r=\e+\delta)= - K_{ij}(\r=\e-\delta)$  due to the $Z_2$-symmetry, 
so the net effect is to get back (\ref{junction}) but with an 
extra factor of two. In the remainder we will work with 
equations (\ref{eom}) and (\ref{junction}) and we will refer to 
(\ref{junction}) as the junction condition.

The usual way \cite{RS} of establishing localisation of gravity on the brane 
is to study small fluctuations around a  given configuration (such as a 
flat brane in AdS space) which solves equations (\ref{eom}). 
The equations for small gravitational fluctuations around the 
solution take the form of a quantum mechanical problem. In terms 
of the effective quantum mechanical problem 
the existence of a localised graviton translates into the existence
of a normalisable zero-mode solution
(this solution is the wave function associated to the 
graviton localised on the brane). In addition to the zero mode
there are additional massive modes. One still has to show 
that these modes do not drastically change the physics, 
i.e. that they yield sub-leading corrections relative to the zero mode.
Note that the question of normalisability of the zero mode
depends on global properties of the gravitational solution.
If the bulk space is asymptotically flat there is still 
a zero-mode but it is not normalisable. 
There may still be a quasi-localisation due to a collection 
of low-energy Kaluza-Klein modes \cite{GRS,CEH,DGP}. 

The analysis just described is at the linearised level. 
It is technically involved in this approach to go beyond the linear 
approximation and demonstrate the full non-linear structure of the 
gravity localised on the brane. In this chapter we use the AdS/CFT duality 
in order to achieve this goal. Previous works that use the AdS/CFT 
duality in the RS context include 
\cite{herman,Gub,GKR,HHR,DuffLiu,ANO,GK,DK}.

It has been shown in \cite{FeffermanGraham,HS}
that given a metric $g_{(0)}$ on the boundary of AdS 
one can obtain an asymptotic expansion of the bulk
metric near the boundary up to certain order in the radial 
co-ordinate (which is regarded as the small parameter in
the expansion). The next order coefficient is 
left undetermined by the bulk field equations \cite{FeffermanGraham}.
This coefficient is determined once a symmetric 
covariantly conserved tensor $T^{\tnnn{CFT}}_{ij}(x)$ with trace 
equal to the holographic Weyl anomaly is supplied.  
The tensor $T^{\tnnn{CFT}}_{ij}(x)$
is the holographic stress tensor of the dual conformal field 
theory \cite{KSS1} (see also \cite{BFRS}). Notice that the CFT stress 
energy tensor encodes global information too. In particular, regularity 
of the bulk solution sometimes uniquely fixes $T^{\tnnn{CFT}}_{ij}(x)$.

Let us consider a brane placed close to the AdS boundary. Then one can solve 
(\ref{eom}) by simply considering the asymptotic solution described in the 
previous paragraph. The junction condition (\ref{junction}) then becomes 
Einstein's equation for the induced metric on the brane. The right-hand side 
in Einstein's equations is equal to the stress-energy tensor due to matter 
localised on the brane plus the CFT stress-energy tensor. In fact, 
irrespectively of the value of the bulk cosmological constant,
Einstein's equations in the bulk plus the junction condition 
effectively impose Einstein's equations on the brane.
This result first appeared in \cite{SMS}.
In particular in all cases the gravitational equations
on the brane involve a ``holographic stress-energy tensor''.
This can be taken to holographically represent the bulk 
space-time.

This chapter is organised as follows. In the next section 
we adopt the results from the AdS/CFT duality to brane-world
scenarios. In particular, we put a brane near the boundary of 
AdS and obtain the equation that the induced metric on the brane 
satisfies. In section \ref{localanalysis} we place a brane at some 
(arbitrarily chosen) position in the bulk and analyse the equations near the 
brane,
i.e. we consider the radial distance from the brane as 
a small parameter. These considerations are valid for
any bulk cosmological constant. In section \ref{AF}
we consider the case of a brane placed near infinity
of an asymptotically flat bulk space-time. Finally, in section \ref{sec5.5} 
we study bulk metrics that are conformally flat.

In this chapter we only perform a local analysis. Global issues
are important and need to be addressed in order to establish 
localisation of the graviton on the brane. This important 
issue is left for future study.

\section{Brane gravity from the asymptotic analysis of AdS 
space}\label{sec5.2}

The asymptotic solutions of the bulk Einstein equation (\ref{eom}) in vacuum 
were worked out in \cite{HS} to sufficiently high order. These solutions are 
best found by writing the bulk metric in the Fefferman-Graham form 
\cite{FeffermanGraham} used throughout the previous chapter (see \eq{coord}):
\be \label{GrFe2}
\dd s^2={l^2\over4\rho^2}\,\dd\rho^2 
+{l^2\over\rho}\,g_{ij}(\rho,x)\dd x^i\dd x^j,
\ee
where the metric $g_{ij}$ has the expansion
\be\label{metric}
g(\rho,x)=g_{(0)}+\rho 
g_{(2)}+\cdots+\rho^{d/2}g_{(d)}+h_{(d)}\rho^{d/2}\log\rho+{\cal 
O}(\rho^{(d+1)/2}).
\ee
$l^2$ is related to the cosmological constant as $\L=-d(d-1)/2l^2$.
Given $g_{(0)}$ all coefficients up to $g_{(d)}$ can be found as 
local functions of $g_{(0)}$. The coefficient $g_{(d)}$ is undetermined from 
the gravity equations, and it is related to the stress-energy tensor of the 
dual CFT:
\be\label{T}
\< T_{ij}\>_{\tnnn{CFT}}={dl^{d-1}\over16\p G_{d+1}}\,g_{(d)ij} 
+X_{ij}^{(d)}[g_{(j)}],
\ee
where $X_{ij}^{(d)}[g_{(j)}]$ is a known function of the lower-order 
coefficients $g_{(j)}, j<d$ \cite{KSS1} (see \cite{Kstrings00} for a review).
The gravitational equations imply that 
$\<T_{ij}\>_{\tnnn{CFT}}$ is covariantly conserved and its
trace reproduces the conformal anomaly of the boundary CFT. 

Let us place a brane close to infinity  at constant $\rho=\epsilon$, 
where $\epsilon$ is small enough for the expansion (\ref{metric}) to be 
a good approximation for the metric in the bulk.
Using the results of chapter \ref{reconstruction} \cite{KSS1}, it is now a 
simple matter 
(using \eq{tij1}-\eq{counterT}) to see that the junction condition gives 
Einstein's equation on the brane. For a 3-brane we get:
\bea\label{braneeq}
R_{ij}[\g]&-&{1\over2}\,\g_{ij}\,(R[\g] -{12 \over l^2}) 
+{1\over4} l^2 \log\e \left({1\over12}\nabla_i\nabla_iR[\g]
-{1\over4}\,\nabla^2R_{ij}[\g] +{1\over24}\, \g_{ij}\nabla^2R[\g]\right.\nonu
&+&\left.{1\over2}\,R^{kl}[\g]R_{ikjl}[\g] -{1\over6}\,R[\g]R_{ij}[\g] 
+{1\over24}\,\g_{ij}\,R^2[\g] 
-{1\over8}\,\g_{ij}\,R^{kl}[\g]R_{kl}[\g]\right)\nonu
&=&-16\p G_5 {1 \over l}\,
(\< T_{ij}[\g]\>_{\tnnn{CFT}} +T_{ij}^{\sm{bdry}}[\g]),
\eea 
where we kept only terms ${\cal O}(R^2)$, and there is an explicit 
dependence on the cut-off through the logarithmic term. 

There are several comments in order here:
\begin{itemize}
\item In deriving (\ref{braneeq}) it was essential that we added no 
counter-terms to the action. Had we added counter-terms, then all the 
curvature terms in the above formula would have been cancelled. Indeed, 
these precisely come from the infrared divergent part of the action.
\item In the effective Einstein equations the bulk space-time
is represented by the holographic stress-energy tensor. In other 
words, the Brane-World has a purely 
$d$-dimensional description where the bulk space-time has been 
replaced by the cut-off CFT. The CFT couples to matter on the brane 
only through gravitational interactions. 
\item The effective Newton's constant is given by
\be \label{newton}
G_4 = {2 G_5 \over l}
\ee 
In the context of the two-sided RS scenario one should divide this 
result by two (see the discussion after (\ref{junction})).
\item The AdS/CFT duality predicts specific $R^2$-terms. The terms 
in (\ref{braneeq}) are derivable from the local 
action: $\int\dd^dx\,a_{(4)}$, where $a_{(4)}$ is the 
holographic trace anomaly in four dimensions. 
\item The original expansion in the cut-off becomes an 
expansion in the brane curvature. 
\end{itemize}

It is straightforward to extend these results to higher
dimensions using the results in chapter \ref{reconstruction}.

In $(2+1)$ dimensions the series in (\ref{metric}) 
terminates at the $\rho^2$-term, and one has the
exact expression \cite{KS}
\begin{equation}
g(x,\rho )=(g_{(0)}+{1\over 2}
g_{(2)}\rho )^2~,~~g_{(2)}={1\over 2}(R\,g_{(0)ij}+t_{ij})~~,
\label{2d}
\end{equation}
where $t_{ij}$ is conserved, $\nabla_{(0)}^j t_{ij}=0$, and its trace is
$\Tr\, t=-R$. It follows that $t_{ij}$ can be identified
as the Liouville stress-energy tensor. 
The holographic stress-energy tensor is equal to 
$\<T_{ij}\>={l \over 16 \p G_3}\, t_{ij}$.

Placing an one-brane at $\rho=\epsilon$ 
and neglecting $\epsilon^2$-terms one finds that 
the junction condition (\ref{junction}) implies
\begin{equation}
\gamma_{ij}=-8\pi G_3 (T^{\sm{bdry}}_{ij}+\<T_{ij}\>)~~,
\label{2dgrav}
\end{equation}
where $\gamma_{ij}={1\over \rho}g_{ij}(x, \rho )$ is the induced metric 
on the brane, and $T^{\sm{bdry}}_{ij}$ is the stress tensor 
of matter on the brane.
Note that in two dimensions there is no dynamical theory for just the metric 
tensor. Gravity induced on the one-brane is of the scalar-tensor type.

In the presence of matter in the bulk, it was shown in the previous chapter 
that the bulk equations can be solved in the same way. In this case, one 
again 
reinterprets the leading in $\epsilon$ terms as giving the terms in the 
action that determine the dynamics on the brane. For bulk scalar fields of 
mass $m^2=(\D-d) \D$, the effective brane action is:
\bea
S[\g,\F]&=&\int\,\dd^dx\sqrt{\g}\,\left[{1 \over 2(2 \D -d -2)}\F(x,\e) 
\Box_\g 
\F(x,\e) \right. \nonu
&&\left. 
+{(d-\D) \over 2} \left(1 +{1 \over 2 (d-1) (2 \D -d -2)} R[\g]\right) 
\F^2(x,\e) \right],
\eea
where again we only show the first few terms in the low energy  
expansion. The $d$-dimensional 
mass receives contributions both from the mass term in $(d+1)$ 
dimensions but also from the bending of the brane. Notice that 
a massless field in $d+1$ dimensions remains massless in $d$ dimensions.

\section{Local analysis}\label{localanalysis}

In the previous section we made use of the asymptotic expansion of the bulk 
AdS metric (\ref{metric}). A similar analysis can be done for a brane 
located anywhere in the bulk by considering the local geometry near the 
brane.

Consider the Einstein equations in the bulk
\begin{equation}
R_{\mu\nu} + {2 \over d-1} \L G_{\mu\nu}=0.
\label{1.1}
\end{equation}
Near the brane one can use Gaussian normal
co-ordinates. In these co-ordinates the bulk metric takes 
the form 
\begin{equation}
\dd s^2=\dd r^2+\gamma_{ij}(r,x)\dd x^i\dd x^j,
\label{2}
\end{equation}
where  $r$ stands for the radial co-ordinate adjusted  so that the brane 
location is  at $r=0$.
Then the $(ij), (rr)$ and $(ri)$ components of 
Einstein equations (\ref{1.1}) read
\begin{equation}
R_{ij}[\gamma ]
+{2 \over d-1} \L \gamma_{ij}+{1\over 2}\partial^2_r\gamma_{ij}-{1\over 2}
(\partial_r\gamma\gamma^{-1}\partial_r\gamma)_{ij} 
+{1\over 4} \partial_r\gamma_{ij}\Tr(\gamma^{-1}\partial_r\gamma )=0
\label{3}
\end{equation}
\begin{equation}
{1\over 2}\partial_r (\Tr (\gamma^{-1}\partial_r \gamma ))
+{1\over4 }\Tr(\gamma^{-1}\partial_r \gamma )^2+
{2 \over d-1} \L =0
\label{4}
\end{equation}
\begin{equation}
\nabla_j [\gamma^{-1}\partial_r\gamma
- \Tr(\gamma^{-1}\partial_r\gamma )]^j_i=0.
\label{5}
\end{equation}
Combining the equations (\ref{3}) and (\ref{4}) we find that
\begin{equation}
R[\gamma ]+2 \L +{1\over 4}\left([\Tr(\gamma^{-1}\partial_r\gamma )]^2
-\Tr(\gamma^{-1}\partial_r\gamma )^2\right)=0.
\label{6}
\end{equation}

Let $\gamma_{ij}(x,r)$ have the following expansion near the brane: 
$$
\gamma=\gamma_{(0)}+\gamma_{(1)}r+\gamma_{(2)}r^2+...
$$
Then solving equations (\ref{3}), (\ref{4}) and (\ref{5}) 
iteratively we find expressions relating
the coefficients $\g_{(k)}$.  From equation (\ref{3}) we find that
\be
\Ric [\g_{(0)}]
+{2 \over d-1} \L
\g_{(0)}+\g_{(2)}-{1\over 2}\g^2_{(1)}+{1\over 4}\g_{(1)}\Tr\g_{(1)}=0.
\label{Ricci}
\ee
Equation (\ref{4}) to leading order gives 
\be
\Tr\g_{(2)}={1\over 4}\Tr \g^2_{(1)}-{2\over d-1}\Lambda.
\label{trace}
\ee
Taking the trace of (\ref{Ricci}) and using (\ref{trace}) 
one finds
\be
R[\g_{(0)}]+2\Lambda -{1\over 4}(\Tr \g^2_{(1)}-(\Tr \g_{(1)})^2)=0.
\label{Riccisc}
\ee
This equation can also be obtained from (\ref{6}).
Equation (\ref{4}) to the first two orders yields
\bea
&&\nabla^j\g_{(1)ij}=\nabla_i\Tr\,\g_{(1)}, \label{nabl1} \\
\nabla^j\g_{(2)ij} &=& \half \nabla_j
[\g_{(1)}^2-{1\over 2}\g_{(1)} 
\Tr \g_{(1)}-{1\over 4}\g_{(0)} (\Tr \g_{(1)}^2-(\Tr \g_{(1)})^2)]^j_i.
\label{divergences}
\end{eqnarray}
Equation (\ref{nabl1}) can be integrated as
\be \label{inte}
\g_{(1)} = t_{(1)} + \g_{(0)} \Tr\,\g_{(1)},
\ee
where $t_{(1)ij}$ is an ``integration constant'' that satisfies
$\nabla^i t_{(1)ij}=0$. 
One can check that (\ref{divergences}) is automatically satisfied
when (\ref{Ricci}) and (\ref{Riccisc}) are satisfied.

Forming the Einstein tensor, we obtain
\be
R_{ij}[\g_{(0)}]-{1\over 2}\g_{(0)ij}R[\g_{(0)}]
=\Lambda \g_{(0)ij}+T_{ij},
\label{Einstein1}
\ee
where 
\be
T_{ij}=-{2 \over d-1} \L \g_{(0)ij}-\g_{(2)ij} + \half \g_{(1)ij}^2 
-{1 \over 4} \g_{(1)ij} \Tr \g_{(1)} -{1 \over 8} \g_{(0)ij}
[\Tr \g_{(1)}^2 - (\Tr \g_{(1)})^2].\nn
\ee
Equation (\ref{divergences}) implies that $T_{ij}$ is covariantly conserved.
In addition, equation (\ref{trace}) determines the trace of $T_{ij}$,
\be
\Tr\,T=
-2\Lambda -{(d-2)\over 8}\left(\Tr t_{(1)}^2-{1\over d-1}
(\Tr t_{(1)})^2\right)
\ee

Let us now consider a physical brane with stress tensor 
$T^{\sm{bdry}}_{ij}$ located
at $r=0$. Then in addition to equations (\ref{3}), (\ref{4}), (\ref{5})
we have the  junction condition (\ref{junction}).
For the metric (\ref{2}) the second fundamental form is equal to 
$K_{ij}={1\over 2}\partial_r\gamma_{ij}$. From the junction 
condition (\ref{junction}) we get using the equation (\ref{nabl1})
\begin{equation}
t_{(1)ij}=16 \p G_{d+1} T^{\sm{bdry}}_{ij}.
\label{8}
\end{equation}
The junction condition thus identifies the 
undetermined covariantly conserved 
tensor $t_{(1)}$ in (\ref{inte}) with the stress tensor of the brane.
Notice that conservation of the boundary stress-energy tensor 
is a necessary condition for this identification.

To summarise, we have shown that Einstein's equations in the 
bulk plus the junction condition lead to Einstein's equations 
on the brane. The effective stress-energy tensor $T_{ij}$ 
represents both the bulk space-time and the matter on the 
brane. Its trace is determined
by the matter stress-energy tensor on the brane.
This is similar to the case discussed in the previous section.
There the effective stress-energy tensor was a sum of 
the stress-energy tensor of matter localised on the brane
of the $\<T_{ij} \>_{\sm{CFT}}$. The latter was taken to 
represent the bulk space-time, and its trace was fixed to be the 
holographic conformal anomaly.
 
The results in this section agree with the results obtained 
in \cite{SMS} for $d=4$. To see this, let 
\be \label{SMS}
\g_{(2)ij}=-E_{ij} + {1 \over 4} \g^2_{(1)ij} -{2 \over d(d-1)} \L 
\g_{(0)ij},
\ee
and also let the boundary stress-energy tensor be equal 
to  $T^{\sm{bdry}}_{ij}=-\l \g^{(0)}_{ij}+\t_{ij}$, where 
$\l$ is the tension and $\t_{ij}$ the matter energy momentum 
tensor on the brane. Equation (\ref{SMS}) defines the 
tensor $E_{ij}$. A short calculation shows that it agrees
with the tensor $E_{\m \n}$ of \cite{SMS}. In particular,
$E_{ij}$ is traceless and its divergence is equal to 
$\nabla^j E_{ij} = K^{jk}(\nabla_i K_{jk} - \nabla_j K_{ik})$.
This agrees with formula (22) of \cite{SMS}. 
One can also verify agreement with (17)-(20) of \cite{SMS}.

Note that the above considerations are
quite general and valid for any value of the bulk cosmological constant.
Note also that when the brane matter consists of only a brane 
cosmological constant the brane geometry has a constant Ricci scalar. 

\section{Asymptotically flat case}\label{AF}

In this section we perform an asymptotic analysis of Einstein's
equations with zero cosmological constant similar to the one 
that has been done for asymptotically AdS spaces in 
\cite{FeffermanGraham,HS}. 

We work in Gaussian normal co-ordinates. The metric takes the form
\be
\dd s^2=\dd r^2+\gamma_{ij}(x,r)\,\dd x^i\dd x^j.
\label{ds2}
\ee
Einstein's equations in this co-ordinate system are given in
equations (\ref{3}), (\ref{4}) and (\ref{5}). We look
for an asymptotic solution near infinity. Assuming
that the leading part of $\g$  near infinity is 
non-degenerate we find that it scales like $r^2$
(to prove this use  (\ref{3})). Restricting ourselves 
to this case, we look for solutions of the form
\be
\gamma (x,r)=r^2 (g_{(0)}+g_{(2)}{1\over r}+g_{(4)}{1\over r^2}+...)~~.
\label{asyF}
\ee
In other words, the bulk metric asymptotes to a cone with $g_{(0)}$
the metric on the base. In general, one can include logarithmic 
terms in (\ref{asyF}). Such more general asymptotic solutions 
have been studied in \cite{BS,B}\footnote{In \cite{BS,B} the authors look for 
solutions whose metric coefficients near infinity is given by an 
expansion in negative powers of the radial co-ordinate. Co-ordinate 
transformations allow one to put the metric in  
the form $\dd s^2=N^2 \dd r^2+\gamma_{ij}(x,r)\dd x^i\dd x^j$,
with $N=1+\s(x)/r$ and $\gamma(x,r)$ as in (\ref{asyF}).
By a further logarithmic transformation one can reach 
Gaussian normal co-ordinates but at the expense of 
introducing logarithmic terms in $\gamma(x,r)$.
Our results for $d=3$ agree with the results of \cite{BS,B} for $\s(x)=0$.
We thank Kirill Krasnov for bringing these papers to our attention.}.
We restrict ourselves to (\ref{asyF}).
      
We solve Einstein's equations
\be
R_{\mu\nu}=0,
\ee
perturbatively in $1/r$.
The leading order equations imply \cite{GPP,SS} that $g_{(0)}$
should satisfy
\be \label{leading}
R_{(0)ij}+(d-1)g_{(0)ij}=0~~.
\ee
This means that the space at infinity is described by an Einstein metric
of constant positive scalar curvature. 
In particular, for Euclidean signature
the standard metric on the unit sphere $S^{d-1}$ satisfies
this equation. Then the leading part of the bulk metric (\ref{ds2}), 
(\ref{asyF}) is just 
Euclidean $R^d$ space. In the Lorentzian signature
equation (\ref{leading}) is solved by the de Sitter space.
Thus, already at leading order, we find an important difference 
between the cases of asymptotically flat space-time and
of asymptotically AdS space-times. Whereas in the latter case one could 
choose the boundary metric arbitrarily, in the former case
the boundary metric has to satisfy (\ref{leading}).

To next order we find
\bea 
&&\nabla^jg_{(2)ij}=\nabla_i\Tr g_{(2)}, \label{na1} \\
&&d g_{(2)}+2 \Ric_{(2)}-g_{(0)}\Tr g_{(2)}=0, \label{Ei1}
\eea
where 
\be
\Ric[\g]=\Ric_{(0)}+{1\over r}\,\Ric_{(2)}+{1\over r^2}\,\Ric_{(4)}
\cdots
\ee
and
\be\label{R2ij}
R_{(2)ij}=-{1\over2}[\nabla_i\nabla_j\Tr g_{(2)}
-\nabla^2 g_{(2)ij} +2(d-1)g_{(2)ij}+2R_{(0)ikjl} g_{(2)}^{kl}],
\ee
where indices raised and lowered by $g_{(0)}$. In deriving 
this equation, (\ref{leading}) and (\ref{na1}) were used.
Then equation (\ref{Ei1}) becomes
\be
\nabla_i \nabla_j \Tr g_{(2)} - \nabla^2 g_{(2) ij} 
+(d-2) g_{(2)ij} + g_{(0)ij} \Tr g_{(2)} 
+ 2 R_{(0)ikjl}\, g_{(2)}^{kl}=0.
\le{Ei12}
Notice that this equation leaves undetermined the trace of $g_{(2)}$.
Let us define
\be \label{deftij}
t_{ij} = g_{(2)ij} - g_{(0)ij} \Tr g_{(2)}.
\ee
It follows from the (\ref{na1}) that $\nabla^i t_{ij}=0$.

To the next order we find the equations
\bea \label{2order1}
g_{(4)}&=&-{1\over2}\Ric_{(4)} +{1\over2}g_{(0)} \Tr g_{(4)} 
-{1\over4} g_{(0)} \Tr g_{(2)}^2 +{1\over4}g_{(2)}^2 
+{1\over8}g_{(2)}\Tr g_{(2)}, \\
\Tr g_{(4)}&=&{1\over4}\Tr g_{(2)}^2, \label{2order2} \\
\nabla^j g_{(4)ij} &=& \half \nabla_j
[g_{(2)}^2-{1\over 2}g_{(2)} 
\Tr g_{(2)}-{1\over 4}g_{(0)} (\Tr g_{(2)}^2-(\Tr g_{(2)})^2)]^j_i,
\label{2order3}
\eea
and 
\bea
R_{(4)ij}&=&{1\over2}[-{1 \over 4} \nabla_i \nabla_j \Tr g_{(2)}^2 
-\nabla^k \nabla_i g_{(4)jk} -\nabla^k \nabla_j g_{(4)ik}
+ \nabla^2 g_{(4)ij} \nonu
&&+g_{(2)}^{kl} 
[\nabla_l \nabla_i g_{(2)jk} +\nabla_l \nabla_j g_{(2)ik} -\nabla_l
\nabla_k g_{(2)ij}] \nonu
&&+{1 \over 2} \nabla^k \Tr g_{(2)} 
(\nabla_i g_{(2)jk} +\nabla_j g_{(2)ik} - \nabla_k g_{(2)ij}) \nonu
&&+\half \nabla_i g_{(2)kl} \nabla_j g_{(2)}^{kl} 
+\nabla_k g_{(2)il} \nabla^l g_{(2)j}{}^{k}
-\nabla_k g_{(2)il} \nabla^k g_{(2)j}{}^{l}].
\eea
It may seem that by taking the trace of 
(\ref{2order1}) and using (\ref{2order2}) and (\ref{2order1})
one obtains a new equation for $g_{(2)}$. However, 
it turns out that the resulting equation is automatically 
satisfied. The same is true when taking the trace of (\ref{Ei1}) and using 
(\ref{R2ij}).

The equations we obtained look similar to the equations one gets in the 
case of asymptotically AdS space-times. There are important 
differences, however. In the case of asymptotically AdS space-times
the equations were algebraic, and they could be solved up to 
order $\r^d$. The coefficient $g_{(d)}$ was undetermined
except for its trace and divergence. In the case at hand the equations
for the coefficients are differential, and it is the trace 
of $g_{(2)}$ which is undetermined.

Let us comment on the logarithmic terms that can be included in our ansatz 
\eq{asyF} and which were considered in \cite{BS,B} for the case $d=3$. We 
start with the metric of \cite{BS},
\be
\dd s^2=N^2\dd \r^2+\r^2f_{ij}(\r,x)\,\dd\f^i\dd\f^j,
\ee
where $N$ and $h$ are given by
\be
N&=&1+{\s(x)\over r},\nn
f_{ij}&=&f_{(0)ij}+{1\over r}\,f_{(2)ij} +\dots
\le{trafo}
The following co-ordinate transformation:
\be
\r&=&r-\s(x)\log r+{1\over r}\,\s^2(x) +\dots\nn
\f^i&=&x^i -{1+\log r\over r}\,\na^i\s +\dots
\le{BSmetric}
brings the metric to the Gaussian normal form:
\be
\dd s^2=\dd r^2+r^2g_{ij}(r,x)\dd x^i\dd x^j.
\ee
However, $g_{ij}$ now has an expansion that includes logarithmic terms: 
\be
g(r,x)=g_{(0)}+{1\over r}\,g_{(2)} +{1\over r}\,\log r\,h_{(2)} +{1\over 
r}g_{(4)}+\dots
\ee
The first few coefficients $g_{(n)}$ and $h_{(2)}$ are related to the ones in 
\eq{trafo} in the following way:
\be
g_{(0)}(x)&=&f_{(0)}(x)\nn
g_{(2)}(x)&=&f_{(2)}(x) -2\na_i\na_j\s\nn
h_{(2)}(x)&=&-2(\s f_{(0)} +\na_i\na_j\s).
\le{d3}
Repeating the analysis above one finds that $h_{(2)}$ satisfies the same 
equations \eq{na1}-\eq{Ei1} as $g_{(2)}$ above, but for $d=3$ its trace is 
zero:
\be
\Tr h_{(2)}=0.
\ee
Filling in the expression for $h_{(2)}$ from \eq{d3}, one finds that for 
$d=3$ \eq{Ei12} reduces to:
\be
(\Box+3)\s=0
\ee 
which agrees with the result in \cite{BS}. Notice that for $d=3$ the 
equations for $g_{(2)}$ remain unchanged. A further co-ordinate 
transformation
\be
\r&=&\bar\r\,(1+\s)+\dots\nn
\f^i&=&\bar\f^i+{1\over\r}\,\na^i\s+\cdots
\ee
maps our $g_{(2)}$ into the tensor $k$ of \cite{BS}, $g_{(2)}=k=f_{(1)}+2\s 
f_{(0)}$. In the following we continue our analysis for general $d$ and 
$\s=0$.

Let us place the brane at a fixed large radius $r=r_0
\gg 1$. 
Then expanding the Einstein tensor 
for the induced metric $\gamma_{ij}$ we find that
\be
R_{ij}[\g]-{1\over2}\,\g_{ij}R[\g]=(d-2)\left({d-1\over2r_0^2}\,\g_{ij} 
+{1 \over 2 r_0}\,t_{ij}\right)+{\cal O}(1/r_0^2)~~,
\label{Einstein}
\ee
where $t_{ij}$ is given in (\ref{deftij}).
On the other hand we have
\be
K_{ij}-\gamma_{ij}K=-{d-1\over r_0}\gamma_{ij}-
{1 \over 2}t_{ij} + {\cal O}(1/r_0^2)~~.
\ee
Notice that this is the Brown-York stress-energy tensor \cite{BrownYork}. 
Thus, up to the leading divergence in $r_0 \to \infty$, $t_{ij}$ is equal to 
the Brown-York stress-energy tensor. This divergence can again be cancelled 
by adding covariant counter-terms, along the lines of the previous chapter 
(see also \cite{KSS1}). The junction condition on the brane gives a relation 
between  $t_{ij}$ and the stress tensor $T^{\sm{bdry}}_{ij}$
of matter fields on the brane. Plugging back to (\ref{Einstein}) we find 
\be
R_{ij}[\gamma ]
-{1\over2}\gamma_{ij} R[\gamma ]
=-{(d-2)(d-1)\over2r_0^2}\,\gamma_{ij} 
-{(d-2)8\pi G_{d+1}\over r_0}\,T_{ij}^{\sm{bdry}}+{\cal O}(1/r_0^2),
\ee
i.e. we get Einstein's equations with negative cosmological 
constant $\Lambda=-{(d-1)(d-2)\over 2 r_0^2}$
and Newton's constant $G_{d}={(d-2) G_{d+1} \over r_0}$. The position
of the brane becomes the AdS radius of gravity on the brane.
Notice also that the formula for $G_d$ is the same with formula 
(\ref{newton}) with $l$ replaced by $r_0$.

\section{Conformally flat metrics}\label{sec5.5}

In the case of asymptotically AdS spaces it was found in \cite{KS} that, 
imposing the vanishing of the bulk Weyl tensor, Einstein's equations could be 
integrated, and the perturbative expansion ended at order $\r^2$ (see 
\eq{KS}). In this case, the (conformally flat) boundary condition on the 
metric was enough to obtain the exact solution as we found in the previous 
chapter. This implied that the boundary stress-energy tensor was completely 
determined by the background. These results can be extended for arbitrary 
value of the cosmological constant. Let us write the Weyl tensor in the 
following way:
\bea
C_{\m\a\n\b}&=&R_{\m\a\n\b} -P_{\m\n}G_{\a\b} -P_{\a\b}G_{\m\n} 
+P_{\m\b}G_{\a\n} +P_{\a\n}G_{\m\b}
\eea
where
\be
P_{\m\n}={1\over d-1}(R_{\m\n}-{1\over2d}\,R\,G_{\m\n}).
\ee
Using Einstein's equations for generic cosmological constant,
\be
R_{\m\n}&=&-{2\L\over d-1}\,G_{\m\n},
\eea
one easily sees that the vanishing of the Weyl tensor implies
\be
R_{\m\n\a\b}=-{2\L\over d(d-1)}(G_{\m\a}G_{\n\b}-G_{\m\b}G_{\n\a}).
\le{W=0}
Hence the bulk space is a maximally symmetric space which is locally dS, AdS 
or flat space. The cosmological constant is $\L={d(d-1)\over2l^2}\,\ve $, 
where $\ve=\pm 1$ or 0  for dS/AdS space or AF space, respectively.

The Gaussian normal co-ordinate system \eq{2} is the most convenient one for 
the AF case. In these co-ordinates, \eq{W=0} gives, in components:
\be
\g''-\half \g'\g^{-1}\g'&=&-{2\ve\over l^2}\,\g\nonu
\nabla_i\g'_{jk}&=&\nabla_k\g'_{ij}\nonu
R_{ikjl}[\g]&=&-{1\over4}(\g'_{ij}\g'_{kl} -\g'_{il}\g'_{jk}) -{\ve\over 
l^2}(\g_{ij}\g_{kl}-\g_{il}\g_{jk}).
\ee
Differentiating the first equation one gets, for $\ve=0$,
\be
\g'''=0,
\ee
and so the expansion stops at order $r^2$. In fact, one finds that the 
induced metric has the same following form as the boundary metric in the case 
of AdS \cite{KS}:
\be
\g=(\g_{(0)}+{r\over2}\,\g_{(1)})\g_{(0)}^{-1}(\g_{(0)}+{r\over2}\,\g_{(1)}).
\ee
It has only two undetermined coefficients: $\g_{(0)}$ and $\g_{(1)}$.
Unlike the AdS case, where the boundary metric $g_{(0)}$ gives a conformally 
flat bulk solution if and only if it is conformally flat, in this case there 
are no restrictions on $\g_{(0)}$. On the other hand, we do have constraints 
on $\g_{(1)}$. It satisfies the following two equations:
\be
\nabla_i\g_{(1)jk}&=&\nabla_k\g_{(1)ij}\nn
R_{ikjl}[\g_{(0)}]&=&-{1\over4}\,(\g_{(1)ij}\g_{(1)kl} 
-\g_{(1)il}\g_{(1)jk}).
\le{c0}
Taking the trace of these equations gives back the constraints found in 
section \ref{localanalysis} for $\L=0$. The conditions \eq{c0}, however, are 
stronger. One also finds an expression for the effective stress-energy tensor 
purely in terms of $t_{(1)}$:
\be
T_{ij}={1\over4}[t_{(1)}^2-{1\over(d-1)}\,t_{(1)}\,\Tr\,t_{(1)}-{1\over2}\g_{
(0)}\,\Tr\,t_{(1)}^2+{1\over2(d-1)}\,\g_{(0)}\,(\Tr\,t_{(1)})^2].
\ee
In the brane-world scenario, this also provides a direct relation between 
$T_{ij}$ and $T^{\sm{bdry}}_{ij}$.

One can also perform the analysis in the co-ordinate system \eq{asyF}, which 
is more convenient to analyse the equations at infinity in the asymptotically 
flat case. We find the following solution:
\be
g=(g_{(0)}+{1\over2r}\,g_{(2)})g_{(0)}^{-1}(g_{(0)}+{1\over2r}\,g_{(2)}),
\ee
so once again the vanishing of the Weyl tensor is enough to solve the 
equations. We also get additional constraints on both coefficients $g_{(0)}$ 
and $g_{(2)}$. The equation for $g_{(0)}$ tells us that the boundary has 
vanishing Weyl tensor:
\be
R_{ikjl}[g_{(0)}]=-g_{(0)ij}g_{(0)kl}+g_{(0)il}g_{(0)jk},
\ee
and so the space is asymptotically de Sitter \cite{SS}. We also get an 
additional differential equation which puts further constraints on $g_{(2)}$. 
Notice that both equations give back the equations in section \ref{AF} when 
contracting two indices, but again the requirements that we find here are 
stronger.

The co-ordinate systems \eq{asyF} and \eq{2} describe the same space and so 
the metric must be the same. A complicated co-ordinate transformation may 
however be required to go from one system to the other.

One can check that the equations can be integrated in the other cases as 
well. The solutions, however, become more involved in the Gaussian 
co-ordinate system and it is better to change the radial co-ordinate. It 
would be interesting to have a more detailed analysis that includes also 
different brane embeddings and dS space.

\newpage

\appendix

\chapter{Shock-Wave Geometries}\label{appA}

\section{More on AdS shock-wave solutions}\label{appA1}

In this appendix we give some details of the geodesics and stress-energy 
tensor of massless particles in AdS, discussed in chapters 
\ref{HEscattering}-\ref{GSM} and their properties. 

We write AdS space in the co-ordinate system \eq{pureAdS}, $y^\m=(u,v,y^i)$, 
$i$ running from 1 to $d-2$. The metric reads:
\be
\dd s^2={4\over\O^2}\,\et_{\m\n}\dd y^\m\dd y^\n,
\ee
where the conformal factor is given by $\O=1-y^2/\ell^2$.

It is well-known that the null geodesics of two conformally related 
space-times are the same, up to a reparametrisation of the geodesic length. 
Therefore, null trajectories in the above co-ordinates will take the same 
form as those in Minkowski space. It is nevertheless convenient for the 
computation of the stress-energy tensor to see explicitly how the affine 
parameter changes.

The geodesic equation and the mass-shell condition give:
\be
{\dd\over\dd\l}\left({\et_{\m\n}\dot z^\n\over\O^2}\right) &=& 
2{\et_{\m\n}z^\n{\cal L}\over\ell^2\,\O}\nn
{\cal L}&=&{1\over\O^2}\,\et_{\m\n}\dot z^\m\dot z^\n=0.
\le{A11b}
${\cal L}$ is the Lagrange density, defined by the second of \eq{A11b}, and 
$\l$ the affine parameter along the geodesic. These equations integrate to
\be
\et_{\m\n}\,\dot z^\n&=&v_\m\,\O^2.
\le{A12}
$v_\m$ is a constant, lightlike vector satisfying $\et^{\m\n}v_\m v_\n=0$ to 
be 
determined by the boundary conditions. This equation also relates the affine 
parameter in AdS to the affine parameter in Minkowski space.

The stress-energy tensor \eq{10} now equals:
\be
T_{\m\n}&=&-p\,\O^d\,v_\m v_\n\int\dd s\,\d^{(d)}(y-z(s)),
\le{A13}
and choosing co-ordinates where momentum is purely in the $v$-direction, this 
reduces to:
\be
T_{uu}&=&-p\,\O^d\,\d(u-u_0)\,\d(\r-\r_0),
\le{A15}
where $\r=\sum_{i=1}^{d-2}y_i^2$. Notice that in order for the metric 
\eq{H-I} to be a solution of Einstein's equations with this stress-energy 
tensor, we need the initial condition $u_0=0$. It is also convenient to take 
$\r_0=0$. Thus we get the stress-energy tensor used in chapter \ref{GSM},
\be
T_{uu}=-p\,\d(u)\d(\r),
\le{A15b}
which gives rise to the delta-function in \eq{shiftads}. This form for the 
stress-energy tensor agrees with the one computed in chapter 
\ref{reconstruction}, equation \eq{AdSstress}, in Poincare co-ordinates, and 
for the case $g_{(0)ij}=\et_{ij}$. One can check this by performing the 
following co-ordinate transformation from $y^\m$ to Poincare co-ordinates 
$x^\m=(r,t,x^i)$:
\be
u&=&{t^2-r^2-\vec{x}^2\over r+t}\nn
v&=&{\ell^2\over r+t}\nn
y^i&=&{\ell x^i\over r+t}\nn
\O&=&{2r\over r+t}\nn
r&=&{1\over2v}\,(\ell^2-uv-\rho^2)\nn
t&=&{1\over2v}\,(\ell^2+uv+\rho^2)\nn
x^i&=&{\ell\over v}\,y^i.
\le{A2}

With \eq{A15b} at hand, one can compute the back-reaction on the AdS metric, 
obtaining the solution found by Horowitz and Itzhaki with the shift functions 
as given in \eq{fsol}. The next step is then to compute the geodesics of a 
test particle in the back-reaction corrected 
metric. The computation goes along the same lines as the one above. We do not 
give the details here since it is a 
straightforward exercise, but give only the results. We concentrate on 
trajectories whose initial velocities are perpendicular to the velocity of 
the shockwave, that is, the geodesics with $v=y^i=0$ before the collision. 
This gives a head-on collision.

It turns out that the geodesic equations can again be exactly integrated, 
and the effect is the same as in Minkowski space: there is a shift in the 
$v$ co-ordinate and a deflection in the $x^i$-plane which nevertheless is 
negligible in the eikonal approximation where the impact parameter is much 
larger than the Planck length. In this approximation, the shift is given by
\be
\d v=-8\pi\GN\,p_u\,F_0\,\th(u),
\le{A1}
where $F_0$ is the shift function before the collision, $F_0=F(u=0)$.

Of course the same results can be found from geodesics in Minkowski space by 
noting that massless geodesics are invariant under conformal transformations 
of the metric.

It is interesting to note that, when one considers only one particle, there 
is no self-interaction, and therefore the present solution to the 
Einstein-matter system with the given boundary conditions is exact. However, 
when considering two particles this is no longer true, and one has to 
restrict oneself to consider a ``soft" test particle in the background of a 
``hard" particle.

\section{The induced two-dimensional Ricci tensor}\label{appA3}

In this Appendix we outline the proof that Einstein's equations with a 
massless source reduce to the conditions \eq{9}-\eq{9a}. We also compute the 
curvature of the transverse part of the metric, equation \eq{53}. This 
computation follows \cite{gnp85}, and for more details we refer to that 
paper.

The ansatz in \cite{gnp85} for the metric is the following:
\be
\dd\^s^2 =2A(\^u,\^v)\,\dd\^v(\dd\^u- \d (v)\dd\^v) +g(\^u,\^v)\, 
h_{ij}(\^x^i) \dd\^x^i\dd\^x^j.
\le{B1}
We also have the unperturbed metric
\be
\dd s^2=2A(u,v)\,\dd u\dd v+g(u,v)\,h_{ij}(x^i)\dd x^i\dd x^j,
\le{B0}
which will be assumed to solve Einstein's equations. \eq{B1} is related to 
\eq{B0} by a shift {\it and} a co-ordinate transformation:
\be
\^u&=&u+\th f\nn
\^v&=&v\nn
\^x^i&=&x^i.
\le{C3}

The metric \eq{B1} should be a solution of Einstein's equations with a 
massless source:
\be
R_{\m\n}[\^G]&=&R_{\m\n}[G]+\d R_{\m\n}[G]=-8\p\GN\^T_{\m\n}\nn
R_{\m\n}[G]&=&0\nn
T^{\^u\^u}&=&4p\,\d (\^v)\,\d (\ti x),
\le{C3b}
so our massless particle travels along the null geodesic $\^v=0$, $\^x^i=0$. 

Let us first work out the vacuum piece of Einstein's equations, 
$R_{\m\n}[G]=0$. We use the formula
\be
R^\m_{\;\;i\m j}[G]=\frac{1}{\@{-G}}\,\pa_\m\left(\@{-G}\, 
\G^\m_{ij}\right)-\pa_i\pa_j\left(\log\@{-G}\right)-\G^\m_{\n i}\G^\n_{j\m},
\le{B3}
and we have:
\be
\@{-G}&=&Ag\,\@{h}\nn
\G^\a_{ij}&=&-\frac{1}{2}\,g^{\a\b}h_{ij}\pa_\b g\nn
\G^i_{j\a}&=&\frac{1}{2g}\,\d ^i_j\pa_\a g\nn
\G^i_{\a j}&=&\frac{1}{2g}\,\d ^i_j\pa_\a g\nn
\G^\a_{\b i}&=&\G_{i\a\b}=0,
\le{B4}
where the indices $\m,\n$ run from 1 to 4, $\a$ and $\b$ take the values $1$, 
$2$, and $i,j$ take the values $3,4$. Plugging this in equation \eq{B3}, we 
get:
\be
R_{ij}[G]&=&R_{ij}[h]-\frac{1}{2Ag}\,h_{ij}\pa_\a\left(Agg^{\a\b}\pa_\b 
g\right)-\G^\a_{ki}\G^k_{j\a}-\G^k_{\a i}\G^\a_{jk}\nn
&=&R_{ij}[h]-\frac{1}{A}\,h_{ij}\pa_u\pa_vg=0.
\le{B5}
Here $R_{ij}[h]$ is the two-dimensional Ricci tensor calculated in the metric 
$h_{ij}$. This gives
\be
R_{ij}[h]&=&{1\over A}\,\pa_u\pa_vg\,h_{ij},
\le{B6}
which gives \eq{53}. 

After some algebra, and using the vacuum solutions, one finds that the 
remaining piece of the metric only contributes the $\^u\^u$-component of the 
Ricci tensor. Einstein's equations, 
\be
\d R_{\m\n}[G]=-8\p\GN\^T_{\m\n},
\ee
are then satisfied provided $R_{\m\n}[G]=0$ and \eq{9}-\eq{9a} hold. More 
details can be found in the appendices of \cite{gnp85}.

\chapter{Scaling and Classical Solutions of the Einstein-Hilbert 
Action}\label{appB}

\section{Scaling of curvature}\label{appB1}

In this section we give the details of the final rescaled (up to lowest order
in $h_{i\a}$, all orders in $\e$) Ricci tensor $R_{\mu\nu}$. From here 
one can simply check the expansion of the Einstein-Hilbert 
action\footnote{For notational simplicity, we denote the transverse metric by 
$g_{ij}$. In chapter \ref{GSM} it is denoted by $h_{ij}$.}. Higher orders in 
$h_{i\a}$ (quadratic at $1/\epsilon^2$ and at $\epsilon^0$) are not necessary 
as we are not going to consider the fluctuations of the metric.

Under the rescaling \eq{rescaling}, the Christoffel symbols transform as:
\be
\G^\g_{\a\b}(G)&=&\G^\g_{\a\b}(\^G) +{1-\e\over2}\,g^{i\g}\pa_ig_{\a\b}\nn
\G^\a_{ij}(G)&=&{1\over\e}\,\G^\a_{ij}(\^G) +{\e-1\over2\e^2}\,g^{\a\b}\pa_\b 
g_{ij}\nn
\G^\a_{\b i}(G)&=& \G^\a_{\b i}(\^G) +{1-\e\over2\e}\,(g^{\a\g}\pa_\b g_{\g 
i}-g^{\a\g}\pa_\g g_{\b i} +g^{\a j}\pa_\b g_{ij})\nn
\G^k_{ij}(G)&=&\G^k_{ij}(\^G) +{\e-1\over2\e}\,g^{k\a}\pa_\a g_{ij}\nn
\G^i_{\a\b}(G)&=&\e\,\G^i_{\a\b}(\^G) 
+{\e(1-\e)\over2}\,g^{ij}\pa_jg_{\a\b}\nn
\G^i_{\a j}(G)&=&\G^i_{\a 
j}(\^G)+{\e-1\over2}\,(g^{i\b}\pa_jg_{\a\b}+g^{ik}\pa_jg_{\a 
k}-g^{ik}\pa_kg_{\a j})
\ee
where $G_{\m\n}$ is the $\e$-dependent metric, whereas $\^G_{\m\n}$ is the 
rescaled metric, which is independent of $\e$.

Working out the curvature components, we get:
\be
R_{\a\b}[G] &=& \epsilon^0 ( R_{\a\b}[\hat G] - \frac{1}{2} \nabla_\b (g^{ik}
\partial_\a g_{ik}) - \frac{1}{4} g^{ij}\partial_\a g_{kj} 
g^{km}\partial_\b g_{im})+\nn
 &+& \epsilon^2 (-\frac{1}{2} \nabla_i(g^{ij}\partial_j g_{\a\b}) 
- \frac{1}{4} g^{\g\rho}\partial_i g_{\g\rho} g^{ij}\partial_j g_{\a\b}+ \nn
&+& \frac{1}{4} g^{\g\rho}\partial_kg_{\b\rho} g^{ki}\partial_i g_{\a\g}
+ \frac{1}{4} g^{\g\rho}\partial_kg_{\a\rho} g^{ki}\partial_i g_{\b\g})
\ee
The leading term in $R_{i\a}$ is at zero order in $h_{i\a}$ which is already 
sufficient for our purposes as it is always multiplied by $h_{i\a}$ in the 
action and this term arises at order $\epsilon^0$
\be
R_{i\a} &=& \epsilon^0 (\frac{1}{2}\nabla_\b (g^{\b\rho}\partial_i 
g_{\a\rho})
 - \frac{1}{2}\nabla_\a (g^{\b\rho}\partial_i g_{\b\rho}) + \frac{1}{2}
\nabla_k(g^{kj}\partial_a g_{ij})\nn &-& \frac{1}{2} \nabla_i
(g^{kj}\partial_\a g_{kj}) + \frac{1}{4} g^{\g\rho}\partial_i g_{\a\rho} 
g^{kj} \partial_\g g_{kj} + \frac{1}{4} g^{km}\partial_\a g_{im} 
g^{\g\b} \partial_k g_{\g\b}\nn &-& \frac{1}{2} 
g^{\b\rho}\partial_\rho g_{ik} g^{kj} \partial_j g_{\a\b})
\ee
$R_{ij}$ is identical to $R_{\a\b}$ under the interchange of Greek and 
Roman indices and $\epsilon\ra\epsilon^{-1}$.
\be
R_{ij}[G] &=& \epsilon^{-2} (-\frac{1}{2} \nabla_\a(g^{\a\b}\partial_\b 
g_{ij}) 
- \frac{1}{4} g^{km}\partial_\a g_{km} g^{\a\b}\partial_\b g_{ij} \nn
&+& \frac{1}{4} g^{km}\partial_\g g_{jm} g^{\g\a}\partial_\a g_{ik}
+ \frac{1}{4} g^{km}\partial_\g g_{im} g^{\g\a}\partial_\a g_{jk})\nn
&+& \epsilon^0 ( R_{ij}[\hat G] - \frac{1}{2} \nabla_j (g^{\a\g}
\partial_i g_{\a\g}) - \frac{1}{4} g^{\a\b}\partial_\a g_{\g\b} 
g^{\g\rho}\partial_j g_{\a\rho})
\ee

\section{Scaling of the exterior curvature}\label{appB2}

The exterior curvature part of the Einstein -- Hilbert action is
\be
S &=& \frac{1}{\Pl^{d-2}}\int\sqrt{\g}\,\nabla_\mu n^\mu 
\le{fundform}
$\g$ is the boundary metric which under rescaling is multiplied 
by $\ell_\parallel^2 \ell_\perp^{2(d-2)}$. The normal $n$ will have 
a non-zero component only in the direction perpendicular to the 
boundary, parallel to the longitudinal scattering plane. Thus as 
the longitudinal metric scales with $\ell_\parallel^2$ the normalisation
condition for $n$ implies that it will also scale with $\ell_\parallel$. 
Thus,
\be
\nabla_\mu n^\mu = \frac{\nabla_\a n^\a}{\ell_\parallel} + 
\frac{\nabla_i n^i}{\ell_\perp}.
\ee
The exterior curvature term of the action becomes
\be
\e^{d-4}S_{\partial M}&=& \frac{1}{\epsilon^2}\int\sqrt{\g}\,\nabla_\a n^\a +  
\frac{1}{\epsilon}\int\sqrt{\g}\,\nabla_i n^i.
\ee
As claimed in the text there is no additional contribution
to the boundary action coming from the exterior curvature. 

\section{Classical solutions}\label{appB3}

In this appendix we give some more details on how to solve the equations 
of motion for the background, coming from the ${1\over\epsilon^2}$ part of 
the 
action \eq{Sperp0}.

We rewrite the action \eq{Sperp} in the following form (now 
concentrating on the two-dimensional covariant part),
\be
S=-{1\over2}\int\@{-g}\left(g^{\a\b}\pa_\a\phi\pa_\b\phi+
\L\phi^2-{1\over2}\phi^2R[g]\right).
\ee
This action belongs to the class of actions considered in \cite{BOL}, 
with Lagrangian of the form
\be
L=\@{-g}\left(g^{\a\b}\pa_\a\phi\pa_\b\phi-\l\phi^{2k}-Q\phi^2R[g]\right),
\ee
with the obvious values $k=1$, $\l=-{(d-2)\over2(d-3)}\L$, 
$Q={(d-2)\over4(d-3)}$. 

As argued in the main text, we consider static metrics of the form 
\eq{longmetric}. The lagrangian (with $k=1$) then reduces to the 
particle Lagrangian
\be
L={1\over g}\left(e{\phi'}^2-4Qe'\phi\phi'\right)-\l ge\phi^2.
\ee
The prime denotes derivatives with respect to $x$. It is obvious that 
the field $g$ does not contribute to the dynamics - the equation of 
motion for $g$ is simply an expression of reparameterization invariance 
in the spatial co-ordinate. In fact, all the 
$g$-depenence disappears from the equations of motion if we define a new 
variable $r=\int_0^x\dd x'\,g(x')$. We then get
\be
-2Q\,{\ddot e\over e} +{\dot\phi\dot e\over\phi e}
+{\ddot\phi\over\phi}+\l &=&0\nn
{\ddot\phi\over\phi}+\left(1+{1\over4Q}\right)
\left({\dot\phi\over\phi}\right)^2-
{\l\over4Q}&=&0\nn
{\dot\phi\over\phi}\left({\dot\phi\over\phi}-4Q{\dot e\over e}\right)+
\l&=&0,
\le{lagreom}
and the dots denote derivatives with respect to $r$.

Substituting this solution one has for the curvature

\be
R[g]=-2a^2 \left[1+{3\gamma\over 4Q}+ {\gamma (\gamma-4Q) \over 16Q^2 } 
\left({A-Be^{-2ar}\over A+Be^{-2ar}}\right)^2    \right]
\le{curvature}
where 
\be
a^2={ \lambda \over 4Q\gamma}.
\ee
We see from \eq{curvature} that among various solutions we also have
the case in which the curvature is constant if either $A=0$ or
$B=0$.

Since both cases differ only by a co-ordinate transformation, we choose 
$B=0$. 
The longitudinal metric $g_{\alpha \beta}$ is then
\be
\dd s^2=-(aCA^{q})^2 e^{ 2aqr}\dd t^2+\dd r^2.
\ee
where
\be
q=1+{ \gamma \over 4Q}
\ee

This is indeed the AdS$_2$ metric with the proper warp factor growing 
linearly in the radial co-ordinate. Our co-ordinates, however, do not 
cover the whole of AdS. One finds global co-ordinates by defining
\be
e^{aqr}=\cos\rho,
\ee
where $0\leq\rho\leq\pi/2$.

The curvature $R[g]=-{2\ddot e\over e}$ obviously simplifies and becomes
\be
R[g]=-{\lambda(4Q+\gamma) \over 8 Q^2 \gamma }
\ee
Furthermore, those solutions with $A$ and $B$ non-zero will be analogous 
in structure 
to AdS$_2$/Schwarzschild geometries, though the metric will have
a different functional form due to the presence of the non-trivial scalar 
field. 

In $d=3$ there are small modifications due to the appearence of several
$(d-3)$ factors in the general solutions. We can easily proceed here as 
follows. The one-dimensional form of the action is:
\be
L={e'{\phi^2}'\over g}-\L eg\phi^2,
\ee
and so the equations of motion reduce to
\be
\ddot\phi^2+\L\phi^2&=&0\nn
\ddot e+\L e&=&0\nn
{\dot\phi^2\dot e\over\phi^2 e}+\L&=&0,
\ee 
after reabsorbing the non-dynamical field $g$ in the definition 
of the parameter $r$, as before. 

{\bf Global structure of the solutions}

The metric
\be
\dd s^2=-e(r)^2\dd t^2+\dd r^2
\ee
has a horizon when $e(r)=0$. There are two possible locations of this 
horizon, depending on the relative sign of the initial conditions $A$ and 
$B$.

For $B/A>0$, $e(r)$ has a simple zero. With the following rescalings 
of the co-ordinates,
\be
r&=&\sqrt{Q\g\over\l}\log B/A+\et\nn
t&=&{4Q\g\over C\l}(4AB)^{-\g/8Q-1/2}\tau
\ee
the metric near the horizon  is simply the Rindler space metric,
\be
\dd s^2=-\et^2\dd\tau^2+\dd\et^2,
\le{Rindler}
and so locally the space is flat.

For $B/A<0$, we rescale the co-ordinates as follows:
\be
r&=&\sqrt{Q\g\over\l}\log|B/A|+\et\nn
t&=&({\l|AB|\over Q\g})^{-\g/8Q-1/2}\tau,
\ee
and we find the metric
\be
\dd s^2=-\et^{\g/2Q}\dd\tau^2+\dd\et^2
\le{metric2}
with curvature $R=-{\g(\g-4Q)\over8Q^2\et^2}$.

\chapter{Einstein Spaces and the Holographic Stress-Tensor}\label{appC}

\section{Asymptotic solution of Einstein's equations} \label{EinSol}

In this appendix we collect the results for the solution of the
equations (\ref{eqn}) up to the order we are interested in.

{}From the first equation in (\ref{eqn}) one determines 
the coefficients $g_{(n)}$, $n \neq d$, in terms of $g_{(0)}$.
For our purpose we only need $g_{(2)}$ and $g_{(4)}$.
There are given by
\bea \label{gexp}
g_{(2)}{}_{ij} & = & \frac{1}{d - 2} \left( R_{ij} - \frac{1}{2 (d - 1)} 
R\, \gzero{}_{ij} \right), \cr
g_{(4)}{}_{ij} & = & \frac{1}{d - 4} \left( - \frac{1}{8 (d - 1)} D_i
D_j R + \frac{1}{4 (d - 2)} D_k D^k R_{ij} \right . \cr
& & - \frac{1}{8 (d - 1) (d - 2)} D_k D^k R \gzero{}_{ij} - \frac{1}{2 (d - 
2)}
R^{kl} R_{ikjl} \cr
& & + \frac{d - 4}{2 (d - 2)^2} R_i{}^k R_{kj} + \frac{1}{(d - 1)(d -
2)^2} R R_{ij} \cr
& & \left. + \frac{1}{4 (d - 2)^2} R^{kl} R_{kl} \gzero{}_{ij} - \frac{3 
d}{16
(d - 1)^2 (d - 2)^2} R^2 \gzero{}_{ij} \right) .
\eea
All curvature expressions and covariant derivatives here are evaluated in the 
metric $g_{(0)}$. Thus, the above coefficients $g_{(n)}$ are functions of 
$g_{(0)}$ through the Riemann tensor and its derivatives. The expressions for 
$g_{(n)}$ are singular when $n=d$. One can obtain the trace and the 
divergence of $g_{(n)}$ for any $n$ from the last two equations in 
(\ref{eqn}). Explicitly, 
\bea
&&\Tr\, g_{(4)} = {1 \over 4}\, \Tr\, g_{(2)}^2, \qquad 
\Tr\, g_{(6)}= {2 \over 3}\, 
\Tr\, g_{(2)} g_{(4)}
-{1 \over 6}\, \Tr\, g_{(2)}^3, \nonu
&&\Tr\, g_{(3)}=0, \qquad  \qquad \ \ \ \Tr\, g_{(5)}=0,
\eea
and 
\bea
&& \nabla^i g_{(2)ij} = \nabla^i A_{(2) ij}, \qquad 
\nabla^i g_{(3)ij} =0, \qquad
\nabla^i g_{(4) ij} = \nabla^i A_{(4) ij} \nonu
&&\nabla^i g_{(5)ij} =0, \qquad
\nabla^i g_{(6) ij}= \nabla^i A_{(6) ij}
+ {1 \over 6} \Tr\, (g_{(4)} \nabla_j g_{(2)})\, ,
\label{gd}
\eea
where 
\bea
A_{(2) ij}&=& g_{(0) ij} \Tr\, g_{(2)}, \label{Ad} \\
A_{(4) ij} &=& - {1 \over 8}[\Tr\, g_{(2)}^2 - (\Tr\, g_{(2)})^2]\, g_{(0) 
ij} 
+ \half (g_{(2)}^2)_{ij} - {1 \over 4}\, g_{(2) ij}\, \Tr\, g_{(2)}, \nonu
A_{(6) ij} &=& {1 \over 3} \left(
2(g_{(2)} g_{(4)})_{ij}+(g_{(4)} g_{(2)})_{ij}-(g_{(2)}^3)_{ij} 
+ {1 \over 8}\,[\Tr\, g_{(2)}^2 - (\Tr\, g_{(2)})^2]\, g_{(2) ij} \right. 
\nonu
&-&\Tr\, g_{(2)}\,[g_{(4)ij} - \half (g_{(2)}^2)_{ij}] \nn
&-&\left.[{1 \over 8} \Tr\, g_{(2)}^2 \Tr\, g_{(2)} - {1 \over 24} (\Tr\, 
g_{(2)})^3
-{1 \over 6} \Tr\, g_{(2)}^3
+{1\over 2} \Tr \, (g_{(2)}g_{(4)})]\,g_{(0) ij} \right)\, . \nonumber
\eea

For even $n=d$ the first equation in (\ref{eqn}) determines  
the coefficients $h_{(d)}$. They are given by
\bea
h_{(2)ij}&=&0  \label{h2}, \\
h_{(4)ij}
&=&{1\over 2}g^2_{(2)ij}-{1\over 8}g_{(0)ij}\Tr \, g^2_{(2)}+{1\over 8}
(\nabla^k\nabla_ig_{(2)jk}+\nabla^k\nabla_jg_{(2)ik}
-\nabla^2 g_{(2)ij}-\nabla_i\nabla_j
\Tr \, g_{(2)}) \label{h4} \\
&=&{1 \over 8} R_{ikjl} R^{kl} + {1 \over 48} \nabla_i \nabla_j R
-{1 \over 16} \nabla^2 R_{ij} -{1 \over 24} R R_{ij} 
+ ({1 \over 96} \nabla^2 R + {1 \over 96} R^2 -{1 \over 32} R_{kl}R^{kl})
g_{(0) ij},\,
\nonu
h_{(6)ij}&=&{2\over 3}(g_{(4)}g_{(2)}+g_{(2)}g_{(4)})_{ij}
-{1\over 3}g^3_{(2)ij}-{1\over 6}g_{(4)ij}\Tr\, g_{(2)}\nn
&+&{1\over 6}g_{(0)ij}(3\Tr g_{(6)}-3\Tr g_{(2)}g_{(4)}+\Tr g^3_{(2)}) 
\nonumber \\
&&-{1\over12}[-{1 \over 4} \nabla_i \nabla_j \Tr g_{(2)}^2 
-\nabla^k \nabla_i g_{(4)jk} -\nabla^k \nabla_j g_{(4)ik}
+ \nabla^2 g_{(4)ij} \nonu
&&+g_{(2)}^{kl} 
[\nabla_l \nabla_i g_{(2)jk} +\nabla_l \nabla_j g_{(2)ik} -\nabla_l
\nabla_k g_{(2)ij}] \nonu
&&+{1 \over 2} \nabla^k \Tr g_{(2)} 
(\nabla_i g_{(2)jk} +\nabla_j g_{(2)ik} - \nabla_k g_{(2)ij}) \nonu
&&+\half \nabla_i g_{(2)kl} \nabla_j g_{(2)}^{kl} 
+\nabla_k g_{(2)il} \nabla^l g_{(2)j}{}^{k}
-\nabla_k g_{(2)il} \nabla^k g_{(2)j}{}^{l}]. \label{h6}
\end{eqnarray}

\section{Divergences in terms of the induced metric}
\label{div-ind}

In this appendix we rewrite the divergent terms of the 
regularised action in terms of the induced metric at $\r=\e$.
This is needed in order to derive the contribution of 
the counter-terms to the stress-energy tensor.

The coefficients $a_{(n)}$ of the divergent terms in the regulated
action (\ref{regaction1}) are given by 
\bea
&&a_{(0)}=2(1-d),\nn
&&a_{(2)}=b_{(2)}(d)\, \Tr\, g_{(2)},\nn
&&a_{(4)}=b_{(4)}(d)\,
[(\Tr\, g_{(2)})^2 - \Tr\, g_{(2)}^2], \nn
&&a_{(6)}=\left( {1 \over 8}\,\Tr\, g_{(2)}^3
-{3 \over 8}\, \Tr\, g_{(2)} \Tr\, g_{(2)}^2
+{1 \over 2}\, \Tr\, g_{(2)}^3
- \Tr\, g_{(2)} g_{(4)} \right),
\eea  
where $a_{(6)}$ is only valid in six dimensions and the numerical 
coefficients in $a_{(2)}$ and $a_{(4)}$ are given by
\be
b_{(2)}(d\neq2)&=&-{(d-4)(d-1) \over d-2},\nn
b_{(2)}(d=2)&=&1,\nn
b_{(4)}(d\neq4)&=&{-d^2 + 9d -16 \over 4 (d-4)}, \nn
b_{(4)}(d=4)&=&\half.
\eea
Notice that the coefficients $a_{(n)}$ are proportional 
to the expression for the conformal anomaly (in terms of $g_{(n)}$)
in dimension $d=n$ \cite{HS}.

The counter-terms can be rewritten in terms of the induced metric
by inverting the relation between $\g$ and $\gzero$ perturbatively in $\e$. 
One finds
\bea \label{convert}
\sqrt{\gzero}&=&
\e^{d/2} \left(1 - \half\, \e\, \Tr\, \gi g_{(2)}
+{1 \over 8}\, \e^2\, [(\Tr\, \gi g_{(2)})^2 + \Tr\, (\gi g_{(2)})^2] 
+ {\cal O}(\e^3) \right)\sqrt{\g}, \nonu
\Tr \,g_{(2)} &=& 
{1 \over 2 (d-1)} {1 \over \e} \left(R[\g] + 
{1 \over d-2} (R_{ij}[\g] R^{ij}[\g] - {1 \over 2(d-1)} R^2[\g])
+{\cal O}(R[\g]^3) \right), \nonu
\Tr\, g_{(2)}^2 &=& {1 \over \e^2} {1 \over (d-2)^2} 
\left(R_{ij}[\g] R^{ij}[\g] + {-3d+4 \over 4(d-1)^2} R^2[\g]
+{\cal O}(R[\g]^3) \right).
\eea
The terms cubic in curvatures in (\ref{convert}) give vanishing 
contribution in (\ref{tij1}) up to six dimensions.

Putting everything together we obtain that the counter-terms, 
rewritten in terms of the induced metric, are given by
\be \label{ct}
S^{\sm{ct}}&=&-{1 \over 16 \p \GN} \int_{\r=\e} 
\sqrt{\g}\left[2(1-d) + {1 \over d-2} R\right.\nn
&-&\left.{1 \over (d-4) (d-2)^2}
(R_{ij} R^{ij} - {d \over 4 (d-1)} R^2) - \log \e\, a_{(d)} + ...\right],
\eea
where all quantities are now in terms of the induced metric, including the 
one in the logarithmic divergence. These are exactly the counter-terms 
in \cite{BK,EJM,KLS} except that these authors did not include the 
logarithmic divergence. Equation (\ref{ct}) should be understood as 
containing only divergent counter-terms in each dimension. This means that 
in even dimension $d=2k$ one should include only the first $k$ counter-terms 
and the logarithmic one. In odd $d=2k+1$, only the first 
$k+1$ counter-terms should be included. The logarithmic counter-terms
appear only for $d$ even. The counter-terms in (\ref{ct})
render the renormalised action finite up to $d=6$. This covers 
all cases relevant for the AdS/CFT correspondence. It is straightforward
but tedious to compute the necessary counter-terms for $d>6$. From 
(\ref{ct}) one straightforwardly obtains (\ref{counterT}).
 
\section{Relation between $h_{(d)}$ and the conformal anomaly $a_{(d)}$}
\label{h-a}

We show in this appendix that the tensor $h_{(d)}$ appearing in the expansion
of the metric in (\ref{coord}) when $d$ is even is a multiple
of the stress tensor derived from the action
$\int a_{(d)}$. ($a_{(d)}$ is, up to a constant, the holographic 
conformal anomaly).

This can be shown by deriving the stress-energy tensor of the 
regulated theory at $\r=\e$ in two ways and then comparing the
results. In the first derivation one starts from 
(\ref{regaction}) and obtains the regulated stress-energy tensor 
as in (\ref{regtij}). Expanding $T_{ij}^{\sm{reg}}[\g]$ in $\e$
(keeping $g_{(0)}$ fixed) we find that there is a logarithmic divergence,
\be
T_{ij}^{\sm{reg}}[\g;\log] = {1 \over 8 \p \GN} \log \e \
({3\over 2 }d-1) h_{(d) ij}.
\eea
On the other hand, one can derive $T_{ij}^{\sm{reg}}[\g]$ starting from 
(\ref{regaction1}). One has to first rewrite the terms in
(\ref{regaction1}) in terms of the induced metric. This is done 
in the previous appendix. Once $T_{ij}^{\sm{reg}}[\g]$ has been
derived, we expand in $\e$. We find the following logarithmic
divergence:
\be
T_{ij}^{\sm{reg}}[\g;\log] = 
{1 \over 8 \p \GN} \log \e \left( (1-d)h_{(d)ij}-T_{ij}^a,
\right),
\eea
where $T_{ij}^a$ is the stress-energy tensor of the action 
$\int \dd^d x\, \sqrt{\det g_{(0)}}\, a_{(d)}$.
If follows that
\be
h_{(d) ij}=-{2 \over d} T_{ij}^a.
\eea
We have also explicitly verified this relation by brute-force computation
in $d=4$.

\section{Asymptotic solution of the scalar field equation}
\label{as-sc}

We give here the first two orders of the solution of
the equation (\ref{phieq})
\bea \label{mattersol}
&&\f_{(2)}= {1 \over 2 (2 \D -d -2)} 
\left(\Box_0 \f_{(0)} + (d-\D) \f_{(0)} \Tr\, g_{(2)}\right), \nonu
&&\f_{(4)}= {1 \over 4 (2 \D -d -4)} 
\left(\Box_0 \f_{(2)} - 2\, \Tr\, g_{(2)} \f_{(2)}
-\half (d-\D) \,[\Tr\, g_{(2)}^2\, \f_{(0)} - 2 \Tr\, g_{(2)}\, \f_{(2)}] 
\right. \nonu
&& \hspace{3.5cm}\left. 
-{1 \over \sqrt{g_{(0)}}}\, \pa_\m (\sqrt{g_{(0)}}\, g_{(2)}^{\m \n} 
\pa_\n \f_{(0)})
+ \half \pa^i \Tr\, g_{(2)} \pa_j \f_{(0)} \right)\, ,
\eea
where in $\Box_0$ the covariant derivatives are with respect to $g_{(0)}$. 

If $2 \D -d -2k=0$ one needs to introduce a logarithmic term
in order for the equations to have a solution, as discussed 
in the main text. For instance, when $\D=\half d +1$, $\f_{(2)}$
is undetermined, but instead one obtains for the coefficient of the 
logarithmic term,
\be
\psi_{(2)}=-{1 \over 4} 
\left(\Box_0 \f_{(0)} + ({d \over 2}-1)\, \f_{(0)} \Tr\, g_{(2)}\right).
\eea


\chapter*{Samenvatting}
\addcontentsline{toc}{chapter}{Samenvatting}

Deze samenvatting is voor een groot deel gebaseerd op \cite{NTvN}.

De wens de theorie\"{e}n van heelal en atoom bij elkaar te brengen geeft al 
enige tijd vorm aan een aanzienlijk deel van de moderne theoretische fysica. 
Het zijn twee theorie\"{e}n die mathematisch geen verband met elkaar lijken 
te hebben -- en vooralsnog elkaar misschien zelfs uitsluiten. Toch is men op 
zoek naar een eenduidige beschrijving van de natuur zoals zij zich zou 
gedragen bij energie\"{e}n (orde $10^{22}$ MeV) waar het onderscheid tussen 
zwart gat en elementair deeltje verdwijnt. Zowel quantum- als 
gravitatie-effecten kunnen bij zulke hoge energie\"en niet verwaarloosd 
worden en dus zal deze theorie waarschijnlijk karakteristieke elementen van 
de quantummechanica en de relativiteitstheorie moeten bevatten. In de juiste 
limiet zou ze deze theorie\"{e}n moeten reproduceren.

In de jaren zeventig werd al vrij snel duidelijk dat een gequantiseerde 
veldentheorie van de zwaartekracht, geformuleerd op de manier waarop ook de 
kerninteracties tussen kerndeeltjes beschreven worden, niet de gewenste 
theorie kon zijn. Dit model bleek niet renormeerbaar te zijn, dat wil zeggen: 
de wiskundige methoden uit de veldentheorie om oneindigheden uit fysische 
voorspellingen te weren, zullen in deze theorie tekortschieten. In de 
snaartheorie, die in deze periode voor het eerst geformuleerd werd, is niet 
het puntdeeltje maar de snaar het fundamentele object, waardoor de 
divergenties vermeden kunnen worden. Snaren verschillen van puntdeeltjes in 
die zin dat ze uitgestrekt zijn over \'{e}\'{e}n dimensie.

Een tweede probleem voor een quantumveldentheorie van gravitatie is dat er 
volgens de klassieke theorie van Einstein objecten bestaan met een horizon: 
zwarte gaten. Wanneer iemand in een zwart gat valt, is de horizon de laatste 
plaats van waaruit hij een noodkreet kan slaken die ons zal bereiken. Voorbij 
de horizon is geen teugkeer mogelijk. Dit althans volgens Einstein, want in 
1974 ontdekte Stephen Hawking dat quantummechanica ervoor zorgt dat deze 
zwarte gaten straling van een zeer lage frequentie uitzenden. De straling 
dankt zijn bestaan aan de horizon, die de ter plaatse zijnde deeltjes en 
anti-deeltjes -- die in paren uit het vacuum ontstaan -- van elkaar scheidt; 
de anti-deeltjes vallen in het gat, terwijl de vrijkomende deeltjes de 
Hawkingstraling vormen. Hawkingstraling is dus een quantumeffect waarbij de 
zwaartekracht direct betrokken is. Merkwaardig is dat het spectrum van deze 
straling thermisch is. Er is geen eenduidige golffunctie te bedenken die de 
toestand van de straling beschrijft, aldus Hawking. Wanneer een zuivere 
golffunctie implodeert tot een zwart gat, dat daarna thermisch gaat stralen, 
zal het systeem vervolgens alleen te beschrijven zijn met een 
dichtheidsmatrix. Deze situatie handhaaft zich als het zwarte gat volledig 
verdampt is, dan resteert immers niets dan thermische straling.

Dit lijkt een voorbeeld van een niet-unitaire evolutie, een meer algemene 
evolutie dan we kennen uit Schr\"{o}dingers vergelijking, want blijkbaar 
maakt het zwarte gat het mogelijk dat een zuivere toestand overgaat in een 
gemengde toestand. Om het anders te zeggen: het zwarte gat vernietigt 
informatie. Bovenop de onzekerheid van Heisenberg kan nu ook niet meer 
voorspeld worden wat de toekomst van een zuivere toestand is (stel dat-ie 
tegen een zwart gat botst...). Dit bracht Hawking tot de conclusie dat alleen 
al de aanwezigheid van een horizon de wetten van de quantummechanica schendt: 
een beschrijving van een wereld met zwaartekracht (en dus met de mogelijkheid 
om horizons te hebben) is alleen mogelijk met toestandsmatrices.

Maar hebben we hier niet gewoon met een thermodynamische limiet te maken? Een 
limiet waarbij bepaalde interacties op microniveau over het hoofd gezien 
worden, zodat we wel bij een niet-unitaire evolutie uit moeten komen? Deze 
mogelijkheid is inderdaad nog steeds open, het is alleen moeilijk na te gaan 
waar precies in Hawkings berekening een middelingsprocedure is uitgevoerd. 
Hij lijkt van twee fundamentele theorie\"en te zijn uitgegaan, zonder dat er 
sprake is van het verwaarlozen van belangrijke wisselwerkingen. Toch heeft 
met name 't Hooft de laatste jaren getracht de vinger te leggen op datgene 
wat Hawking veronachtzaamd heeft.

Vrijwel alle beschrijvingen van zwarte gaten gaan ervan uit dat de entropie 
van het gat te identificeren is met de oppervlakte van zijn horizon. 
Snaartheorie heeft als enige deze entropieformule uit een microscopische 
beschrijving weten af te leiden. 

Snaartheorie is een theorie die zwaartekracht met quantummechanica tracht te 
verenigen door aan te nemen dat verschillende deeltjes trillingstoestanden 
zijn van een fundamentele snaar. Oorspronkelijk is snaartheorie ontstaan als 
een poging om te verklaren waarom quarks dicht op elkaar kunnen zitten. 
Alleen kwam men er al gauw achter dat de snaren bij veel hogere energie\"en 
moeten leven dan we in onze versnellers kunnen bereiken. In het spectrum van 
de snaren zit bijvoorbeeld ook het graviton, dat alleen zichtbaar is bij de 
Planckenergie.

De theorie bevatte niet alleen het graviton; zij voorspelde ook een deeltje 
waar men van af wilde, het tachyon: een deeltje dat zich sneller dan het 
licht voortbeweegt. De aanname van supersymmetrie, die deeltjes van 
verschillende spin aan elkaar relateert, elimineert dit deeltje uit het 
spectrum. Ook reduceert deze symmetrie het aantal dimensies van de theorie 
van 26 naar 10. Deze 10 dimensies zouden dan zo opgerold zijn dat onze 
vierdimensionale wereld overblijft, maar met welk mechanisme dit precies 
gebeurt, is nog steeds een open vraag. Recent heeft men begrepen dat een zeer 
interessante mogelijkheid ontstaat als men aaneemt dat de extra dimensies 
groot zijn (dus niet opgerold), maar een bijzondere geometrie hebben. Dit 
soort scenario's heten ``warped compactifications", kromgetrokken 
compactificaties of simpelweg gekromde compactificaties.

In snaartheorie bestaat een zwart gat uit $p$-branen. Dit zijn 
meer-dimensionale objecten waar snaren op kunnen eindigen (de $p$ slaat op de 
dimensie: een punt deeltje is dus een 0-braan, een snaar is een 1-braan, een 
membraan is een 2-braan, etc.). In de limiet waarbij de massadichteid van de 
snaren die op de branen vastgepind zijn naar oneindig gaat, terwijl hun massa 
constant blijft (ze worden dus uiterst kort), heeft de geometrie in de buurt 
van de braan de vorm van een zogenaamde anti-de Sitter ruimte (AdS): een lege 
ruimte waar de kosmologische constante negatief is (een contraherend heelal 
dus). De braan kan gezien worden als de rand van het anti-de Sitter heelal. 
Het verband van Maldacena zegt dat een veldentheorie die op de braan 
gedefinieerd is equivalent is met snaartheorie in de anti-de Sitter ruimte.

Voor lage energie\"en reduceert de snaartheorie tot de relativiteitstheorie 
van Einstein. Er is dus sprake van een identificatie tussen Einsteins 
gravitatietheorie in een contraherend heelal en een quantummechanische 
theorie op de vlakke rand van dit heelal. Met andere woorden, de variabelen 
die de zwaartekracht beschrijven kunnen op zo'n manier met elkaar 
gecombineerd en opgeschreven worden dat ze een quantummechanische 
veldentheorie beschrijven. Men spreekt dan ook over een ``woordenboek" die de 
twee theorie\"en aan elkaar relateert: als je de elementaire bouwstenen van 
de ene theorie weet, dan kun je ook door gebruik te maken van dit woordenboek 
een vertaling maken naar de variabelen in de andere theorie. Het handige van 
Maldacena's voorstel is dat berekeningen die in de veldentheorie moeilijk 
zijn, nu eenvoudiger berekend kunnen worden door ze in snaartheorie in AdS 
uit te voeren, en vice versa. Op deze manier kan de ene theorie 
voorspellingen doen over de fysica van de andere theorie.

Op het eerste gezicht lijkt het verband te gek om waar te kunnen zijn. Er 
worden twee theorie\"en aan elkaar gerelateerd die in verschillende dimensies 
leven, zoals een Yang-Millstheorie (de theorie van quarks is ook een 
Yang-Millstheorie) in vier dimensies en snaartheorie in tien dimensies. 
Bovendien bevat snaartheorie gravitatie terwijl de andere theorie op een 
vlakke ruimte leeft. In 1993 stelde 't Hooft dat een van de kenmerken van een 
theorie van quantumgravitatie moet zijn dat het aantal dimensies gereduceerd 
wordt. Zo zou het oppervlak van de horizon van een zwart gat alle informatie 
bevatten over wat zich in het volume binnen de horizon afspeelt. Het is dus 
niet verbazingwekkend dat de entropie, die een maat is voor de informatie die 
schuilgaat in een zwart gat, evenredig is met de oppervlakte en niet met het 
volume van het gat. Een theorie op de horizon van een gat zou dan opgevat 
kunnen worden als een holografische projectie van de theorie die nodig zou 
zijn om de fysica achter de horizon te kunnen beschrijven. Het verband van 
Maldacena (ook AdS/CFT-verband genaamd, CFT staat voor de ``conformal field 
theory" op de rand) stelt nu dat de informatie bevat door snaartheorie in de 
anti-de Sitter-ruimte evengoed weergegeven kan worden door een veldentheorie 
op de rand van zo'n ruimte. Het voorstel van Maldacena is dus ook een 
voorbeeld van een holografische theorie. Het begrijpen van hoe en waarom dit 
principe werkt is daarom een uiterst belangrijke kwestie.

Holografie is echter niet alleen in snaartheorie aanwezig. Toch zijn er 
behalve het AdS/CFT verband van Maldacena niet veel meer voorbeelden van 
holografische theorie\"{e}n. Een van deze voorbeelden betreft de 
eigenschappen van deeltjes die in de buurt van een zwart gat wisselwerken. 
Voor het beschrijven van deze deeltjes kan men volstaan met de zogenaamde 
eikonale benadering. In deze benadering wordt aangenomen dat deeltjes 
frontaal en op extreem hoge energie\"{e}n tegen elkaar botsen. Kenmerkend 
voor dit soort botsingen is dat de zwaartekracht de dominante kracht wordt, 
en alle andere krachten verwaarloosd kunnen worden. 't Hooft heeft aangetoond 
dat de theorie die men in deze benadering krijgt, een 2-dimensionale theorie 
is. E. en H. Verlinde hebben laten zien dat dit resultaat begrepen kan worden 
vanuit een vereenvoudiging van het actieprincipe dat deze wisselwerkingen 
beschrijft. In het eikonale regime is de typische longitudinale lengteschaal 
(langs de as van de bosting) klein, van de orde van de Plancklengte, terwijl 
de transversale fluctuaties veel groter zijn en processen in deze richting 
veel langzamer verlopen. In deze benadering kan men dan laten zien dat de 
theorie van Einstein tot een topologische theorie reduceert. E. en H. 
Verlinde hebben ook aangetoond dat de amplitudes die men met deze theorie 
krijgt, overeenkomen met de door 't Hooft eerder verkregen resultaten. Een 
van de interessante eigenschappen van de theorie op de rand is dat de 
co\"{o}rdinaten tussen verschillende deeltjes niet-commutatief zijn.

Dit proefschrift richt het vizier op een aantal holografische eigenschappen 
van zowel klassieke als quantumgravitatie en snaartheorie. 

In hoofdstuk \ref{HEscattering} van dit proefschrift bestuderen we diverse 
eigenschappen van het model van 't Hooft, zoals covariantie. We laten zien 
dat het mogelijk is de transversale effecten mee te nemen, die in de eikonale 
benadering verwaarloosd worden. Dit geeft een interessante niet-commutatieve 
algebra tussen operatoren. In 2+1 dimensies kan men bovendien laten zien dat 
het meenemen van transversale effecten equivalent is met het covariant 
formuleren van de theorie. We hebben de implicaties van de zwaartekracht voor 
de tweede quantisatie van deeltjes bestudeerd, en gevonden dat ook velden die 
veel deeltjes beschrijven niet-commutatief worden, dat wil zeggen, de 
volgorde waarin deze fysische grootheden gemeten worden maakt uit, op 
dezelfde wijze als in de gewone quantummechanica metingen van plaats en 
impuls elkaar be\"invloeden. Dit komt overeen met eerdere resultaten van E. 
en H. Verlinde in de context van zwarte gaten, maar het mechanisme waardoor 
de niet-commutativiteit ontstaat is verschillend.

De onderliggende motivatie voor het werk gepresenteerd in hoofdstuk \ref{GSM} 
is dat men graag de holografisch duale theorie\"{e}n van 't Hooft en van 
Maldacena met elkaar zou willen vergelijken. Dit lijkt belangrijk voor een 
goed begrip van beide theorie\"{e}n. Het ligt daarom voor de hand om het 
eikonale regime van gravitatie te beschouwen in ruimtes met een negatieve 
kosmologische constante. In dat hoofdstuk wordt een veralgemenisering gegeven 
van de afleiding van E. en H. Verlinde dat gravitatie in de eikonale limiet 
topologisch wordt. We hebben gevonden dat de theorie van Einstein met 
willekeurige waarde van de kosmologische constante inderdaad topologisch is 
in de eikonale limiet. De oplossingen van de theorie op de rand zijn ook 
gerelateerd aan de schokgolven gevonden door Horowitz en Itzhaki. Het zou 
zeer interessant zijn als een expliciet verband gelegd zou kunnen worden 
tussen de zogenaamde ``lichtkegel toestanden" die duaal zijn aan een 
schokgolf in AdS, en de duale theorie die wij in dit proefschrift bespreken.

In hoofdstuk \ref{reconstruction} bestuderen we holografie in het AdS/CFT 
verband. We scherpen het bovengenoemde ``woordenboek" tussen de twee 
theorie\"en aan. We laten zien op welke manier de informatie over de 
geometrie van de anti-de Sitter ruimte en de andere velden die erop leven, 
gecodeerd is in de CFT op de rand van AdS. We ontwikkelen ook een 
systematische methode om de actie te regulariseren en te renormeren.

Deze resultaten worden gebruikt in hoofdstuk \ref{warped}, waar we gekromde 
compactificaties bestuderen. We laten zien dat de $(d{+}1)$-dimensionale 
Einstein vergelijkingen samen met een verbindingsvoorwaarde de 
$d$-dimensionale Einstein vergelijkingen op de braan opleveren met een 
specifieke energie-impulstensor. Dit resultaat is geldig voor willekeurige 
waarde van de kosmologische constante. Voor ruimtes die asymptotisch AdS 
zijn, is de waarde van deze energie-impulstensor gelijk aan de 
energie-impulstensor op de braan plus die van een CFT die op de braan leeft. 
Door de resultaten van hoofdstuk \ref{reconstruction} toe te passen krijgen 
we ook specifieke voorspellingen voor de hogere-orde correcties op de 
Einsteinvergelijkingen.

De afgelopen jaren zijn er snelle ontwikkelingen gekomen op het gebied van 
holografie en hebben we veel meer inzicht gekregen in dit kennelijk 
fundamentele beginsel. Toch hebben we nog geen antwoord op vragen zoals: wat 
is de onderliggende reden waarom de dualiteit werkt? Hoe kan causaliteit 
gerespecteerd worden bij de projectie van een $(d{+}1)$-dimensionale naar een 
$d$-dimensionale theorie en welke rol speelt de zwaartekracht hierin? Naast 
deze fundamentele vragen zijn er uiteraard nog veel open vragen van meer 
technische aard. Het is duidelijk dat veel meer onderzoek nodig is om al deze 
vragen naar tevredenheid te beantwoorden.

\chapter*{Dankwoord}
\addcontentsline{toc}{chapter}{Dankwoord}

Ik wil hier een aantal mensen bedanken zonder wiens steun dit proefschrift 
nooit het daglicht zou hebben gezien, en ook anderen die hebben bijgedragen 
aan het tot stand komen van dit proefschrift.\\
\\
Allereerst gaat mijn dank uit naar mijn promotor, Gerard 't Hooft. Ik heb 
veel geleerd van je kritische opmerkingen en van je vermogen om direct tot de 
kern van een probleem door te dringen. Ik wil je ook bedanken voor de 
vrijheid die je me bij het onderzoek geboden hebt en tegelijk ook omdat je me 
aangespoord hebt om mij in de stringtheorie te verdiepen. Ook wil ik de 
andere hoogleraren van het ITF/Spinoza Instituut bedanken voor de prettige 
wetenschappelijke en sociale sfeer, ondanks de tijdelijke niet-localiteit van 
ons instituut... Gelukkig zijn we nu met de laatste verhuizing weer in een 
eigentoestand van de plaatsoperator terechtgekomen... Heel in het bijzonder 
wil ik bedanken mijn collegae AIO's, OIO's en postdocs van het ITF/Spinoza, 
in het bijzonder Bartjan en Vladimir met wie ik de kamer heb gedeeld, Ivo 
voor handige tips aangaande onder meer reizen en hotels, en Zoltan die voor 
sportieve ontspanning zorgde. Ook Biene, Natasja, Geertje en Leonie wil ik 
danken voor hun inzet en enthousiasme. Jullie hebben er allemaal toe 
bijgedragen dat de sfeer op ons instituut echt aangenaam is. Het plezier 
waarmee ik in deze jaren in Utrecht heb gewerkt zal ik dan ook niet gauw 
vergeten.\\
\\
Special thanks to my collaborators: Giovanni Arcioni, Martin O'Loughlin, 
Annamaria Sinkovics, Kostas Skenderis and Sergey Solodukhin, without whom 
this thesis would have looked very different. Gio and Martin, I am especially 
grateful to you for your enthusiasm in the long-winded road (as someone 
said...) of our collaboration which led us through many nice and unexplored 
woods, rivers and -- why not -- desolate places, to a nice mountain whose top 
we just caught the sight of. Also I learned much from your ability to come up 
with new ideas every time. Kostas, I have learned a lot working with you, and 
later also with Sergey. I thank you both for your patience with all my 
questions and for the fact that most times you knew the right answer... I 
remember the period that you spent at the Spinoza Institute as a particularly 
nice time.  Ani, I have also learned very much from your insight and 
precision, though lately I was not always able to keep up with your working 
pace...\\
\\
Ook wil ik in het bijzonder Erik en Herman Verlinde danken voor de vele 
discussies en suggesties op verschillende momenten in de loop van mijn 
promotie. Met je onbegrensde enthousiasme weten jullie studenten en 
promovendi altijd weer voor de fysica warm te doen lopen.\\
\\
Also special thanks to Soo-Jong Rey for his invitation to visit Seoul 
National University, and for the very interesting discussions we had there. I 
also thank Tsunehide Kuroki for the nice discussions, and the string theory 
group for the hospitality during my stay in Seoul.\\
\\
I am grateful to a number of other people for interesting discussions at 
various stages of this thesis: Vijay Balasubramanian, Steve Giddings, Sunny 
Itzhaki, and Dan Kabat.\\
\\
Jeroen van Dongen wil ik danken voor boeiende discussies op het raakvlak van 
theoretische natuurkunde en filosofie en voor de prettige samenwerking 
tijdens het schrijven van onze gezamenlijke artikelen voor NTvN (en ik heb 
dankbaar gebruik gemaakt van \'{e}\'{e}n daarvan, bij de Nederlandse 
samenvatting van dit proefschrift...). Ik hoop dat je interesse voor 
holografie en snaartheorie niet zal verminderen...\\
\\
Alle inwoners en oud-inwoners van Studentenhuis Lepelenburg wil ik in het 
bijzonder danken voor de altijd gezellige en ontspannende sfeer, die 
onontbeerlijk is wil de promovendus niet onder het zware gewicht van de 
wetenschap bezwijken... Verder zijn er natuurlijk een heleboel vrienden aan 
wie ik dank schuldig ben voor hun steun in de afgelopen vier jaar. In het 
bijzonder wil ik noemen mijn paranimfen, Jasper Berben en Machiel Kleemans. 
Jasper, we hebben met veel lol veel dingen samen gedaan: musiceren, drinken, 
vergaderen, organiseren, discussi\"{e}ren, schrijven, repeteren en noem maar 
op... Machiel, ik heb bijzonder veel plezier beleefd aan onze discussies over 
filosofie, natuurkunde en over van alles en nog wat... Ook wil ik iedereen 
die meegedaan heeft aan de musical bedanken, en in het bijzonder degenen die 
er vroeg bij waren: Jasper, Wilmer, Jan Jaap, Tom, Robert, Ester, Liedewij, 
Bart, Guy en Anna. {\it Op de goede afloop heffen wij het glas... Dat het U 
wel moge bekomen...!}\\
\\
A last word for my parents, and my family, for all their care and 
encouragement. My father has contributed to this thesis in a special way with 
his ever stimulating and experienced academic advice.

\chapter*{Curriculum Vitae}
\addcontentsline{toc}{chapter}{Curriculum Vitae}

De auteur werd geboren op 23 november 1973 te Barcelona. Hij bezocht het 
Viar\'{o} College in Sant Cugat del Vall\`{e}s (Barcelona) waar hij in 1991 
het Spaanse VWO- en het International Baccalaureate-diploma behaalde.\\
In augustus 1991 verhuisde hij naar Nederland en begon hij de met studie 
natuurkunde aan de Universiteit Utrecht. In 1996 studeerde hij af bij de 
vakgroep Theoretische Fysica met een scriptie over zwarte gaten onder 
begeleiding van Prof. G. 't Hooft.\\
Aansluitend daarop zette hij in 1997 zijn onderzoek naar zwarte gaten en 
stringtheorie voort in de vorm van een promotieonderzoek, eveneens onder 
begeleiding van Prof. G. 't Hooft. Dit onderzoek heeft geleid tot de 
dissertatie die u voor u ziet. Grote delen van dit onderzoek zijn verricht in 
samenwerking met onderzoekers uit buitenlandse instituten. Tijdens zijn 
promotie heeft de auteur geassisteerd bij de werkcolleges van de vakken 
Quantummechanica II, Voortgezette Klassieke Mechanica en Thermische en 
Statistische Fysica II. Hij heeft ook deelgenomen aan internationale 
congressen, workshops en scholen en daarbij ook diverse voordrachten gehouden 
over zijn eigen onderzoek. Ook heeft hij voordrachten gegeven in onder andere 
Boston, Princeton, New York, Philadelphia, Chicago, Berkeley, Santa Barbara 
en Los Angeles. In 2000 heeft hij de stringtheorie groep van Seoul National 
University (Seoul, Korea) bezocht en daar ook een reeks lezingen gehouden 
over ``High-Energy Scattering". Tijdens zijn studie- en promotieperiode is 
hij actief geweest in het organiseren en verzorgen van diverse culturele en 
wetenschappelijke activiteiten, waaronder een lezing over zwarte gaten die 
hij voor Studium Generale Delft heeft gehouden. \\
In september 2001 vertrekt hij naar Los Angeles waar hij als postdoc zijn 
onderzoek zal voortzetten aan de University of California at Los Angeles.

\end{document}